\documentclass[twocolumn,twocolappendix,trackchanges]{aastex631}
\submitjournal{ApJ}

\begin{document}

\title{Constraining the Time of Gravitational Wave Emission from Core-Collapse Supernovae}

\newcommand{\Harvard}{Center for Astrophysics \textbar{} Harvard \& Smithsonian, 60 Garden Street, Cambridge, MA 02138-1516, USA}
\newcommand*{\PAMSU}{Department of Physics and Astronomy, Michigan State University, East Lansing, MI 48824, USA}
\newcommand*{\CMSE}{Department of Computational Mathematics, Science, and Engineering, Michigan State University, East Lansing, MI 48824, USA}
\newcommand*{\NSCL}{National Superconducting Cyclotron Laboratory, Michigan State University, East Lansing, MI 48824, USA}
\newcommand*{\JINA}{Joint Institute for Nuclear Astrophysics-Center for the Evolution of the Elements, Michigan State University, East Lansing, MI 48824, USA}
\newcommand*{\ER}{Embry Riddle University, 3700 Willow Creek Road, Prescott Arizona, 86301, USA}
\newcommand*{\UF}{University of Florida, Gainesville, FL 32611, USA}
\newcommand*{\UA}{Steward Observatory, University of Arizona, 933 North Cherry Avenue, Tucson, AZ 85721-0065, USA}

\author{K. Gill} \affiliation{\Harvard}
\author{G. Hosseinzadeh} \affiliation{\UA}
\author{E. Berger}    \affiliation{\Harvard}
\author{M. Zanolin} \affiliation{\ER}
\author{M. Szczepa\'nczyk} \affiliation{\UF}

%\date{\today}

\begin{abstract}
The advent of sensitive gravitational wave (GW) detectors, coupled with wide-field, high cadence optical time-domain surveys, raises the possibility of the first joint GW-electromagnetic (EM) detections of core-collapse supernovae (CCSNe). For targeted searches of GWs from CCSNe optical observations can be used to increase the sensitivity of the search by restricting the relevant time interval, defined here as the GW search window (GSW). The extent of the GSW is a critical factor in determining the achievable false alarm probability (FAP) for a triggered CCSN search. The ability to constrain the GSW from optical observations depends on how early a CCSN is detected, as well as the ability to model the early optical emission. Here we present several approaches to constrain the GSW, ranging in complexity from model-independent analytical fits of the early light curve, model-dependent fits of the rising or entire light curve, and a new data-driven approach using existing well-sampled CCSN light curves from {\it Kepler} and the Transiting Exoplanet Survey Satellite (TESS). We use these approaches to determine the time of core-collapse and its associated uncertainty (i.e., the GSW). We apply our methods to two Type II SNe that occurred during LIGO/Virgo Observing Run 3: SN\,2019fcn and SN\,2019ejj (both in the same galaxy at $d=15.7$ Mpc). Our approach shortens the duration of the GSW and improves the robustness of the GSW compared to techniques used in past GW CCSN searches.
\end{abstract}

\section{Introduction}
\label{sec:intro}

Following the detection of gravitational wave (GW) emission from compact object binary mergers \citep{abbott:16a, 2016PhRvL.116m1103A, 2016ApJ...818L..22A,2017PhRvL.119p1101A}, an additional known GW source yet to be detected are core-collapse SNe (CCSNe), representing the deaths of stars more massive than $\approx 8$ M$_\odot$.  While CCSNe are volumetrically much more common than binary mergers, their GW emission is expected to be weaker \citep{snsearch,2017nuco.confb0109I,Evans2017,Morozova_2018,oosnsearch}, and they are thus detectable in a much more limited volume. Still, the GW detection of CCSNe will provide a critical view of the core-collapse process itself, which is not directly accessible through electromagnetic (EM) observations.  In particular, since GW emission is produced by the quadrupole distribution of energy and mass, it can probe the degree of asymmetry and angular momentum in the core-collapse process \citep{2009sf2a.conf..143F,10.1093/ptep/pts067, janka_16, vartanyan_19, 10.1093/mnras/stz990}.

GW detections of CCSNe present distinct challenges compared to the detections of binary mergers, which have well defined waveforms that enable a search based on template matching \citep{2006CQGra..23.5477B, 2009PhRvD..80b4009V, 2016CQGra..33u5004U, PhysRevD.99.024048, PhysRevD.99.123022}. Although there are some critical GW features in time-frequency space that carry imprints of the underlying core-collapse physics, the nature of the waveforms are expected to be predominately stochastic \citep{janka_12,janka_16,kuroda_16,andresen_17,oconnor_18,hayama_18,kawahara_18,takiwaki_18,andresen_19,powell_19,radice_19,vartanyan_19}. Moreover, the energy conversion in GWs is expected to vary depending on the progenitor and the details of the explosion mechanism \citep{2009A&A...496..475M, 2009ApJ...707.1173M, 2009CQGra..26f3001O, 2009ApJ...704..951K, 2010CQGra..27s4005Y,2010CQGra..27k4101S, 2012PhRvD..86b4026O, 2012ApJ...755...11K, 2013ApJ...770...66H,2013ApJ...766...43M, 2013CRPhy..14..318K, 2013ApJ...768..115O, 2013ApJ...775...35C,2014ApJ...793...45N,2015ApJ...808...70A,PhysRevD.92.084040,  PhysRevD.92.122001,Kuroda_2016, 10.1093/mnras/stx618, PhysRevD.96.063005, Kuroda_2017, Morozova_2018, 10.1093/mnrasl/sly008, 2018MNRAS.477L..96H,Kawahara_2018, Radice_2019, 10.1093/mnras/stz990}.  

Even the most favorable core-collapse GW emission models point to a detection horizon that does not extend much beyond the local universe (D $\lesssim 25$ Mpc). For current generic GW burst searches at the design sensitivity of Advanced LIGO/Virgo for slowly rotating progenitors, only CCSNe within the Galaxy are expected to be detectable \citep{2015CQGra..32b4001A,snsearch,oosnsearch,  Szczepanczyk:2021bka}. The rate of CCSNe is about 1 in 50 years in the Galaxy, and grows to a few per year within a few Mpc \citep{1987ApJ...323...44V, vandenbergh:91, cappellaro:93, 1993A&A...273..383C, 1994ApJS...92..487T,1997A&A...322..431C,  cappellaro:99, karachentsev:04,ando:05,li:11,botticella:11,mattila:12, 2012ApJ...757...70D, 2015A&A...584A..62C, Rozwadowska:2020nab}. 

Given the expected complexity of the GW signal, and the limited detection range, it is essential to use all avenues to improve the chances of detection. These include detector hardware improvements, continued development of reliable core-collapse simulations, the inclusion of progenitor properties as input to detection algorithms, and the use of optical data for the purpose of pinpointing the location and distance, as well as for minimizing the GW search window (GSW).
% \color{red}
% It is worth pointing out that a GSW might not reach $100\%$ probability to include GW emission. We discuss this limitation regarding the EOM (described in ...) and point out how to mitigate some of the shortcomings. However, other issues that prevent this from reaching $100\%$, such as extcintion effects, may be unavoidable. However, the point of any targeted search is not to maximize the detection probability for $100\%$ of the potential candidates but to increase the probability of detection for a larger fraction of candidates. This line of operation has also been used in previous GW searches as part of the tuning methodologies. 
% \color{black}
Here, we specifically focus on the latter point. Estimating the GSW depends on two steps: (i) determining the time delay between core-collapse and the start of optical emission (shock breakout; SBO), $\Delta t_{\rm cc}\equiv t_{\rm SBO} - t_{\rm cc}$; and (ii) determining the time of SBO, $t_{\rm SBO}$, using available optical data.  We explore several approaches for the latter, and use the results of numerical simulations for the former \citep{2021arXiv210201118B}.  By way of example, we apply our methodology to Type II SNe that occurred during LIGO/Virgo Observing Run 3 (O3), with the specific requirements of: (i) a distance of $\lesssim 20$ Mpc (to have a non-zero GW detection efficiency); and (ii) sufficient optical data to constrain the GSW to $\lesssim {\rm week}$ in order to improve the search sensitivity and statistical confidence relative to an unconstrained search \citep{2010PhRvD..81j2001A, 2018APS..APRH14002L, 2019arXiv190503457T}. Two CCSNe that match these requirements\footnote{There were five additional CCSNe within $\approx 20-32$ Mpc: SN2019hsw (Type II) \citep{2019TNSTR1030....1S}, SN2019ehk (Type Ib) \citep{2019TNSTR.666....1G}, SN2019gaf (Type IIb) \citep{2019TNSTR.862....1T}, SN2020fqv (Type II) \citep{2020TNSTR.914....1F}, and SN2020oi (Type Ic) \citep{2020TNSTR..67....1F}. However, SN2019ehk and SN2020oi were Type I CCSNe and for the rest we had insufficient early-time data, which disqualified them from the methods presented here.} are SN\,2019fcn and SN\,2019ejj.

The paper is organized as follows. In \S\ref{sec:obs} we present the two CCSNe and our analysis of their available photometry and spectroscopy. In \S\ref{sec:lvc} we briefly review existing methodology of calculating the GSW as used by the LIGO/Virgo Collaboration (LVC). In \S\ref{sec:gsw} we introduce four new methodologies to more robustly estimate the time of SBO from the SN data, and use theoretical models to estimate of the delay between core-collapse to SBO. In \S\ref{sec:results} we present the results of these methodologies as applied to the two CCSNe considered here. We discuss the implications of these results in \S\ref{sec:disc} and conclude in \S\ref{sec:conc}.

\section{Discovery and Observations of SN\,2019\lowercase{ejj} and SN\,2019\lowercase{fcn}}
\label{sec:obs}

SN\,2019ejj was discovered by the Asteroid Terrestrial-impact Last Alert System (ATLAS) on 2019 May 2.26 UT (2458605.76) \citep{2019TNSTR.687....1T} and classified as a Type II SN by the extended Public ESO Spectroscopic Survey for Transient Objects (ePESSTO) with a spectrum taken on 2019 May 2.98 \citep{2019TNSCR.700....1N}. 
SN\,2019fcn was discovered by the All-Sky Automated Survey for Supernovae (ASAS-SN) on 2019 May 5.96 \citep{2019TNSTR.766....1S} and classified as a Type II SN by ePESSTO with a spectrum taken on 2019 May 14.03 (2458612.46) \citep{2019TNSCR.782....1N}. Both of these SNe exploded within a few days of each other in the same galaxy, ESO~430-G~020, which has a Tully-Fisher distance of $\approx 15.7$ Mpc \citep{2007A&A...465...71T}. 

Both SNe 2019ejj and 2019fcn were observed in the \textit{UBVgri} filters with the Las Cumbres Observatory's global network of 1\,m telescopes, equipped with Sinistro cameras \citep{2013PASP..125.1031B}, as part of the Global Supernova Project. Because SN~2019fcn exploded in the same galaxy as SN~2019ejj, we serendipitously observed SN~2019fcn on 2019 May 5.06, 21.7 hours (2458611.57) before it was discovered by ASAS-SN. We treat this earlier observation as the discovery in the following analysis.

We subtracted \textit{BVgri} references images, also taken with a Sinistro camera on 2020 March 8 after both SNe had faded, from the earlier observations using PyZOGY \citep{david_guevel_2017_1043973}, a Python implementation of the image subtraction algorithm of \citet{2016ApJ...830...27Z}. We extracted PSF photometry using \texttt{lcogtsnpipe} \citep{2016MNRAS.459.3939V} and calibrated the photometry relative to the Pan-STARRS1 $3\pi$ catalog \citep{2016arXiv161205560C}. For $B$ and $V$, we transformed the catalog magnitudes according to \citet{lupton_transformations_2005}.

We corrected the resulting magnitudes for Galactic extinction using the dust maps of \citet{2011ApJ...737..103S} and the extinction law of \citet{1989ApJ...345..245C}. We note that there is relatively high Galactic extinction in the direction of ESO~430-G~020, with $A_V=1.22$ mag. This, combined with the galaxy's low redshift, makes it difficult to discern whether either of these SNe suffer from additional host galaxy extinction. Our spectra of SN~2019fcn show very strong ($W_\lambda \approx 0.29$ nm) \ion{Na}{1D} absorption at $z=0$. This is beyond the regime where equivalent width correlates with dust extinction.  Given the large Galactic extinction, we assume that any host galaxy extinction is small in comparison. Finally, we converted to absolute magnitudes using the Tully-Fisher-based distance modulus $\mu=31.0\pm 0.4$ mag \citep{2007A&A...465...71T}. The photometry is available in the online journal.  

We also obtained two spectra of SN\,2019ejj and four spectra of SN\,2019fcn with the FLOYDS spectrographs on the Las Cumbres Observatory's 2\,m telescopes \citep{2013PASP..125.1031B} using a $2''$ slit oriented at the parallactic angle, and reduced them using the \texttt{floydsspec} pipeline \citep{2014MNRAS.438L.101V}. These spectra are available from the Weizmann Interactive Supernova Data Repository \citep{2012PASP..124..668Y}.

\section{Calculating the GSW using the LVC Early Observation Method (EOM)}
\label{sec:lvc}
\defcitealias{oosnsearch}{LVC20}

\begin{figure}[t!] 
\centering 
\includegraphics[width=0.5\textwidth]{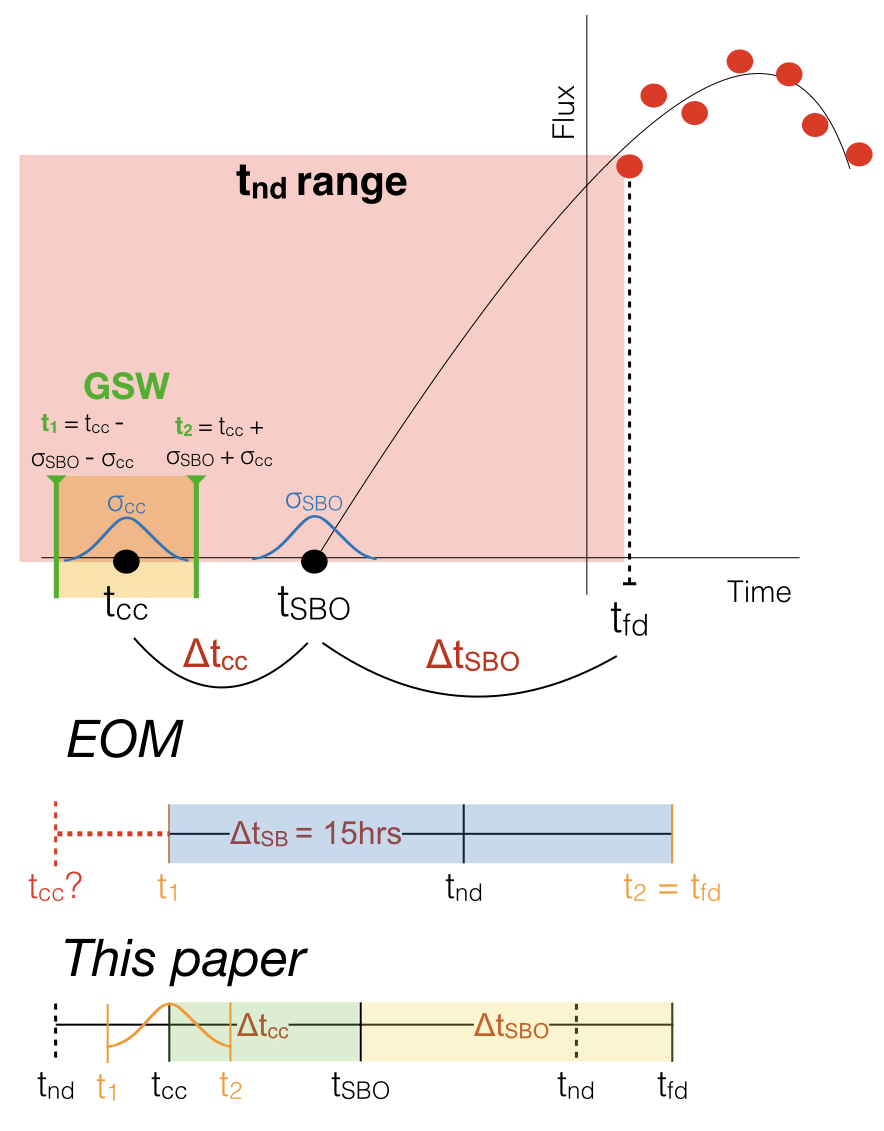}
\caption{{\it Top:} A schematic view of how to calculate the GSW (yellow shaded region) from a supernova optical light curve (red points; \citealt{2017ApJS..233....6H}). We constrain the GSW by fitting the light curve with several models to determine $t_{\rm SBO}$, and then use theoretical estimates to determine $t_{\rm cc}$. {\it Middle:} The EOM approach adopted by \citetalias{oosnsearch} uses the first optical detection ($t_{\rm fd}$) and previous non-detection (including a fixed shock breakout timescale: $t_{\rm nd}-15\,{\rm hr}$) to determine the GSW. Depending on the details of the optical data, the EOM may actually miss $t_{\rm cc}$.  {\it Bottom:} Our approach avoids the pitfalls of the EOM by not strictly depending on specific data points, but instead considers the rising part or the full optical light curve to directly fit for $\Delta t_{\rm SBO}$.  For $\Delta t_{\rm cc}$ we use a much wider range of possible values to account for progenitor uncertainties.}
\label{fig:finalGSW} 
\end{figure} 

\citet[hereafter \citetalias{oosnsearch}]{oosnsearch} calculate the GSW (referred to as the \textit{on-source window}), using the Early Observation Method (EOM) for four CCSN candidates. The EOM defines the GSW as the time range between $t_1=t_{\rm nd} - \Delta t_{\rm SB}$ and $t_2=t_{\rm fd}$, where $t_{\rm fd}$ is the time of first optical detection, $t_{\rm nd}$ is the time of the last observation without the SN present, and $\Delta t_{\rm SB}$ accounts for the shock propagation travel time between core-collapse and SBO; \citetalias{oosnsearch} employ a range of $15-24$ hr for the latter \citep{SNLS:2008bsx,Dessart:2017pfi} to account for the unknown progenitor star properties. However, when reporting the \textit{on-source windows}, \citetalias{oosnsearch} actually use a fixed $\Delta t_{\rm SB}=15$ hr.  A schematic view of the EOM from \citetalias{oosnsearch} is shown in Figure~\ref{fig:finalGSW}. 

\begin{deluxetable*}{lccccccccc}[t!]
\tablecaption{GSW using the Early Observation Method (EOM)}
\tablehead{ &
\multicolumn{3}{c}{\citetalias{oosnsearch} EOM} & \multicolumn{3}{c}{Corrected \citetalias{oosnsearch} EOM} & \multicolumn{3}{c}{Updated EOM} \\[-8pt]
\colhead{} & 
\multicolumn{3}{c}{------------------------------------------------} & 
\multicolumn{3}{c}{------------------------------------------------} & 
\multicolumn{3}{c}{------------------------------------------------} \\[-10pt]
\colhead{CCSN} & 
\colhead{$t_\mathrm{1}$} &
\colhead{$t_\mathrm{2}$} & 
\colhead{GSW}& 
\colhead{$t_\mathrm{1, 1}$} & \colhead{$t_\mathrm{2, 2}$} & 
\colhead{GSW} &
\colhead{$t_\mathrm{1, updated}$} & \colhead{$t_\mathrm{2, updated}$} & \colhead{GSW} \\ [-5pt]
\colhead{} &
\multicolumn{2}{c}{(JD)} & 
\colhead{(days)} &
\multicolumn{2}{c}{(JD)} & 
\colhead{(days)} &
\multicolumn{2}{c}{(JD)} &
\colhead{(days)} 
} 
\startdata
{SN~2019ejj} & 2458599.16 & 2458605.76 & 6.60   &  2458598.78  &  2458605.13
   &       6.35 & 2458596.78 & 2458604.36                                  & 7.58    \\\hline
{SN~2019fcn} & 2458608.89 & 2458612.46 & 3.57   &   2458608.52    &   	2458611.84  &     3.32    &2458606.52                                       & 2458611.06                                 & 4.54   \\
\enddata 
\tablecomments{Determinations of the GSW for SNe 2019ejj and 2019fcn using the EOM method. {\it Left:} $t_1=t_{\rm nd} - 15\,{\rm hr}$ and $t_2=t_{\rm fd}$. {\it Middle:} $t_{\rm 1,1}= t_{\rm nd} - 24\,{\rm hr}$ and $t_{\rm 2,2}=t_{\rm fd} - 15\,{\rm hr}$. {\it Right:} $t_{\rm 1,updated}=t_{\rm nd} - 3\,{\rm d}$ and $t_{\rm 2,updated}=t_{\rm fd} - 1.4\,{\rm d}$. Here $t_{\rm fd}$ is the first optical detection of a SN %(where for the case of SN\,2019fcn, it is taken from the publicly announced reported time) 
and $t_{\rm nd}$ is its previous non-detection. Our updated EOM is more robust and realistic than the one from \citetalias{oosnsearch}.}
\label{tab:EOM}
\end{deluxetable*}

The EOM approach suffers from four potential shortcomings. First, it assumes that the time of core-collapse ($t_{\rm cc}$), and hence GW emission, is bracketed by $t_1$ and $t_2$, but this cannot be guaranteed as it depends critically on the relative sensitivity of the non-detection at $t_{\rm nd}$ compared to the brightness of the SN at $t_{\rm fd}$. Second, allowing the GSW to extend all the way to $t_{\rm fd}$ makes the GSW too wide since clearly $t_{\rm cc}$ cannot be at $t_{\rm fd}$. Third, since optical time-domain surveys have a range of cadences and sensitivities, and are further affected by weather, it is possible that the time interval between $t_1$ and $t_2$ will in fact contain $t_{\rm cc}$, but will be much wider than necessary. Fourth, existing studies \citep{1992ApJ...394..599C, 2004MNRAS.351..694C,2011ApJ...729L...6C, 2017hsn..book..967W, Kistler_2013, Matzner:1998mg, muller:2017, morozova:2015, Davies2017, 2021arXiv210201118B} suggest a much wider range for $\Delta t_{\rm SB}$ than the single value of 15 hr used by \citetalias{oosnsearch}. Thus, depending on the specific circumstances of each CCSN, the GSW (as defined by the EOM) may actually miss the time of GW emission, or alternatively may be much wider than needed.

It is also worth pointing out that the GSWs reported in this paper might not also reach $100\%$ probability to include GW emission. We discuss this limitation in regards to the EOM and point out how to eliminate three of the drawbacks and mitigate the fourth one in this paper. However, the point of any targeted GW search is not to maximize the detection probability to $100\%$ but to increase the overall probability of detection for a larger fraction of GW candidates. This line of operation has also been used in previous GW searches as part of the tuning methodologies.

For the purpose of comparison with the EOM implementation of \citetalias{oosnsearch}, in Table~\ref{tab:EOM} we list the resulting GSW set by the EOM for the two SNe considered here. The left column ({\citetalias{oosnsearch} EOM}) uses $t_1=t_{\rm nd} -15$ hr and $t_2=t_{\rm fd}$. The middle column (Corrected \citetalias{oosnsearch} EOM) applies an additional shift to the GSW (to correctly account for $\Delta t_{\rm SB}$ as defined by \citealt{oosnsearch}) using $t_{\rm 1,1}=t_{\rm nd} - 24$ hr and $t_{\rm 2,2}=t_{\rm fd} - 15$ hr. In the right column (Updated EOM), we use updated shock breakout delay estimates based on the recent work of \citet{2021arXiv210201118B} (their Figure 6) and the observed range of Type IIP SN progenitors in the local universe (see \S\ref{subsec:tcc}). The updated EOM defines $t_{\rm 1, updated}= t_{\rm nd}-3$ d and $t_{\rm 2, updated}=t_{\rm fd} - 1.4$ d.  In the context of the EOM approach, we consider this latter estimate to be the most robust.

\begin{deluxetable*}{lcl}
\tablecolumns{3}
\tablecaption{Model Parameters for Methods used to Calculate $\Delta t_{\rm SBO}$. \label{tab:inputresults}}
\tablehead{
\colhead{Method} & 
\colhead{Input Data} & 
\colhead{Parameters} }
\startdata
Quadratic & Early Photometry & $\Delta t_{\rm SBO}$, $\alpha$, $\beta$ \\\hline
Shock Cooling & Early Photometry & $\Delta t_{\rm SBO}$, $v_\mathrm{s*}$, $M_\mathrm{env}$, $R$, $M$, $ f_\rho$ \\\hline
Empirical Fits & Early Photometry & $\Delta t_{\rm SBO}$, $\eta$, $\epsilon$ \\\hline 
\texttt{PP15} & Entire Photometry \& Spectroscopy & $t_0\equiv \Delta t_{\rm SBO}$, $t_P$, $t_w$, $E(B-V)$, $\omega$, $\mu$\\
\enddata
\tablecomments{A description of the parameters is given in \S\ref{sec:gsw}.}
\label{tab:inputparam} 
\end{deluxetable*}

\section{New Methods for Calculating the GSW}
\label{sec:gsw}

The EOM approach described in the previous section, including our updates to the LVC calculation, does not make use of all of the available information from optical SN data and modeling.  In this section we introduce and explore several methods to better determine the GSW. As illustrated in Figure~\ref{fig:finalGSW}, estimating the GSW requires a determination of both $\Delta t_{\rm SBO}=t_{\rm fd}-t_{\rm SBO}$ and $\Delta t_\mathrm{cc} = t_{\rm SBO} - t_{\rm cc}$, and their associated uncertainties.

\subsection{Estimating the Time to Shock Breakout and $\Delta t_{\rm SBO}$}

\begin{figure*}[t!] 
\centering 
\includegraphics[width=0.7\textwidth]{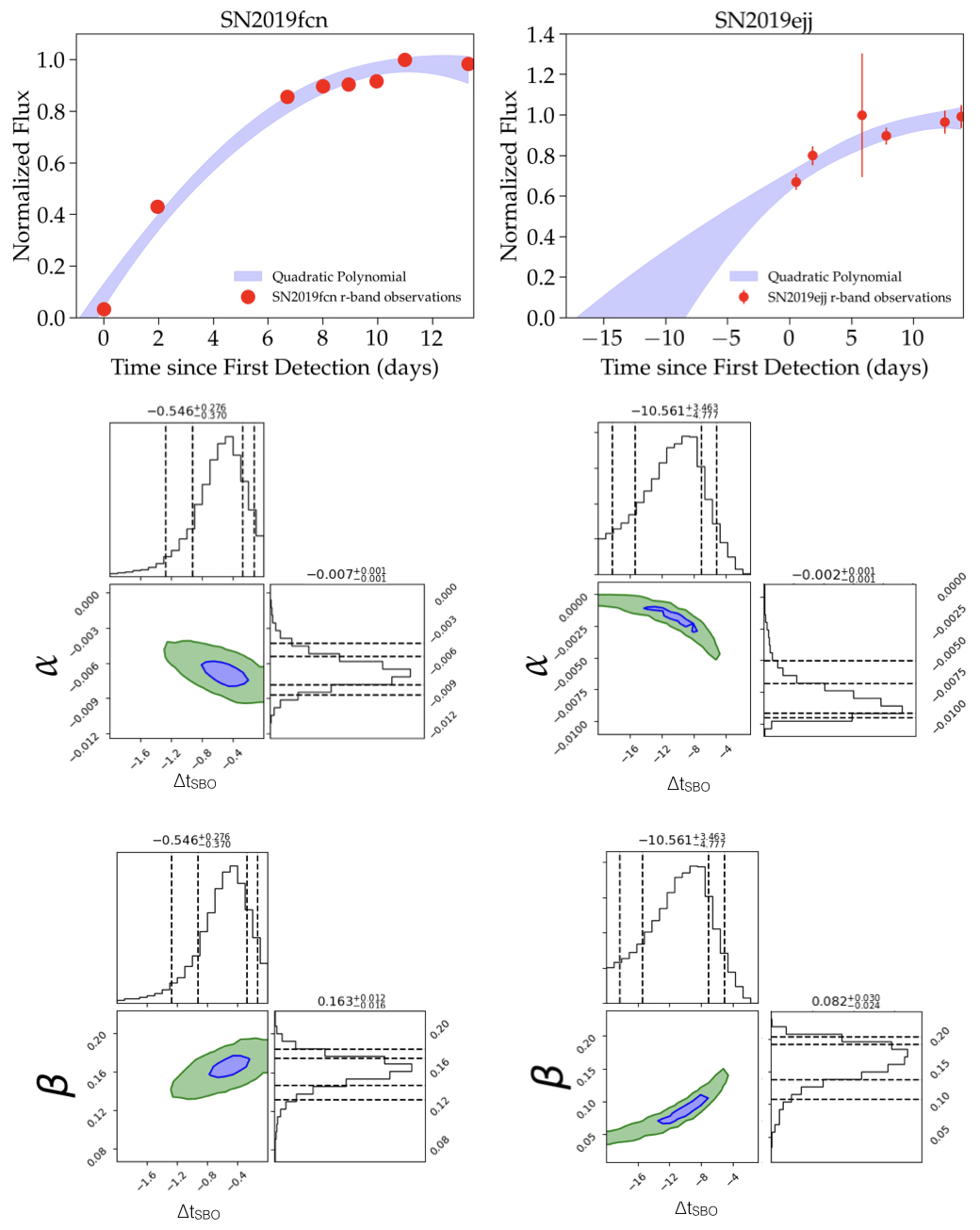}
\caption{{\it Top Panels:} Second-order polynomial fits (blue curves) to the optical $r$-band light curves (red points) of SN\,2019fcn ({\it Left}) and SN\,2019ejj ({\it Right}). The intersection of the blue region with zero flux represents the $1\sigma$ range on $\Delta t_{\rm SBO}$. As expected, a less well-sampled early light curve (SN\,2019ejj) leads to a larger uncertainty on $\Delta t_{\rm SBO}$. {\it Middle and Bottom Panels:} Two dimensional posterior distributions of $\Delta t_{\rm SBO}$, $\alpha$, and $\beta$ for SN\,2019fcn ({\it Left}) and SN\,2019ejj ({\it Right}). The dashed lines in the projected histograms mark the $5\%$, $16\%$, $84\%$, and $95\%$ quantile ranges. The blue and green contours mark the appropriate confidence regions.}
\label{fig:quadfit} 
\end{figure*} 

In this paper, we model either the early SN light curves (i.e., the rise to peak), or the full light curves, to determine $t_{\rm SBO}$ and its associated uncertainty. Our general approach is shown schematically in Figure~\ref{fig:finalGSW}. We define zero time as the time of first detection ($t_{\rm fd}$). We use four different approaches with varying levels of assumptions and complexity: (i) a simple quadratic polynomial fit, which we assume for the early rise of the SN light curve and which does not assume any specific physics; (ii) an analytical shock cooling model, based on the formulation of \citet{2017ApJ...838..130S}; (iii) a novel data-driven approach in which we use high cadence early light curves of Type II SNe observed with {\it Kepler} \citep{2016ApJ...820...23G} and the Transiting Exoplanet Survey Satellite (TESS;  \citealt{2021MNRAS.500.5639V}) as input models; and (iv) using the full data set of photometry and spectroscopy to model the entire light curve using the method of \citet{2015ApJ...799..215P,2015ApJ...806..225P}. We describe each method in detail below.

\subsubsection{Quadratic Polynomial Fit}
\label{sec:QP}

The rising phase of SNe empirically follows a low-order polynomial, so here we fit the early photometry with a quadratic polynomial of the form\footnote{Two additional fits using a power-law and a fourth-order polynomial (quartic) are found in the Appendix.}
\begin{equation}
F(t) = \alpha (t - t_\mathrm{SBO})^2 + \beta (t - t_\mathrm{SBO}),
\end{equation}
where $t_{\rm SBO}$ is the time of first light (i.e., zero flux) measured in days and $\alpha$ and $\beta$ are polynomial coefficients. We use a Markov-chain Monte Carlo (MCMC) routine using \texttt{pymc3} \citep{2015arXiv150708050S} to fit for the three free parameters. For $\alpha$ and $\beta$ we use a log-uniform prior with a wide range of [$-10$, 10] to avoid biasing the posterior. For $t_{\rm SBO}$, we use a uniform prior spanning [$-20$, 0] days. We also fit for an intrinsic scatter parameter, $\sigma$, which we add in quadrature to the observational uncertainties of each data point ($\sigma_i' = \sqrt{\sigma_i^2 + \sigma^2}$), with a half-Gaussian prior peaking at 0 and with a standard deviation of 0.1 in units of normalized flux. Figure~\ref{fig:quadfit} shows the resulting fits to the optical $r$-band light curves of SNe 2019fcn and 2019ejj, where the resulting best-fit regions allowed by the data indicate the uncertainty in $\Delta t_{\rm SBO}$.

\subsubsection{Shock Cooling Model}  
\label{sec:sc}

The early light curves of Type II SN are dominated by shock cooling emission before hydrogen recombination becomes an important source of heating. We employ the shock cooling model of \citet{2017ApJ...838..130S}, which modelled the expanding and cooling ejecta of a CCSNe following SBO. The model used a red supergiant (RSG) progenitor fit with a broken power law density profile. The envelope is defined as the mass above the radius at which the density profile steepens. The relevant model parameters of the model are the shock velocity ($v_\mathrm{s*}$), envelope mass ($M_\mathrm{env}$), progenitor radius ($R$), the ejecta mass ($M$), an order-unity factor describing the inner envelope structure ($f_\rho$), and the time of shock breakout ($t_0 \equiv t_{\rm SBO}$).

\begin{figure*}
    \centering
    \includegraphics[width=\textwidth]{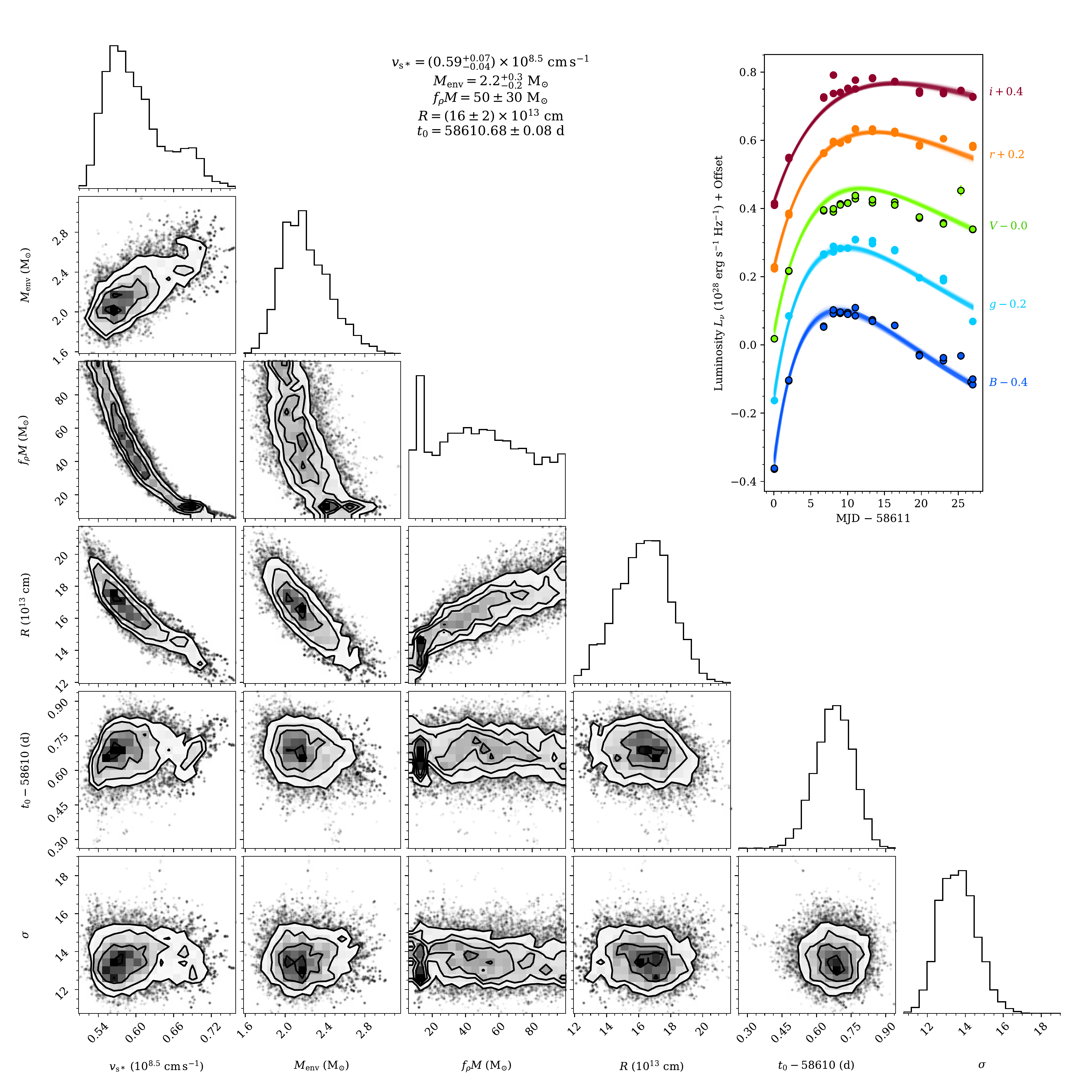}
    \caption{Results of the \cite{2017ApJ...838..130S} shock cooling model fits to the early light curve of SN\,2019fcn ({\it Top Right}). The corner plot shows the posterior probability distributions for $v_\mathrm{s*}$, $M_\mathrm{env}$, $f_\rho M$, $R$, $t_0=t_{\rm SBO}$, and $\sigma$. The $1\sigma$ credible intervals for each parameter, centered on the median, are listed at the top.}
    \label{fig:shockcooling}
\end{figure*}

\begin{figure*}
    \centering
    \includegraphics[width=\textwidth]{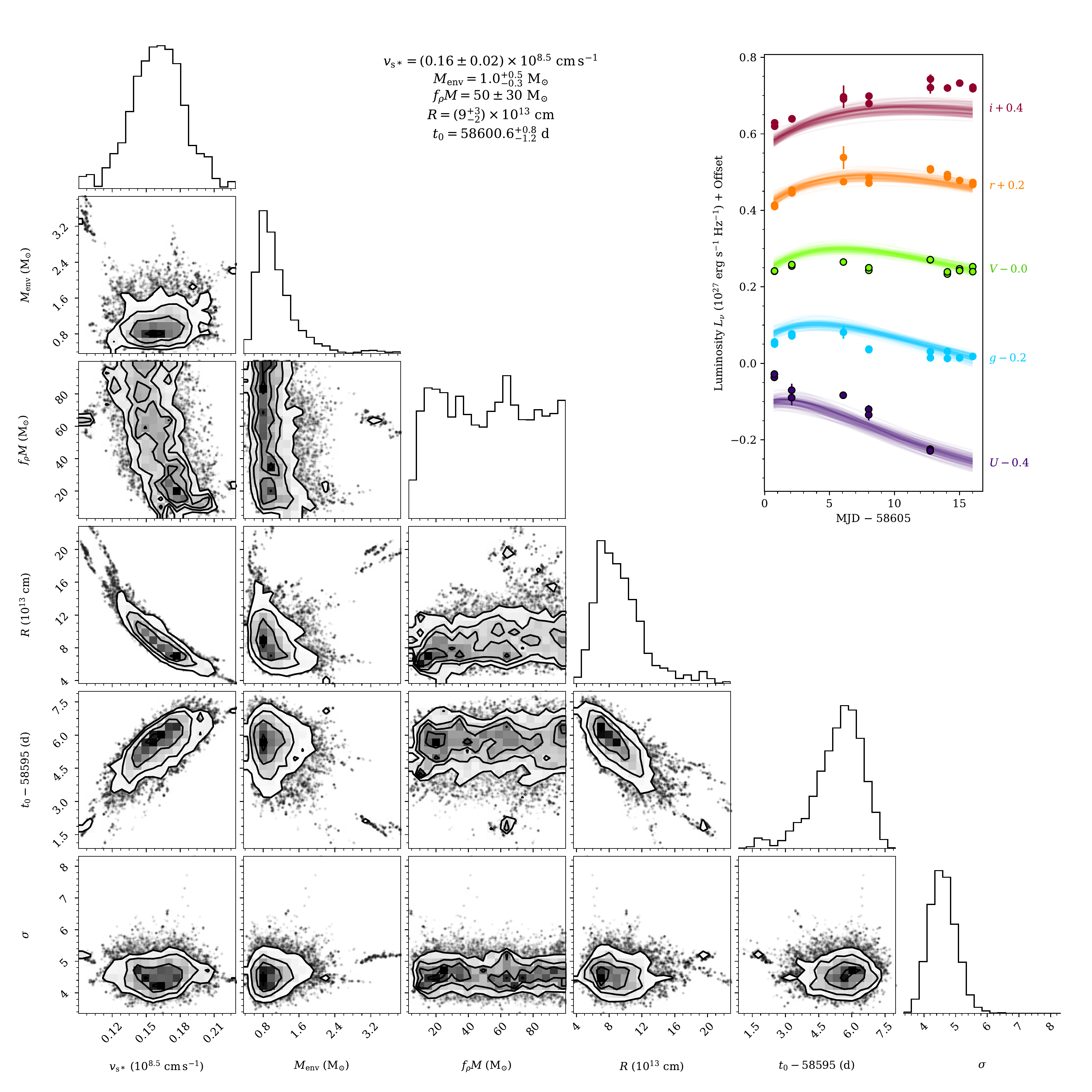}
    \caption{Same as Figure~\ref{fig:shockcooling} but for SN\,2019ejj.}
    \label{fig:shockcoolingejj}
\end{figure*}

We fit this model to the early portion of our  multi-band light curves using the MCMC routine implemented in the Light Curve Fitting package \citep{griffin_hosseinzadeh_2020_4312178}, including a multiplicative intrinsic scatter term ($\sigma_i' = \sigma_i \sqrt{1 + \sigma^2}$) such that the resulting reduced $\chi^2\approx 1$ \citep{2010arXiv1012.3754A}. The model holds for temperatures $>0.7$ eV (8120 K), above which hydrogen recombination effects may be neglected and the approximation of constant opacity holds. As such, we did not fit points below this temperature. Figures~\ref{fig:shockcooling} and \ref{fig:shockcoolingejj} show the results for SNe 2019fcn and 2019ejj, respectively.

\subsubsection{Empirical Fits using Kepler and TESS Type II SN Light Curves}
\label{sec:emp}

\begin{figure*}[t] 
\centering 
\includegraphics[width=\textwidth]{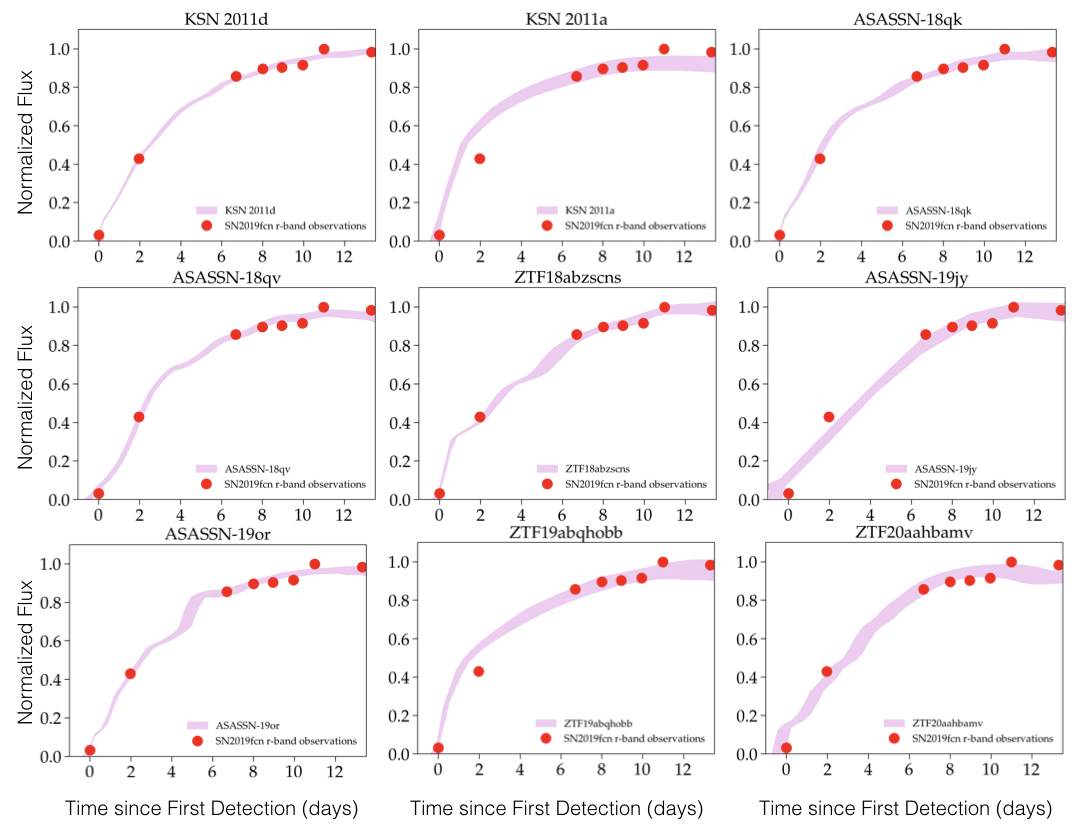} 
\caption{Empirical Light Curve fits for SN~2019fcn. The intersection of the pink region with the x-axis represents the 1$\sigma$ range on $t_{\mathrm{SBO}}$ using the empirical light curve templates constructed from \S\ref{sec:emp}.} 
\label{fig:fcnempfits} 
\end{figure*} 

\begin{figure*}[t] 
\centering 
\includegraphics[width=\textwidth]{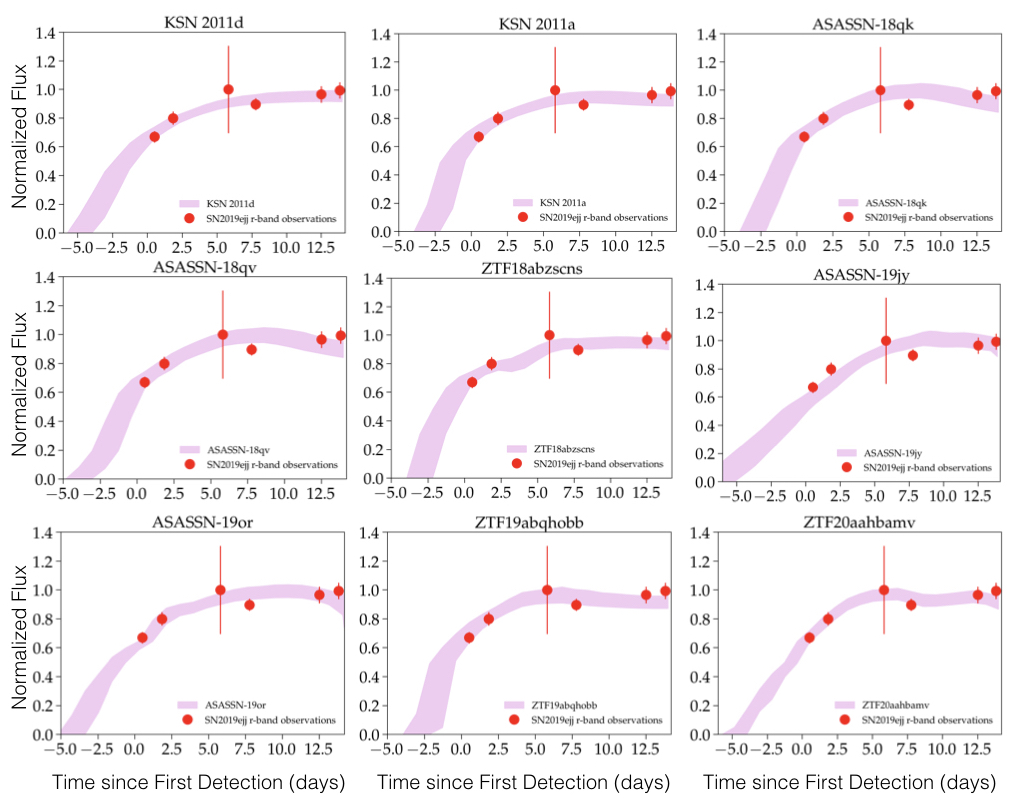}
\caption{Same as Figure~\ref{fig:fcnempfits} but for SN\.2019ejj.}
\label{fig:ejjempfits} 
\end{figure*} 

\begin{figure*}[t] 
\centering 
\includegraphics[width=\textwidth]{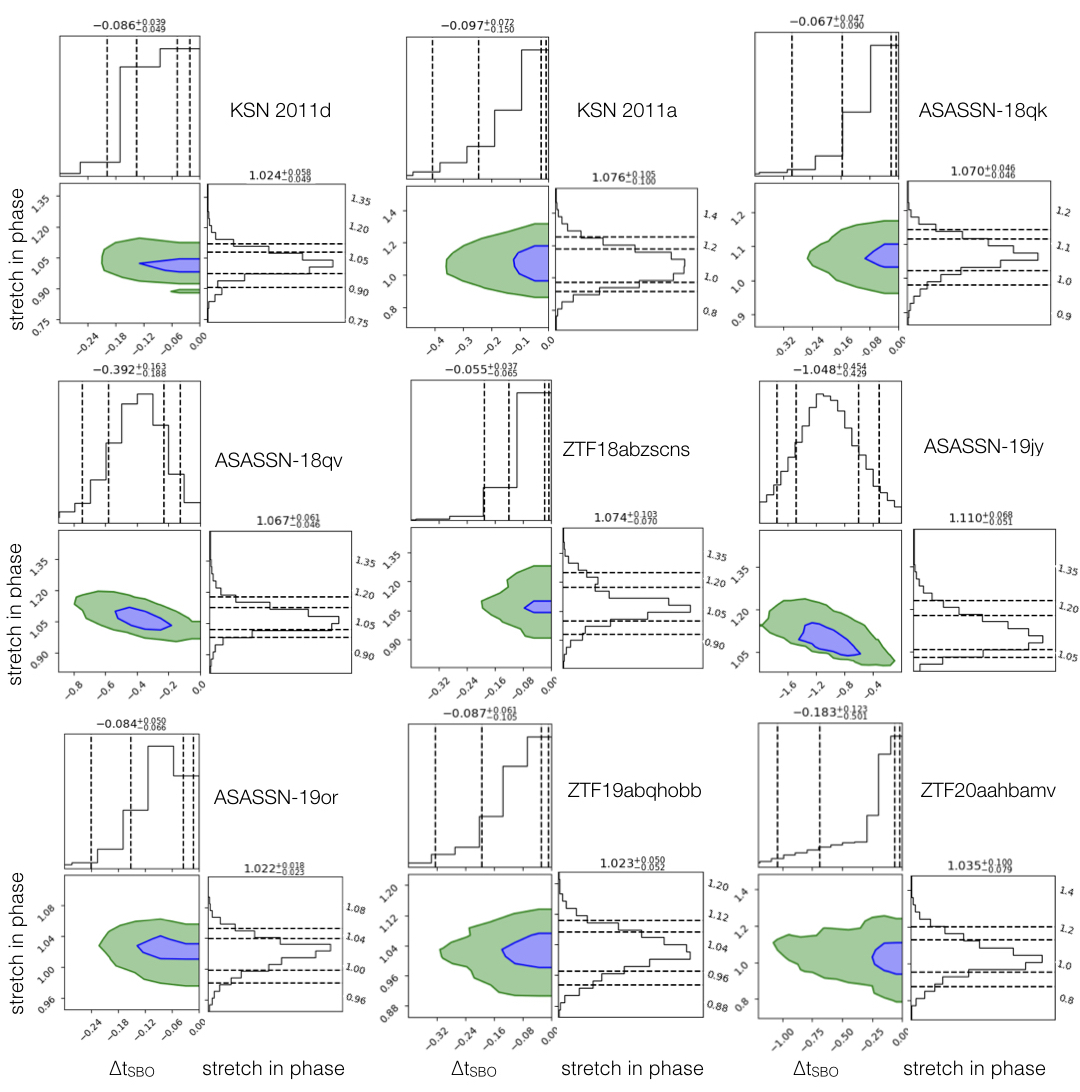} 
\caption{Two dimensional posterior distributions of $\Delta t_{\mathrm{SBO}}$ and the stretch in phase for SN\,2019fcn using the empirical {\it Kepler} and TESS light curves. The dashed lines in the projected histograms mark the $5\%$, $16\%$, $84\%$, and $95\%$ quantile ranges. The blue and green contours mark the appropriate confidence regions.} 
\label{fig:fcnconfits} 
\end{figure*} 

\begin{figure*}[t] 
\centering 
\includegraphics[width=\textwidth]{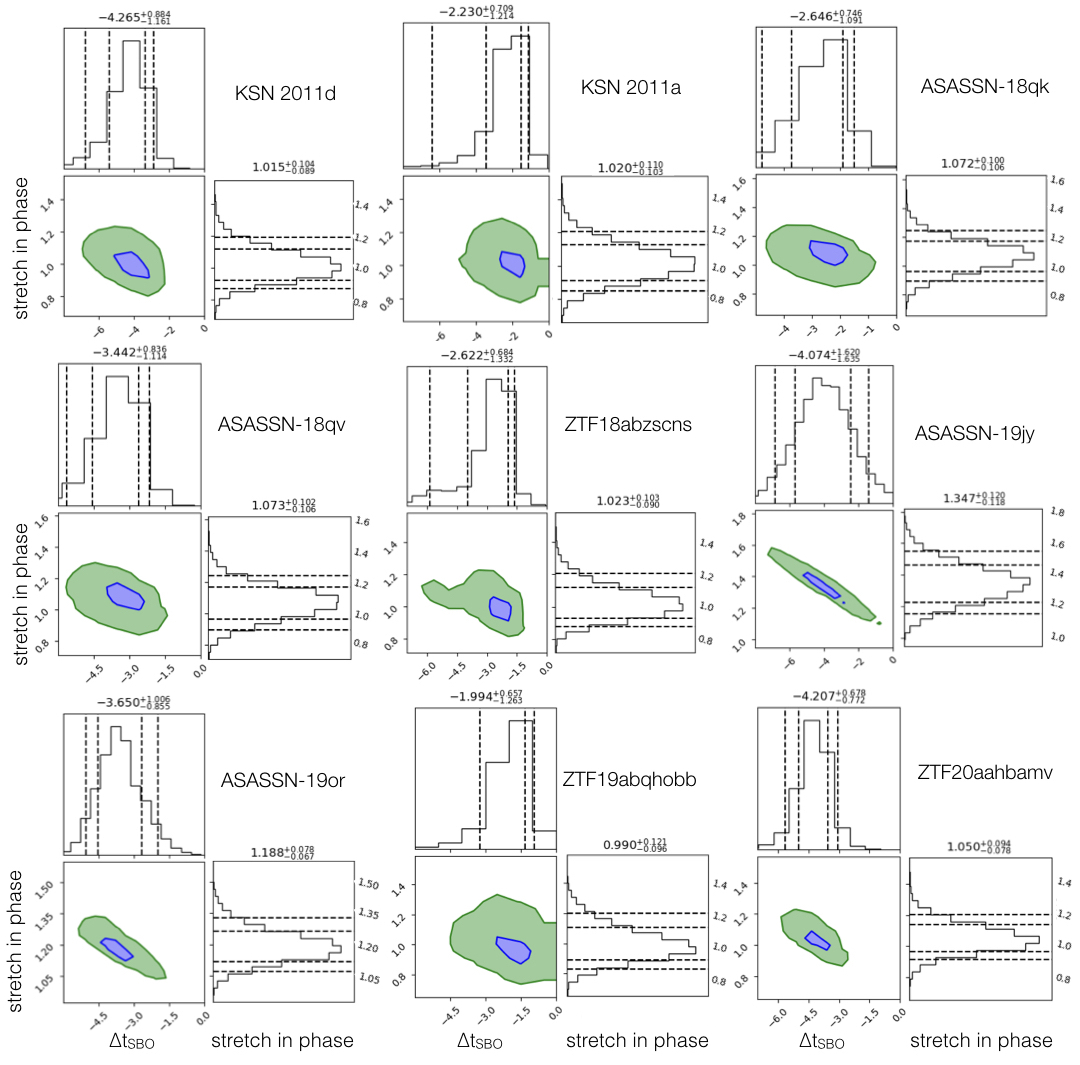} 
\caption{Same as Figure~\ref{fig:fcnconfits} but for SN\,2019ejj.} 
\label{fig:ejjconfits} 
\end{figure*} 

Over the past few years, high cadence observations with {\it Kepler} and TESS have led to detailed early light curves of several Type IIP SNe \citep{2016ApJ...820...23G,2021MNRAS.500.5639V}.  Here we use this library (where naming conventions for the {\it Kepler} and TESS SNe are taken from \citet{2021MNRAS.500.5639V}) of well measured early light curves to empirically fit for $\Delta t_{\rm SBO}$ in SNe 2019ejj and 2019fcn.

From the 14 Type II SNe observed by TESS \citep{2021MNRAS.500.5639V} and the 2 SNe observed by {\it Kepler} \citep{2016ApJ...820...23G}, we choose the 9 highest quality light curves as empirical templates, discarding events with large scatter. We adopt the $t_{\rm SBO}$ values for these events from \citet{2021MNRAS.500.5639V} and \citet{2016ApJ...820...23G}. We dynamically bin the 3-hour sampled data in intervals spanning $0.3-1$ days. We normalize the flux, such that 1 is peak in binned flux. We interpolate our templates to the observed times of SNe 2019ejj and 2019fcn during the fit, which applies the stretch in flux and offset in phase, in order to calculate the likelihood for a given model. 

We use an MCMC routine in \texttt{pymc3} \citep{2015arXiv150708050S} to fit for three free parameters: $\Delta t_{\rm SBO}$, and multiplicative ``stretch'' factors in phase and flux. The stretch in phase has a log-normal prior centered at 1 with a standard deviation of $0.1$, (corresponding to a stretch in phase within 10$\%$), in order to minimize non-physical and/or artificial distortions to the model light curves.\footnote{For poorly sampled light curves (see middle right panel in Figure 7) there is a correlation between $\Delta t_{\rm SBO}$ and the stretch in phase. Since we aim to use the {\it Kepler} and TESS light curves as empirical models of Type II SN light curves we allow for a mild stretch, but constrain it using the log-normal prior.} The stretch in flux has a uniform prior centered at 1 spanning $[0.5,2]$. For $\Delta t_{\rm SBO}$ we use a uniform prior spanning $[-20,0]$. We fit for the intrinsic scatter parameter, $\sigma$, as in \S\ref{sec:QP}. We apply the fitting to only the $r$-band data since it is the closest to the broadband photometry provided by both {\it Kepler} and TESS. 

In Figures~\ref{fig:fcnempfits} and \ref{fig:ejjempfits} we show the empirical model light curve fits, and in Figures~\ref{fig:fcnconfits} and \ref{fig:ejjconfits} we show the resulting two-dimensional posteriors for $\Delta t_{\rm SBO}$ and the stretch in phase, for SNe 2019fcn and 2019ejj, respectively.

\subsubsection{Pejcha \& Prieto 2015 Model}
\defcitealias{2015ApJ...799..215P}{PP15}

We employ the method of \citet[hereafter \citetalias{2015ApJ...799..215P}]{2015ApJ...799..215P,2015ApJ...806..225P}, which uses a combination of photometric and velocity measurements, to provide an estimate of $t_{\rm SBO}$. This estimate is obtained from fitting the apparent angular radius versus the expansion velocity (obtained through spectroscopic measurements) through the application of the expanding photosphere method (EPM) \citep{1994ApJ...432...42S,2007AIPC..937..394V,2011AAS...21733721E, 2014ApJ...782...98B}. 

The velocity measurements fed into the model are obtained by measuring the photospheric velocity as a function of time by fitting the H$\alpha$ feature in the four individual spectra (i.e, taken at four different epochs) of SN\,2019fcn and the two spectra of SN\,2019ejj. We used the interactive Spectrum Fitting package \citep{griffin_hosseinzadeh_2020_specfit} to define this continuum in time, fit a P~Cygni profile, and derive the velocity from the minimum of the absorption.

\begin{figure*}[t] 
\centering 
\includegraphics[width=\textwidth]{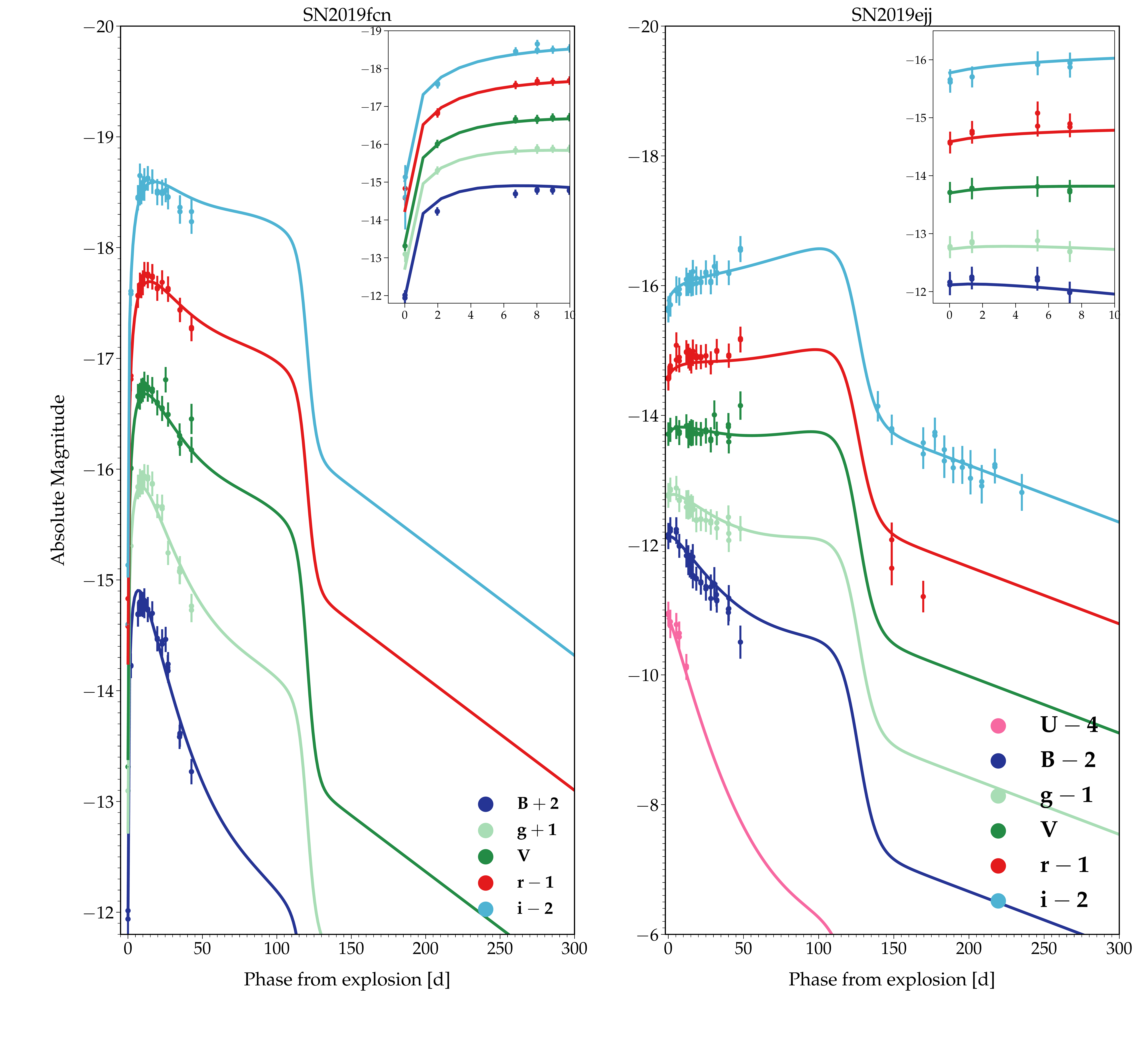} 
\caption{Light curve model fits for SN\,2019fcn ({\it Left}) and SN\,2019ejj ({\it Right}) using the \citetalias{2015ApJ...799..215P} model.  The insets focus on the early phase of the light curves and show that the model can capture the rising behavior, especially for photometrically well-sampled candidates.} 
\label{fig:PP15LC} 
\end{figure*}

\citetalias{2015ApJ...799..215P} also construct a model of the multi-band light curves. The multi-band light curve model is dependent on the construction of the spectral energy distribution (SED). We then use the SED to derive both the photospheric radius and temperature variations. The model light curves shown in Figure~\ref{fig:PP15LC} begin at an estimated $t_{\rm SBO}$ when the optical flux rises due to photospheric expansion and cooling where the peak of the SED is in the optical bands. 

The relevant parameters of the model are the total reddening ($E(B-V)$), the expansion velocity power law exponent ($\omega$), the plateau duration ($t_P$), the plateau transition width ($t_w$), the distance modulus ($\mu$), and the so-called explosion time ($t_0$), which is equivalent to $t_{\rm SBO}$. 

We re-normalize the error bars by multiplying them by a factor of 2.70 for SN\,2019fcn and a factor of 1.38 for SN\,2019ejj in order to reduce $\chi^2$ = 1 and account for intrinsic scatter present in the photometric measurements. Figure~\ref{fig:PP15LC} shows the resulting best fit models for SNe 2019fcn and 2019ejj using the \citetalias{2015ApJ...799..215P} method.

\subsection{The Delay Between Core-Collapse and Shock Breakout, $\Delta t_{\rm cc}$}
\label{subsec:tcc}

In the previous section, we addressed the determination of the time of shock breakout, $t_{\rm SBO}$.  To determine the entire length of the GSW,  we also need to include the time delay between core-collapse and shock breakout, $\Delta t_{\rm cc}$. This delay is progenitor dependent due to the travel time of the SN shock through the progenitor star. Since the progenitor properties are difficult to ascertain in detail, and since there is currently no empirical information on $\Delta t_{\rm cc}$, we resort to using theoretical estimates.  In particular, we use the recent work of \citet{2021arXiv210201118B}, where estimated ranges for $\Delta t_{\rm cc}$ are given for a range zero-age main sequence masses. We focus on the case of Type II SNe, which arise from red supergiant (RSG) stars.

Assuming the progenitor mass is generally $\lesssim 20$ M$_\odot$ \citep{2016IAUFM..29B.220N, 2017RSPTA.37560270D, 2017RSPTA.37570219G, 2018MNRAS.474.2116D}, we apply a \textit{conservative}
range of $\Delta t_{\rm cc}\approx 1.4-3$ days, covering the range of masses, as well as theoretical assumptions and uncertainties. 

We thus define the GSW by shifting the time window relative to $t_{\rm SBO}$ by this range; namely, we shift and extend the time range defined by $t_{\rm SBO}$ and its associated uncertainty to 3 days earlier from the lower end of the range and 1.4 days earlier from the upper end of the range (see Figure~\ref{fig:finalGSW}, where this shift is applied in both GSWs resulting from \textit{Updated EOM} and \textit{Method from this paper}.).

We note that for some SNe, with a well determined value of $t_{\rm SBO}$, the theoretical uncertainty in $\Delta t_{\rm cc}$ may become the dominant source of uncertainty in the GSW. In the absence of more firm constraints on SN shock propagation, and a reliable determination of progenitor properties, we advocate for a more conservative range for $\Delta t_{\rm cc}$.

\section{Results for SN\,2019\lowercase{ejj} and SN\,2019\lowercase{fcn}}
\label{sec:results}

\begin{figure*}[t] 
\centering 
\includegraphics[width=\textwidth]{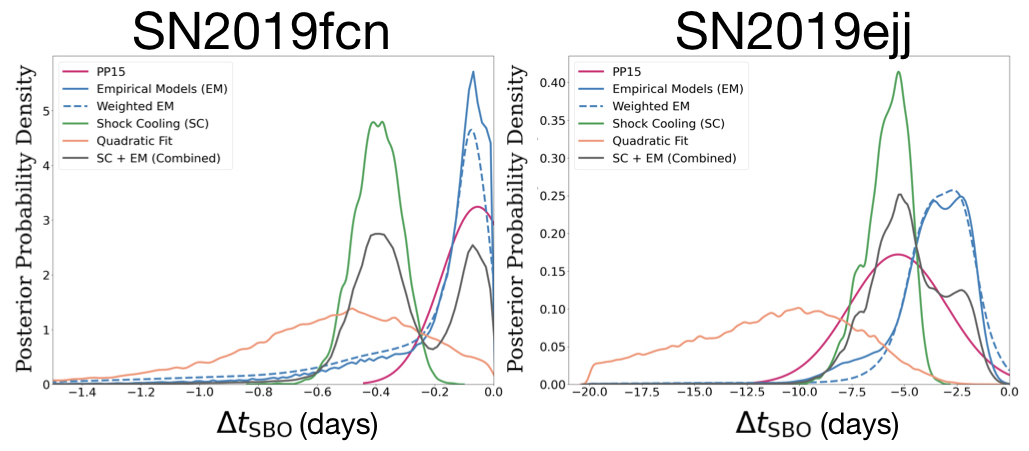} 
\caption{The posterior distributions of $\Delta t_{\mathrm{SBO}}$ for the four methods used in this paper, as well as a combined posterior based on the two most robust models (shock cooling and empirical models).} 
\label{fig:JPD} 
\end{figure*}

In the previous section we presented several methods to determine $\Delta t_{\rm SBO}$ with increasing model complexity and level of physical accuracy. Here we summarize the results of the model fits to SNe 2019ejj and 2019fcn, and determine $t_{\rm SBO}$ for both events. The results from all methods, along with their GSWs, are summarized in Table~\ref{tab:results} and shown in Figure~\ref{fig:JPD}. 

We show the results of the simple quadratic polynomial fitting approach in Figure~\ref{fig:quadfit}. As expected, the earlier detection of SN\,2019fcn, with reasonable subsequent time sampling during the rise to peak, lead to $\Delta t_{\rm SBO}=-0.54^{+0.27}_{-0.37}$ d.  For SN\,2019ejj the uncertainties are significantly larger due to the later detection of the SN, $\Delta t_{\rm SBO}= -10.56^{+3.46}_{-4.77}$ d. As shown in Figure~\ref{fig:JPD}, the posterior distributions for $\Delta t_{\rm SBO}$ from the quadratic fit overlap the posteriors of the other methods, but tends to be broader since the model is quite simplistic (due to the curvature staying constant, see the discussion of the bias in the Appendix) and less physically informative. 

The more physically motivated models provide tighter constraints on $\Delta t_{\rm SBO}$.  In Figures~\ref{fig:shockcooling} and \ref{fig:shockcoolingejj} we show the results of the shock cooling model fits for SN\,2019fcn and SN\,2019ejj, respectively.  We find that the model provides a good fit to the data, especially in the case of the rising portion of SN\,2019fcn. We note that at later time (near peak) there are some color mismatches between the data and models, likely due to the more complex spectrum of the SNe compared to the model spectral energy distribution.  We find that $\Delta t_{\rm SBO}=-0.39^{+0.13}_{-0.14}$ d for SN\,2019fcn and $-5.71^{+1.32}_{-2.04}$ d for SN\,2019ejj. The shock cooling model also provides an estimate of the progenitor radius: $(2.3\pm 0.3)\times 10^3$ R$_\odot$ for SN\,2019fcn and $(1.3^{+0.4}_{-0.3})\times 10^3$ R$_\odot$ for SN\,2019ejj.  The envelope masses are constrained to be $2.2^{+0.3}_{-0.2}$ M$_\odot$ and $1.0^{+0.5}_{-0.3}$ M$_\odot$ for SN\,2019fcn and SN\,2019ejj, respectively.  Finally, we note that the model does not constrain the total ejecta mass ($f_pM$); the posterior is approximately equal to the prior.

The data-driven approach, where we use empirical SN light curves as input with some allowance for a stretch, provide results that are comparable to the shock cooling models, but tend to give systematically smaller values of $\Delta t_{\rm SBO}$. Here we take the conservative approach of combining the posteriors of all 9 input SNe, without down-weighting or discarding poorer fits\footnote{We explore the effect on the estimation of $\Delta t_{\rm SBO}$ from the scatter present in the light curve library by introducing a \textit{weighted combined posterior} discussed in \S\ref{sec:disc}.}. Using the un-weighted joint posterior from the 9 input SNe, we find $\Delta t_{\rm SBO}=-0.11^{+0.07}_{-0.04}$ d for SN\,2019fcn and $-3.31^{+1.32}_{-1.46}$ d for SN\,2019ejj.  In Figures~\ref{fig:fcnempfits} (SN\,2019fcn) and \ref{fig:ejjempfits} (SN\,2019ejj) we show the posterior distributions for $\Delta t_{\rm SBO}$ and the stretch in phase for each of the 9 input SNe.  We find that for the less well-sampled optical light curve (discovered at later times post $t_{\rm SBO}$) of SN\,2019ejj there is some correlation between the stretch factor and $\Delta t_{\rm SBO}$, such that a stretch larger than unity leads to a larger value and uncertainty associated with $\Delta t_{\rm SBO}$.  This is the case for input SN light curves that are a poorer match to those of SNe 2019ejj and 2019fcn (and their respective light curves).  For example, the fits from KSN 2011a, ASASSN-19jy and ZTF19abqhobb would have benefited from an additional stretch (allowed in phase space), especially in regards to the fit of the second observed datum for SN~2019fcn with the trade-off being a larger uncertainty assigned to $t_{\mathrm{SBO}}$.

Our final approach using \citetalias{2015ApJ...799..215P}, shown in Figure~\ref{fig:PP15LC}, provides good fits to the full light curves of both SNe, including in particular the rising phase of the light curves (see insets in Figure~\ref{fig:PP15LC}).  The resulting posterior distributions of $\Delta t_{\rm SBO}$ overlap with the shock cooling and empirical models; in the case of SN\,2019fcn we find a better match to the empirical models, while in the case of SN\,2019ejj there is overlap with both the shock cooling and empirical model posteriors. Specifically, we find $\Delta t_{\rm SBO}=-0.06^{+0.01}_{-0.03}$ d for SN\,2019fcn and $-5.30\pm 2.31$ d for SN\,2019ejj. 

Finally, we construct a joint weighted posterior for $\Delta t_{\rm SBO}$, using the posteriors of the shock cooling and empirical models.  We choose these particular models since they are more informative than the simplistic quadratic model and focus on the early part of the light curve (unlike the \citetalias{2015ApJ...799..215P} which inherently leads to a compromise between matching the early and late parts of the light curve).  At the present, we don't strongly advocate for preferring the empirical model over the shock cooling model given their slight systematic offsets. However, in the Appendix (see A.2), we perform a test to quantify the potential bias of the empirical model, which appears to be negligible. This seems to indicate that the shock cooling model tends to potentially predict an earlier $t_{\rm SBO}$. 

We use a combined posterior (using both the empirical and shock cooling models), which gives an equal weight to both models.  This combined posterior on $\Delta t_{\rm SBO}$ is shown in Figure~\ref{fig:JPD} and listed in
Table~\ref{tab:results}:  $\Delta t_{\rm SBO}=-0.33^{+0.32}_{-0.31}$ d  for SN\,2019fcn and $-4.81^{+2.21}_{-1.56}$ d for SN\,2019ejj.

\begin{figure*}[t] 
\centering 
\includegraphics[width=0.9\textwidth]{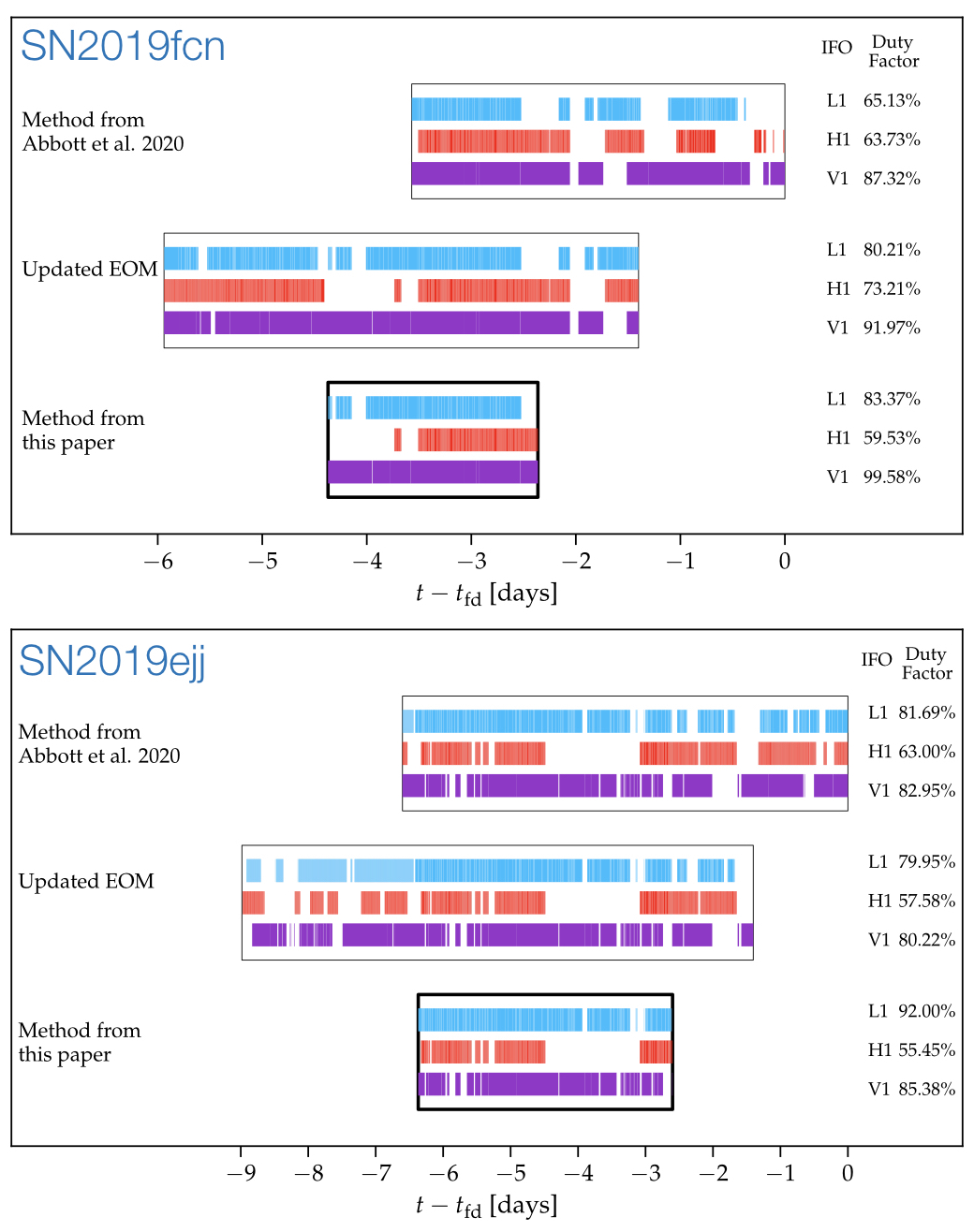} 
\caption{GSWs using the method presented in this work, compared to the EOM methodology of \citetalias{oosnsearch} and our update to that method. Our GSWs are narrower than those from the \citetalias{oosnsearch} method.  Equally important, in the case of SN\,2019fcn, we find significant non-overlapping region, suggesting that the \citetalias{oosnsearch} method may miss the time of GW emission. The horizontal colored regions within each GSW indicate the on-time of the three GW detectors (LIGO-Livingston, LIGO-Hanford, and Virgo), and the percentages on the right-hand-side indicate the resulting duty factor.  For both SNe there is nearly 100\% coverage with at least one GW detector during our calculated GSW.}
\label{fig:duty} 
\end{figure*}

To define the GSW, we shift $\Delta t_{\rm SBO}$ by a range of $\Delta t_{\rm cc}=[1.4, 3]$ d, as we defined in \S\ref{subsec:tcc}. The final GSW durations computed using both the EOM from \citetalias{oosnsearch} and the methods presented in this paper are shown in Figure~\ref{fig:duty}.

\begin{deluxetable*}{lCC|CR}
\tablecolumns{3}
\tablecaption{$\Delta t_{\rm SBO}$ and Resulting GSW for SN\,2019fcn and SN\,2019ejj.  \label{tab:results}}
\tablehead{
\colhead{Method} & 
\colhead{SN~2019fcn} &
\colhead{GSW (MJD)} &
\colhead{SN~2019ejj} &
\colhead{GSW (MJD)}} 
\startdata
Quadratic & -0.54^{+0.27}_{-0.37} & [2458607.66, 2458609.90] &  -10.56^{+3.46}_{-4.77}  & [2458587.43, 2458597.26] \\\hline
Shock Cooling & -0.39^{+0.13}_{-0.14} & [2458608.04, 2458609.91] &-5.71^{+1.32}_{-2.04} & [2458595.01, 2458599.97] \\\hline
Empirical Models & -0.11^{+0.07}_{-0.04} & [2458608.42, 2458610.13] & -3.31^{+1.32}_{-1.46} & [2458597.99, 2458605.68]\\\hline
\citetalias{2015ApJ...799..215P} & -0.06^{+0.01}_{-0.03} & [2458608.48, 2458610.12] & -5.30\pm 2.31  & [2458595.15, 2458601.37]\\\hline\hline
Combined (SC + EM) & -0.33^{+0.32}_{-0.31} & [2458607.93, 2458610.16] & -4.81^{+2.21}_{-1.56} & [2458596.39, 2458601.76]\\
\enddata
\tablecomments{The listed GSW MJD values correspond to $1\sigma$ ranges.}
\label{tab:fcnejjtsbofinal} 
\end{deluxetable*}

\begin{figure*}[t!] 
\centering 
\includegraphics[width=0.9\textwidth]{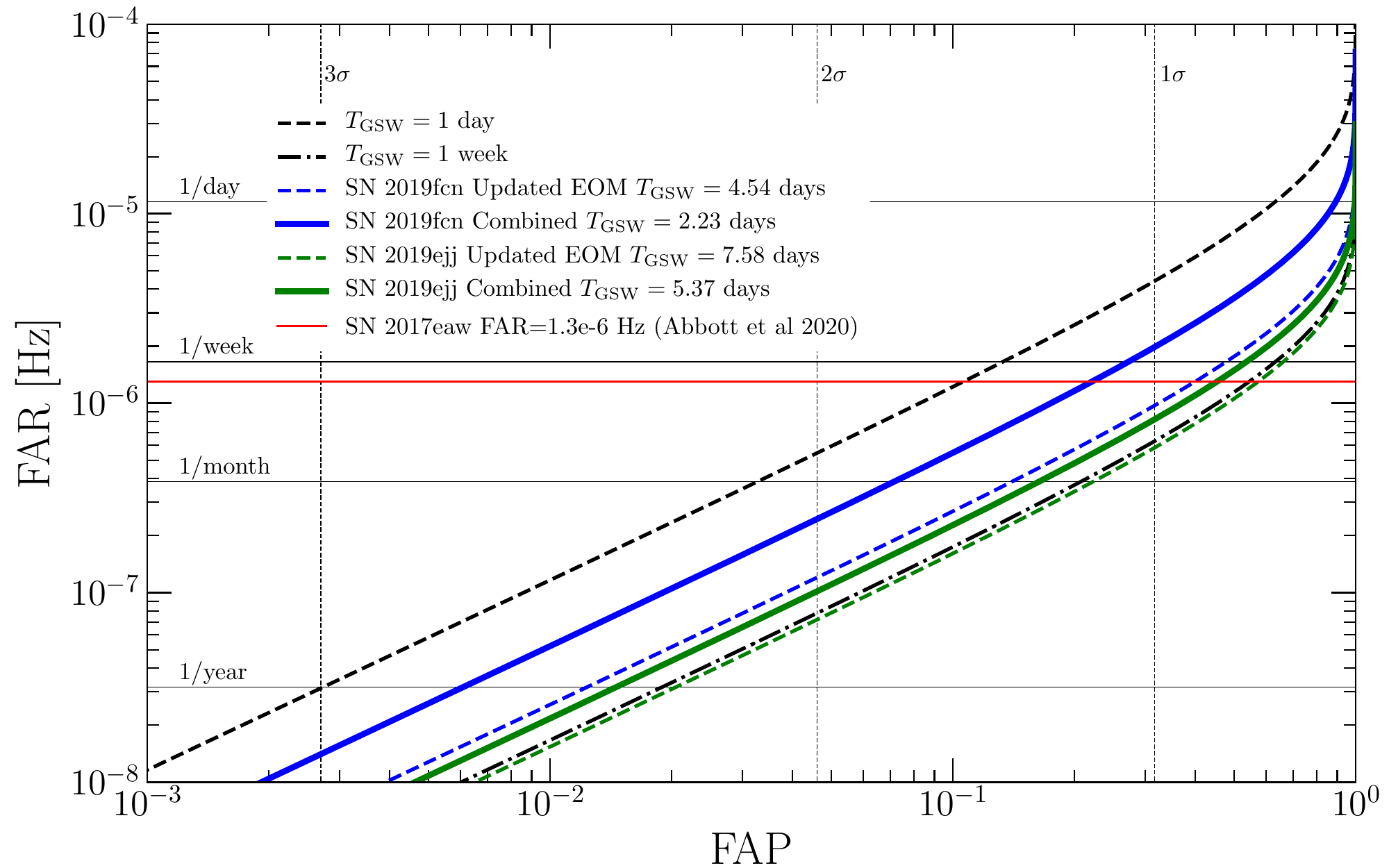}
\caption{False Alarm Rate (FAR) as a function of False Alarm Probability (FAP), along with the GSW durations for SN\,2019fcn (blue) and SN\,2019ejj (green).  Also marked are fiducial GSW durations of 1 day (black dashed line) and 1 week (black dot-dashed line).  The vertical lines indicate the corresponding statistical significance of a detection at a given FAP.  For SN\,2019fcn, which has the narrowest GSW, we find that a detection with ${\rm FAR}\lesssim 1$ yr$^{-1}$ will correspond to $\sim 3\sigma$.}
\label{fig:FARFAP} 
\end{figure*} 

\section{Discussions}
\label{sec:disc}

From the four methods presented in this work, the quadratic fit provides the most conservative (i.e., the widest window) GSW primarily because it is the least informative/physics-based model. Especially in the case of delayed and poor sampling of the rising light curve, the posterior distribution of $\Delta t_{\rm SBO}$ from this model can be mostly non-informative (i.e., SN\,2019ejj).  

The physically motivated models provide much tighter constraints on $\Delta t_{\rm SBO}$, but we note two possible trends based on the two SNe considered here. First, the shock cooling model leads to an earlier estimate of $t_{\rm SBO}$ compared to the empirical model approach; this difference is about 0.3 d for SN\,2019fcn and about 2.4 d for SN\,2019ejj (although there is more significant overlap in the posterior distribution in the case of SN\,2019ejj). Second, we find that the \citetalias{2015ApJ...799..215P} model gives consistent results with the other two physical models; in the case of SN\,2019fcn it overlaps well with the posterior of the empirical models, while in the case of SN\,2019ejj, its bounds are in agreement with the posteriors of both the shock cooling and empirical models.

We also note that both the shock cooling and empirical model approaches can be improved.  In the case of the shock cooling model, better sampled early-time light curves will help to refine the model, resulting in more robust estimates of the various parameters, including the time of shock breakout.  The shock cooling model also has the advantage that it can provide a determination of the progenitor radius, which in turn may help to constrain the progenitor mass and the range of GW signals expected. We note however, that this translation is not trivial as the inferred radii may be affected by the presence of circumstellar (CSM) matter around the progenitor \citep{2017ApJ...838...28M, 2017hsn..book..843B, 2017ApJ...851..138D, 2018ApJ...858...15M, 2019ApJ...882...68S, 2020MNRAS.497.5395N}.\footnote{If the presence of CSM bias' the estimation of the inferred radii, this may lead to a potential biased offset in the estimation of $t_{\rm SBO}$. Although we explore the effect of any bias present in the empirical model in the Appendix, the exploration of this effect in the shock cooling model is beyond the scope of this paper.}

In the case of the empirical model approach, we anticipate a growing library of events from TESS that will help to better explore the parameter space of Type II SN early light curves. In this paper, we explore the effect on the estimation of $t_{\rm SBO}$ from the scatter present in the library of light curves. We account for this additional correction by creating individual kernel density estimates (KDE) for each light curve model, which essentially acts as a gaussian PDF estimate. We then weight each model by the prior probability in both scatter and the stretch in phase. We normalize the combined weighted KDE estimate then by dividing by the sum of the weights. This weighted combined posterior is shown in Figure \ref{fig:JPD}. For the case of SN\,2019fcn, the weighted posterior is in the same relative range as the empirical models, but with a smoothed single peak. For SN\,2019ejj, the weighted posterior removes the bump at higher $t_{\rm SBO}$. 

Finally, estimates of $t_{\rm SBO}$ using the phenomenological models from \citetalias{2015ApJ...799..215P} could be further improved with detailed spectroscopic data during the rising phase of the light curve.

\subsection{GW Detection Benefits}

The statistical confidence of a GW candidate with given false alarm rate (FAR) is quantified through the false alarm probability (FAP):
\begin{equation}
    \mathrm{FAP} = 1-e^{-\Delta T_{\rm GSW}\times \mathrm{FAR}} \approx \Delta T_{\rm GSW}\times \mathrm{FAR},
\end{equation}
where $\Delta T_{\rm GSW}$ is the GSW duration. The reduction of the GSW by a factor of 2 allows a factor of 2 reduction in the FAP.  In Figure~\ref{fig:FARFAP} we show the impact of the GSW duration on the FAP as a function of FAR.  We see that the method presented in the paper leads to a substantial improvement over the GSW durations from the \citetalias{oosnsearch} method. For example, for SN\,2019fcn a detection at ${\rm FAR}\lesssim 1$ yr$^{-1}$ will have a significance of $\sim 3\sigma$ with our GSW.

\section{Conclusions}
\label{sec:conc}

In this work, we presented an astrophysically-motivated approach for constraining the time of gravitational wave emission in targeted searches of CCSNe. The approaches presented here range in complexity from model-agnostic polynomial fits, to model-dependent fits, and a data-driven approach.  We find general agreement between all approaches, with the most constraining results coming from the empirical model and the shock cooling model. Using the combined posteriors of these models we find an uncertainty on $t_{\rm SBO}$ of $\approx 0.3$ d for SN\,2019fcn and $\approx 1.9$ d for SN\,2019ejj.

Thus, for SNe with well-sampled early light curves (such as SN\,2019fcn) the resulting uncertainty on the time of shock breakout can be a few hours.  In those cases, the dominant uncertainty on the final GSW (i.e., the time of core-collapse and GW emission) is entirely influenced by current theoretical uncertainties in $\Delta t_{\rm cc}$ -- the time between core-collapse and shock breakout.  Here we take a conservative range of $1.4-3$ days to encompass model uncertainties and the unknown progenitor mass. Extracting information about the progenitor from the optical light curves and spectra may serve to reduce this uncertainty.  

Our analysis continues to emphasize the importance of early discovery and detailed follow-up of nearby CCSNe to further reduce the GSW for individual events that overlap LVK science runs.  For the present, we stress that the GSW results provided here for SNe 2019ejj and 2019fcn should be used in searches of LIGO/Virgo Observing Run 3 for GW emission.  
%from these two SNe.

\begin{acknowledgements}
We thank Haakon Andresen, Sebastian Gomez, Evan O'Connor, Ondrej Pejcha, Tomas Prieto, David Sand, and Peter Williams for insightful discussions. The Berger Time-Domain Group at Harvard is supported by NSF and NASA grants. This research has made use of data, software and/or web tools obtained from the Gravitational Wave Open Science Center, a service of LIGO Laboratory, the LIGO Scientific Collaboration and the Virgo Collaboration. MS and MZ are grateful for computational resources provided by the LIGO Laboratory. MS is supported by National Science Foundation Grants PHY-0757058 and PHY-0823459. MZ is supported by NSF Grant PHY-1806885. 
\end{acknowledgements}

\appendix

\section{Assessing Methodology Reliability}

One important item to assess is the reliability of each of the methodologies (described in \S\ref{sec:disc}) when estimating $t_{\rm SBO}$. We can assess this reliability by understanding the \textit{true error} associated with each methodology. We define this as, ${\rm Error} = \sqrt{\rm MSE} =\sqrt{(\rm SV)^2 + (\rm B)^2}$, where $MSE$ is the mean squared error, $SV$ is the statistical variance, and $B$ is the bias associated with each methodology. In this paper, we report the statistical variance (the uncertainty) associated with each posterior. It is important to note that all 1$\sigma$ ranges reported in this work assume bias = 0.

\label{sec:TM}

\subsection{Bias' Present in the Quadratic Model}
We removed 1-5 days worth of data and used the remaining light curve to estimate $t_{\rm SBO}$ for KSN 2011d using the approach described in \ref{sec:QP}. We require that $t_{\rm SBO}$ occur earlier than $t_{\rm fd}$, which we define as zero days. Table \ref{tab:quadtest} lists the resulting $t_{\rm SBO}$ estimates. 

The parameters of the polynomial model, $t_{\rm SBO}$, $\alpha$ and $\beta$, give an approximation of the curvature associated with the early part of the light curve. We can define the rise time as $-\beta/(2\alpha)$. Having observations closer to the peak of the light curve forces the interpolation to be dominated by the curvature at the peak of the light curve, defined by $\alpha$. If the magnitude of $\alpha$ is smaller than the average curvature of the light curve\footnote{$\beta$ has no role in the curvature since it has a linear dependence with time}, then the rise time will be overestimated. If the prior on $\alpha$ overfits the peak of the light curve, this also leads to an underestimation of $t_{\rm SBO}$.

\begin{deluxetable}{lccr}[h!]
\tablecaption{Quadratic SBO estimates for KSN2011d. \label{tab:DQ}}
\tablehead{\colhead{Days removed from $t_{fd}$} & \colhead{$t_{\rm SBO}$} & \colhead{$\chi^2$} & \colhead{$\chi^2_{red}$}}
\startdata
1                           & 1.26 $\pm$ 0.12 & 74                           & 1.00                                   \\
2                           & 3.12 $\pm$ 0.34 & 113                          & 1.52                                   \\
3                           & 3.01 $\pm$ 0.36 & 42                           & 0.57                                   \\
4                           & 3.75 $\pm$ 0.35 & 38                           & 0.51                                   \\
5                           & 4.16 $\pm$ 0.39 & 34                           & 0.46                       \\      
\enddata 
\tablecomments{We show that the later the light curve is discovered past SBO, the more the quadratic underestimates $t_{\rm SBO}$. }
\label{tab:quadtest}
\end{deluxetable}

\subsection{The Efficacy of Empirical Model}

We discuss the efficacy of our empirical model fitting if there isn't early photometry available for our nearby SNe. We estimate $t_{\rm SBO}$ for two different scenarios: one with respect to a light curve that is complete (and therefore have access to $t_{\rm SBO}$), while the other one has only an incomplete, low cadence light curve available.

\begin{figure*}[h!] 
\centering 
\includegraphics[width=1\textwidth]{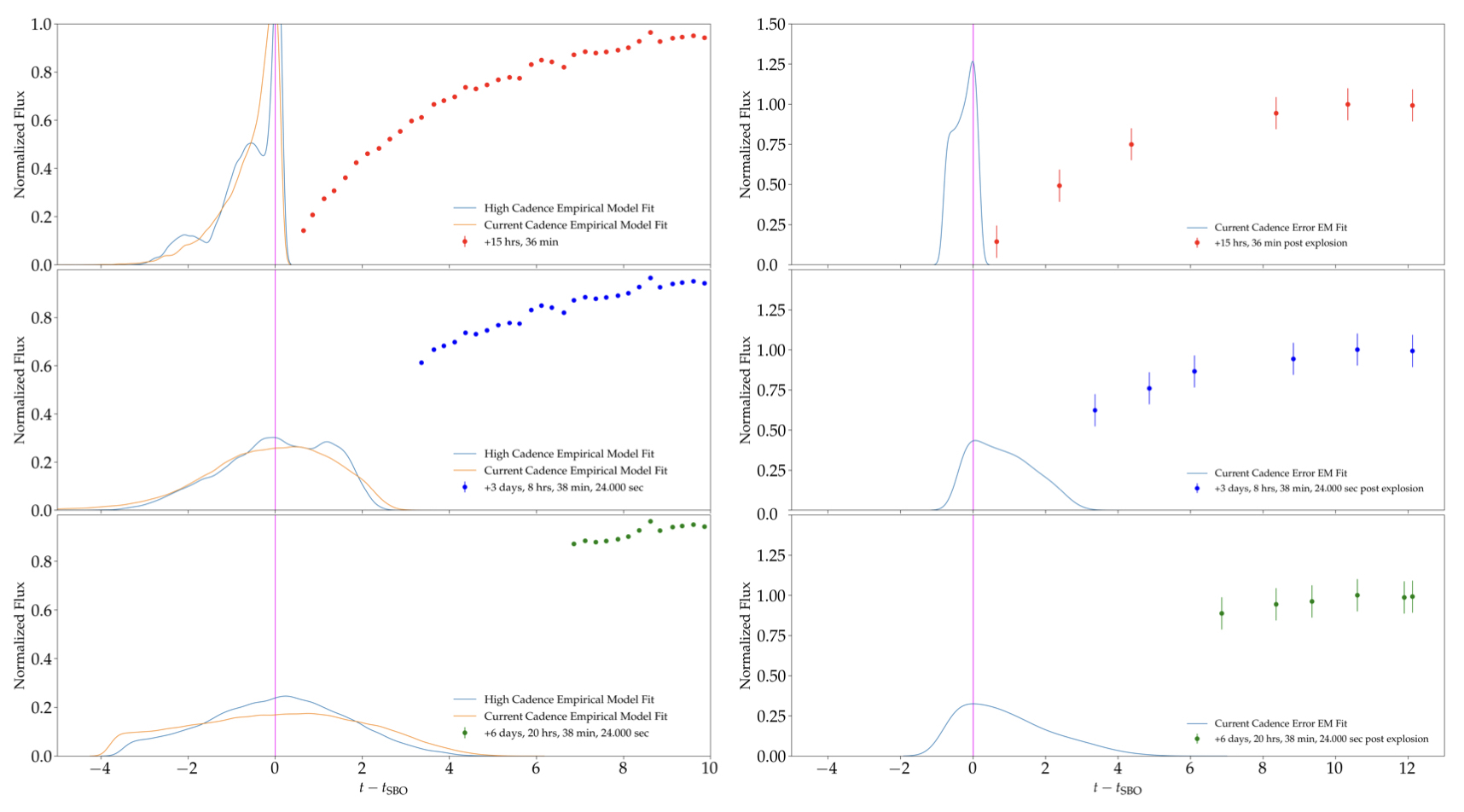}
\caption{We show the resulting posteriors for our Kepler test. The \textit{Left} panel shows the resulting posterior for $t_{\rm SBO}$ using all 9 with the original Kepler photometric points with both \textit{High} and \textit{Current} error bars. The \textit{Right} panel uses all 9 light curve models fitted to 6 randomized observed points from the light curve of KSN 2011d with \textit{Current} cadence error bars.}
\label{fig:KLC} 
\end{figure*}

We replicate the methodology described in \ref{sec:emp} and use KSN 2011d as a test case to estimate $t_{\rm SBO}$. We removed 15 hours, 3 and 6 days worth of data and used the remaining light curve to estimate $t_{\rm SBO}$. We randomized which observational points were fed into the MCMC. We also rescale the associated error bars in order to understand the variance in the quality of both present and planned surveys \citep{2016arXiv161205560C, 2018AAS...23124511S, 10.1093/mnras/stz073, 10.1093/mnras/stx1544, 2021MNRAS.500.5639V}. \textit{High Cadence} represents space based surveys with \textit{Kepler}-like cadence with error bars on the order of $10^{-4}$ whereas \textit{Low Cadence} represents ground based survey cadence with error bars on the order of $10^{-2}$. This is shown in Figure \ref{fig:KLC}.

We note a few things from Figure \ref{fig:KLC}. First, both \textit{High} and \textit{Low} cadence distributions are asymmetric. Secondly, the difference between the \textit{High} and \textit{Low} cadence resolution were reflected their respective posterior shapes. Thirdly, as expected, the posterior distribution became wider with decreasing rise information being fed into the MCMC. However, all posteriors were centered around the known value of $t_{\rm SBO}$.

\section{Additional Models}
\label{sec:PMM}

\subsection{Power Law}
\label{sec:PLF}
\begin{figure*}[b] 
\centering 
\includegraphics[width=1\textwidth]{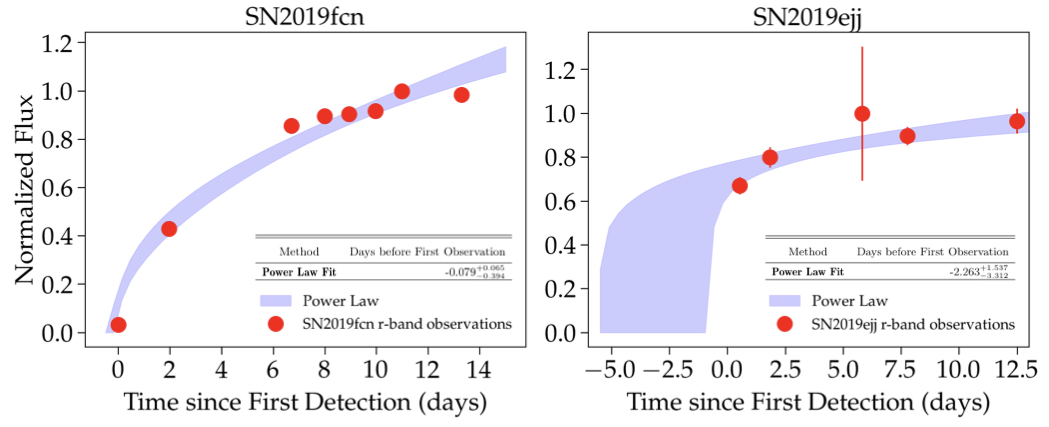}
\caption{We show the power law fits for both SN~2019fcn and SN~2019ejj.}
\label{fig:powerfit} 
\end{figure*}

We fit the rising phase of SNe following a dominating power-law of the form,
\begin{equation}
F(t) = \alpha (t - t_\mathrm{SBO})^n,
\end{equation}
where $t_\mathrm{SBO}$ is the time of first light (i.e., zero flux) and $\alpha$ and $n$ are coefficients with no particular physical meaning. $\alpha$ is the flux on day 1 and $n$ is the power-law index. We use a Markov-chain Monte Carlo (MCMC) routine to fit for the three free parameters ($\alpha$, $n$ and $t_{\rm SBO}$). For $t_{\rm SBO}$, we use a uniform prior set before first detection. For $\alpha$ and $n$, we use a uniform prior between two arbitrary bounds. We also fit for an intrinsic scatter parameter, $\sigma$, which we add in quadrature to the observational uncertainties, with a half-Gaussian prior peaking at 0 with a standard deviation of 0.1 in units of flux. Figure~\ref{fig:powerfit} shows the resulting fits.

\begin{figure*}[t] 
\centering 
\includegraphics[width=1\textwidth]{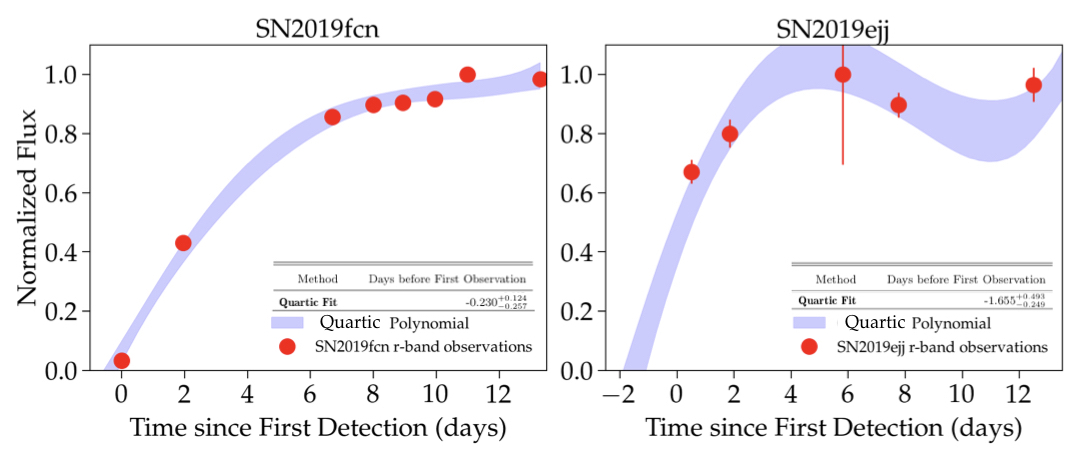}
\caption{We show the quartic fits from \S\ref{sec:PMM} for both SN~2019fcn and SN~2019ejj. The intersection of the blue region with the x-axis represents the 1$\sigma$ range on $t_{\rm SBO}$ for SN~2019fcn and SN~2019ejj. Since the rise was not well sampled in the case of SN~2019ejj, the quartic estimation of $t_{\rm SBO}$ is underestimated and is much closer to the time of first observation.
} 
\label{fig:quarticfit} 
\end{figure*}

A well-sampled rise, as in the case of SN2019fcn is fit relatively well with minimal associated uncertainties, although $t_{\rm SBO}$ is still underestimated. However, for the case for SN2019ejj $t_{\rm SBO}$ is underestimated by more than a few days.
\subsection{Quartic Polynomial}
We also fit a fourth order polynomial (quartic) to SN~2019fcn and SN~2019ejj. The same problems still persist with a higher degree polynomial: The curve lands before the initial value of $t_{\rm SBO}$ because the data from [0, 2] days is narrower than data for days [2, 12]. Therefore, the $t_{\rm SBO}$ is underestimated. This is especially the case for SN~2019ejj, as shown in Figure~\ref{fig:quarticfit}.

\bibliography{osw}

\begin{thebibliography}{}
\expandafter\ifx\csname natexlab\endcsname\relax\def\natexlab#1{#1}\fi
\providecommand{\url}[1]{\href{#1}{#1}}
\providecommand{\dodoi}[1]{doi:~\href{http://doi.org/#1}{\nolinkurl{#1}}}
\providecommand{\doeprint}[1]{\href{http://ascl.net/#1}{\nolinkurl{http://ascl.net/#1}}}
\providecommand{\doarXiv}[1]{\href{https://arxiv.org/abs/#1}{\nolinkurl{https://arxiv.org/abs/#1}}}

\bibitem[{{Abadie} {et~al.}(2010){Abadie}, {Abbott}, {Abbott}, {Accadia},
  {Acernese}, {Adhikari}, {Ajith}, {Allen}, {Allen}, {Amador Ceron}, {Amin},
  {Anderson}, {Anderson}, {Antonucci}, {Arain}, {Araya}, {Arun}, {Aso},
  {Aston}, {Astone}, {Aufmuth}, {Aulbert}, {Babak}, {Baker}, {Ballardin},
  {Ballmer}, {Barker}, {Barone}, {Barr}, {Barriga}, {Barsotti}, {Barsuglia},
  {Barton}, {Bartos}, {Bassiri}, {Bastarrika}, {Bauer}, {Behnke}, {Beker},
  {Belletoile}, {Benacquista}, {Betzwieser}, {Beyersdorf}, {Bigotta},
  {Bilenko}, {Billingsley}, {Birindelli}, {Biswas}, {Bizouard}, {Black},
  {Blackburn}, {Blackburn}, {Blair}, {Bland}, {Blom}, {Boccara}, {Bock},
  {Bodiya}, {Bondarescu}, {Bondu}, {Bonelli}, {Bonnand}, {Bork}, {Born},
  {Bose}, {Bosi}, {Bouhou}, {Braccini}, {Bradaschia}, {Brady}, {Braginsky},
  {Brau}, {Breyer}, {Bridges}, {Brillet}, {Brinkmann}, {Brisson}, {Britzger},
  {Brooks}, {Brown}, {Budzy{\'n}ski}, {Bulik}, {Bullington}, {Bulten},
  {Buonanno}, {Burmeister}, {Buskulic}, {Buy}, {Byer}, {Cadonati}, {Cagnoli},
  {Cain}, {Calloni}, {Camp}, {Campagna}, {Cannizzo}, {Cannon}, {Canuel}, {Cao},
  {Capano}, {Carbognani}, {Cardenas}, {Caudill}, {Cavagli{\`a}}, {Cavalier},
  {Cavalieri}, {Cella}, {Cepeda}, {Cesarini}, {Chalermsongsak}, {Chalkley},
  {Charlton}, {Chassande-Mottin}, {Chatterji}, {Chelkowski}, {Chen},
  {Chincarini}, {Christensen}, {Chua}, {Chung}, {Clark}, {Clark}, {Clayton},
  {Cleva}, {Coccia}, {Colacino}, {Colas}, {Colla}, {Colombini}, {Conte},
  {Cook}, {Corbitt}, {Cornish}, {Corsi}, {Coulon}, {Coward}, {Coyne},
  {Creighton}, {Creighton}, {Cruise}, {Culter}, {Cumming}, {Cunningham},
  {Cuoco}, {Dahl}, {Danilishin}, {D'Antonio}, {Danzmann}, {Dattilo}, {Daudert},
  {Davier}, {Davies}, {Daw}, {Day}, {Dayanga}, {de Rosa}, {Debra}, {Degallaix},
  {Del Prete}, {Dergachev}, {Desalvo}, {Dhurandhar}, {di Fiore}, {di Lieto},
  {di Paolo Emilio}, {di Virgilio}, {D{\'\i}az}, {Dietz}, {Donovan}, {Dooley},
  {Doomes}, {Drago}, {Drever}, {Driggers}, {Dueck}, {Duke}, {Dumas}, {Edgar},
  {Edwards}, {Effler}, {Ehrens}, {Etzel}, {Evans}, {Evans}, {Fafone},
  {Fairhurst}, {Faltas}, {Fan}, {Fazi}, {Fehrmann}, {Ferrante}, {Fidecaro},
  {Finn}, {Fiori}, {Flaminio}, {Flasch}, {Foley}, {Forrest}, {Fotopoulos},
  {Fournier}, {Franc}, {Frasca}, {Frasconi}, {Frede}, {Frei}, {Frei}, {Freise},
  {Frey}, {Fricke}, {Friedrich}, {Fritschel}, {Frolov}, {Fulda}, {Fyffe},
  {Galimberti}, {Gammaitoni}, {Garofoli}, {Garufi}, {Gemme}, {Genin}, {Gennai},
  {Ghosh}, {Giaime}, {Giampanis}, {Giardina}, {Giazotto}, {Goetz}, {Goggin},
  {Gonz{\'a}lez}, {Go{\ss}ler}, {Gouaty}, {Granata}, {Grant}, {Gras}, {Gray},
  {Greenhalgh}, {Gretarsson}, {Greverie}, {Grosso}, {Grote}, {Grunewald},
  {Guidi}, {Gustafson}, {Gustafson}, {Hage}, {Hallam}, {Hammer}, {Hammond},
  {Hanna}, {Hanson}, {Harms}, {Harry}, {Harry}, {Harstad}, {Haughian},
  {Hayama}, {Hayau}, {Hayler}, {Heefner}, {Heitmann}, {Hello}, {Heng},
  {Heptonstall}, {Hewitson}, {Hild}, {Hirose}, {Hoak}, {Hodge}, {Holt},
  {Hosken}, {Hough}, {Howell}, {Hoyland}, {Huet}, {Hughey}, {Husa}, {Huttner},
  {Ingram}, {Isogai}, {Ivanov}, {Jaranowski}, {Johnson}, {Jones}, {Jones},
  {Jones}, {Ju}, {Kalmus}, {Kalogera}, {Kandhasamy}, {Kanner}, {Katsavounidis},
  {Kawabe}, {Kawamura}, {Kawazoe}, {Kells}, {Keppel}, {Khalaidovski},
  {Khalili}, {Khan}, {Khazanov}, {Kim}, {King}, {Kissel}, {Klimenko},
  {Kokeyama}, {Kondrashov}, {Kopparapu}, {Koranda}, {Kowalska}, {Kozak},
  {Kringel}, {Krishnan}, {Kr{\'o}lak}, {Kuehn}, {Kullman}, {Kumar}, {Kwee},
  {Lam}, {Landry}, {Lang}, {Lantz}, {Lastzka}, {Lazzarini}, {Leaci}, {Lei},
  {Leindecker}, {Leonor}, {Leroy}, {Letendre}, {Li}, {Lin}, {Lindquist},
  {Littenberg}, {Lockerbie}, {Lodhia}, {Lorenzini}, {Loriette}, {Lormand},
  {Losurdo}, {Lu}, {Lubi{\'n}ski}, {Lucianetti}, {L{\"u}ck}, {Lundgren},
  {Machenschalk}, {Macinnis}, {Mageswaran}, {Mailand}, {Majorana}, {Mak},
  {Maksimovic}, {Man}, {Mandel}, {Mandic}, {Mantovani}, {Marchesoni}, {Marion},
  {M{\'a}rka}, {M{\'a}rka}, {Markosyan}, {Markowitz}, {Maros}, {Marque},
  {Martelli}, {Martin}, {Martin}, {Marx}, {Mason}, {Masserot}, {Matichard},
  {Matone}, {Matzner}, {Mavalvala}, {McCarthy}, {McClelland}, {McGuire},
  {McIntyre}, {McKechan}, {Mehmet}, {Melatos}, {Melissinos}, {Mendell},
  {Men{\'e}ndez}, {Mercer}, {Merill}, {Meshkov}, {Messenger}, {Meyer}, {Miao},
  {Michel}, {Milano}, {Miller}, {Minenkov}, {Mino}, {Mitra}, {Mitrofanov},
  {Mitselmakher}, {Mittleman}, {Miyakawa}, {Moe}, {Mohan}, {Mohanty},
  {Mohapatra}, {Moreau}, {Moreno}, {Morgado}, {Morgia}, {Mors}, {Mosca},
  {Moscatelli}, {Mossavi}, {Mours}, {Mowlowry}, {Mueller}, {Mukherjee},
  {Mullavey}, {M{\"u}ller-Ebhardt}, {Munch}, {Murray}, {Nash}, {Nawrodt},
  {Nelson}, {Neri}, {Newton}, {Nishida}, {Nishizawa}, {Nocera}, {Ochsner},
  {O'Dell}, {Ogin}, {Oldenburg}, {O'Reilly}, {O'Shaughnessy}, {Ottaway},
  {Ottens}, {Overmier}, {Owen}, {Page}, {Pagliaroli}, {Palladino}, {Palomba},
  {Pan}, {Pankow}, {Paoletti}, {Papa}, {Pardi}, {Parisi}, {Pasqualetti},
  {Passaquieti}, {Passuello}, {Patel}, {Pathak}, {Pedraza}, {Pekowsky}, {Penn},
  {Peralta}, {Perreca}, {Persichetti}, {Pichot}, {Pickenpack}, {Piergiovanni},
  {Pietka}, {Pinard}, {Pinto}, {Pitkin}, {Pletsch}, {Plissi}, {Poggiani},
  {Postiglione}, {Prato}, {Principe}, {Prix}, {Prodi}, {Prokhorov}, {Puncken},
  {Punturo}, {Puppo}, {Quetschke}, {Raab}, {Rabeling}, {Rabeling}, {Radkins},
  {Raffai}, {Raics}, {Rakhmanov}, {Rapagnani}, {Raymond}, {Re}, {Reed}, {Reed},
  {Regimbau}, {Rehbein}, {Reid}, {Reitze}, {Ricci}, {Riesen}, {Riles},
  {Roberts}, {Robertson}, {Robinet}, {Robinson}, {Robinson}, {Rocchi}, {Roddy},
  {R{\"o}ver}, {Rolland}, {Rollins}, {Romano}, {Romano}, {Romie},
  {Rosi{\'n}ska}, {Rowan}, {R{\"u}diger}, {Ruggi}, {Ryan}, {Sakata}, {Salemi},
  {Sammut}, {Sancho de La Jordana}, {Sandberg}, {Sannibale}, {Santamar{\'\i}a},
  {Santostasi}, {Saraf}, {Sarin}, {Sassolas}, {Sathyaprakash}, {Sato},
  {Satterthwaite}, {Saulson}, {Savage}, {Schilling}, {Schnabel}, {Schofield},
  {Schulz}, {Schutz}, {Schwinberg}, {Scott}, {Scott}, {Searle}, {Seifert},
  {Sellers}, {Sengupta}, {Sentenac}, {Sergeev}, {Shapiro}, {Shawhan},
  {Shoemaker}, {Sibley}, {Siemens}, {Sigg}, {Sintes}, {Skelton}, {Slagmolen},
  {Slutsky}, {Smith}, {Smith}, {Smith}, {Somiya}, {Sorazu}, {Sperandio},
  {Stein}, {Stein}, {Steplewski}, {Stochino}, {Stone}, {Strain}, {Strigin},
  {Stroeer}, {Sturani}, {Stuver}, {Summerscales}, {Sung}, {Susmithan},
  {Sutton}, {Swinkels}, {Szokoly}, {Talukder}, {Tanner}, {Tarabrin}, {Taylor},
  {Taylor}, {Thorne}, {Thorne}, {Th{\"u}ring}, {Titsler}, {Tokmakov},
  {Toncelli}, {Tonelli}, {Torres}, {Torrie}, {Tournefier}, {Travasso},
  {Traylor}, {Trias}, {Trummer}, {Turner}, {Ugolini}, {Urbanek}, {Vahlbruch},
  {Vajente}, {Vallisneri}, {van den Brand}, {van den Broeck}, {van der Putten},
  {van der Sluys}, {Vass}, {Vaulin}, {Vavoulidis}, {Vecchio}, {Vedovato}, {van
  Veggel}, {Veitch}, {Veitch}, {Veltkamp}, {Verkindt}, {Vetrano}, {Vicer{\'e}},
  {Villar}, {Vinet}, {Vocca}, {Vorvick}, {Vyachanin}, {Waldman}, {Wallace},
  {Wanner}, {Ward}, {Was}, {Wei}, {Weinert}, {Weinstein}, {Weiss}, {Wen},
  {Wen}, {Wessels}, {West}, {Westphal}, {Wette}, {Whelan}, {Whitcomb},
  {Whiting}, {Wilkinson}, {Willems}, {Williams}, {Williams}, {Willke},
  {Wilmut}, {Winkelmann}, {Winkler}, {Wipf}, {Wiseman}, {Woan}, {Wooley},
  {Worden}, {Yakushin}, {Yamamoto}, {Yamamoto}, {Yeaton-Massey}, {Yoshida},
  {Yvert}, {Zanolin}, {Zhang}, {Zhang}, {Zhao}, {Zotov}, {Zucker}, {Zweizig},
  {LIGO Scientific Collaboration}, \& {Virgo
  Collaboration}}]{2010PhRvD..81j2001A}
{Abadie}, J., {Abbott}, B.~P., {Abbott}, R., {et~al.} 2010, \prd, 81, 102001,
  \dodoi{10.1103/PhysRevD.81.102001}

\bibitem[{{Abbott} {et~al.}(2016{\natexlab{a}}){Abbott}, {Abbott}, {Abbott},
  {et~al.}}]{abbott:16a}
{Abbott}, B.~P., {Abbott}, R., {Abbott}, T.~D., {et~al.} 2016{\natexlab{a}},
  Phys. Rev. Lett., 116, 061102, \dodoi{10.1103/PhysRevLett.116.061102}

\bibitem[{{Abbott} {et~al.}(2016{\natexlab{b}}){Abbott}, {Abbott}, {Abbott},
  {et~al.}}]{snsearch}
---. 2016{\natexlab{b}}, Phys. Rev. D, 94, 102001,
  \dodoi{10.1103/PhysRevD.94.102001}

\bibitem[{{Abbott} {et~al.}(2016{\natexlab{c}}){Abbott}, {Abbott}, {Abbott},
  {Abernathy}, {Acernese}, {Ackley}, {Adams}, {Adams}, {Addesso}, {Adhikari},
  {Adya}, {Affeldt}, {Agathos}, {Agatsuma}, {Aggarwal}, {Aguiar}, {Aiello},
  {Ain}, {Ajith}, {Allen}, {Allocca}, {Altin}, {Anderson}, {Anderson}, {Arai},
  {Araya}, {Arceneaux}, {Areeda}, {Arnaud}, {Arun}, {Ascenzi}, {Ashton}, {Ast},
  {Aston}, {Astone}, {Aufmuth}, {Aulbert}, {Babak}, {Bacon}, {Bader}, {Baker},
  {Baldaccini}, {Ballardin}, {Ballmer}, {Barayoga}, {Barclay}, {Barish},
  {Barker}, {Barone}, {Barr}, {Barsotti}, {Barsuglia}, {Barta}, {Bartlett},
  {Bartos}, {Bassiri}, {Basti}, {Batch}, {Baune}, {Bavigadda}, {Bazzan},
  {Behnke}, {Bejger}, {Bell}, {Bell}, {Berger}, {Bergman}, {Bergmann}, {Berry},
  {Bersanetti}, {Bertolini}, {Betzwieser}, {Bhagwat}, {Bhand are}, {Bilenko},
  {Billingsley}, {Birch}, {Birney}, {Biscans}, {Bisht}, {Bitossi}, {Biwer},
  {Bizouard}, {Blackburn}, {Blair}, {Blair}, {Blair}, {Bloemen}, {Bock},
  {Bodiya}, {Boer}, {Bogaert}, {Bogan}, {Bohe}, {Bojtos}, {Bond}, {Bondu},
  {Bonnand}, {Boom}, {Bork}, {Boschi}, {Bose}, {Bouffanais}, {Bozzi},
  {Bradaschia}, {Brady}, {Braginsky}, {Branchesi}, {Brau}, {Briant}, {Brillet},
  {Brinkmann}, {Brisson}, {Brockill}, {Brooks}, {Brown}, {Brown}, {Brown},
  {Buchanan}, {Buikema}, {Bulik}, {Bulten}, {Buonanno}, {Buskulic}, {Buy},
  {Byer}, {Cadonati}, {Cagnoli}, {Cahillane}, {Calder{\'o}n Bustillo},
  {Callister}, {Calloni}, {Camp}, {Cannon}, {Cao}, {Capano}, {Capocasa},
  {Carbognani}, {Caride}, {Casanueva Diaz}, {Casentini}, {Caudill},
  {Cavagli{\`a}}, {Cavalier}, {Cavalieri}, {Cella}, {Cepeda}, {Cerboni
  Baiardi}, {Cerretani}, {Cesarini}, {Chakraborty}, {Chalermsongsak},
  {Chamberlin}, {Chan}, {Chao}, {Charlton}, {Chassande-Mottin}, {Chen}, {Chen},
  {Cheng}, {Chincarini}, {Chiummo}, {Cho}, {Cho}, {Chow}, {Christensen}, {Chu},
  {Chua}, {Chung}, {Ciani}, {Clara}, {Clark}, {Cleva}, {Coccia}, {Cohadon},
  {Colla}, {Collette}, {Cominsky}, {Constancio}, {Conte}, {Conti}, {Cook},
  {Corbitt}, {Cornish}, {Corsi}, {Cortese}, {Costa}, {Coughlin}, {Coughlin},
  {Coulon}, {Countryman}, {Couvares}, {Cowan}, {Coward}, {Cowart}, {Coyne},
  {Coyne}, {Craig}, {Creighton}, {Cripe}, {Crowder}, {Cumming}, {Cunningham},
  {Cuoco}, {Dal Canton}, {Danilishin}, {D'Antonio}, {Danzmann}, {Darman},
  {Dattilo}, {Dave}, {Daveloza}, {Davier}, {Davies}, {Daw}, {Day}, {DeBra},
  {Debreczeni}, {Degallaix}, {De Laurentis}, {Del{\'e}glise}, {Del Pozzo},
  {Denker}, {Dent}, {Dereli}, {Dergachev}, {DeRosa}, {De Rosa}, {DeSalvo},
  {Dhurandhar}, {D{\'\i}az}, {Di Fiore}, {Di Giovanni}, {Di Lieto}, {Di Pace},
  {Di Palma}, {Di Virgilio}, {Dojcinoski}, {Dolique}, {Donovan}, {Dooley},
  {Doravari}, {Douglas}, {Downes}, {Drago}, {Drever}, {Driggers}, {Du},
  {Ducrot}, {Dwyer}, {Edo}, {Edwards}, {Effler}, {Eggenstein}, {Ehrens},
  {Eichholz}, {Eikenberry}, {Engels}, {Essick}, {Etzel}, {Evans}, {Evans},
  {Everett}, {Factourovich}, {Fafone}, {Fair}, {Fairhurst}, {Fan}, {Fang},
  {Farinon}, {Farr}, {Farr}, {Favata}, {Fays}, {Fehrmann}, {Fejer}, {Ferrante},
  {Ferreira}, {Ferrini}, {Fidecaro}, {Fiori}, {Fiorucci}, {Fisher}, {Flaminio},
  {Fletcher}, {Fournier}, {Franco}, {Frasca}, {Frasconi}, {Frei}, {Freise},
  {Frey}, {Frey}, {Fricke}, {Fritschel}, {Frolov}, {Fulda}, {Fyffe}, {Gabbard},
  {Gair}, {Gammaitoni}, {Gaonkar}, {Garufi}, {Gatto}, {Gaur}, {Gehrels},
  {Gemme}, {Gendre}, {Genin}, {Gennai}, {George}, {Gergely}, {Germain},
  {Ghosh}, {Ghosh}, {Giaime}, {Giardina}, {Giazotto}, {Gill}, {Glaefke},
  {Goetz}, {Goetz}, {Gondan}, {Gonz{\'a}lez}, {Gonzalez Castro}, {Gopakumar},
  {Gordon}, {Gorodetsky}, {Gossan}, {Gosselin}, {Gouaty}, {Graef}, {Graff},
  {Granata}, {Grant}, {Gras}, {Gray}, {Greco}, {Green}, {Groot}, {Grote},
  {Grunewald}, {Guidi}, {Guo}, {Gupta}, {Gupta}, {Gushwa}, {Gustafson},
  {Gustafson}, {Hacker}, {Hall}, {Hall}, {Hammond}, {Haney}, {Hanke}, {Hanks},
  {Hanna}, {Hannam}, {Hanson}, {Hardwick}, {Haris}, {Harms}, {Harry}, {Harry},
  {Hart}, {Hartman}, {Haster}, {Haughian}, {Heidmann}, {Heintze}, {Heitmann},
  {Hello}, {Hemming}, {Hendry}, {Heng}, {Hennig}, {Heptonstall}, {Heurs},
  {Hild}, {Hoak}, {Hodge}, {Hofman}, {Hollitt}, {Holt}, {Holz}, {Hopkins},
  {Hosken}, {Hough}, {Houston}, {Howell}, {Hu}, {Huang}, {Huerta}, {Huet},
  {Hughey}, {Husa}, {Huttner}, {Huynh-Dinh}, {Idrisy}, {Indik}, {Ingram},
  {Inta}, {Isa}, {Isac}, {Isi}, {Islas}, {Isogai}, {Iyer}, {Izumi}, {Jacqmin},
  {Jang}, {Jani}, {Jaranowski}, {Jawahar}, {Jim{\'e}nez-Forteza}, {Johnson},
  {Jones}, {Jones}, {Jonker}, {Ju}, {Kalaghatgi}, {Kalogera}, {Kandhasamy},
  {Kang}, {Kanner}, {Karki}, {Kasprzack}, {Katsavounidis}, {Katzman}, {Kaufer},
  {Kaur}, {Kawabe}, {Kawazoe}, {K{\'e}f{\'e}lian}, {Kehl}, {Keitel}, {Kelley},
  {Kells}, {Kennedy}, {Key}, {Khalaidovski}, {Khalili}, {Khan}, {Khan}, {Khan},
  {Khazanov}, {Kijbunchoo}, {Kim}, {Kim}, {Kim}, {Kim}, {Kim}, {Kim}, {King},
  {King}, {Kinzel}, {Kissel}, {Kleybolte}, {Klimenko}, {Koehlenbeck},
  {Kokeyama}, {Koley}, {Kondrashov}, {Kontos}, {Korobko}, {Korth}, {Kowalska},
  {Kozak}, {Kringel}, {Kr{\'o}lak}, {Krueger}, {Kuehn}, {Kumar}, {Kuo},
  {Kutynia}, {Lackey}, {Land ry}, {Lange}, {Lantz}, {Lasky}, {Lazzarini},
  {Lazzaro}, {Leaci}, {Leavey}, {Lebigot}, {Lee}, {Lee}, {Lee}, {Lee}, {Lenon},
  {Leonardi}, {Leong}, {Leroy}, {Letendre}, {Levin}, {Levine}, {Li}, {Libson},
  {Littenberg}, {Lockerbie}, {Logue}, {Lombardi}, {Lord}, {Lorenzini},
  {Loriette}, {Lormand}, {Losurdo}, {Lough}, {L{\"u}ck}, {Lundgren}, {Luo},
  {Lynch}, {Ma}, {MacDonald}, {Machenschalk}, {MacInnis}, {Macleod},
  {Maga{\~n}a-Sandoval}, {Magee}, {Mageswaran}, {Majorana}, {Maksimovic},
  {Malvezzi}, {Man}, {Mandel}, {Mandic}, {Mangano}, {Mansell}, {Manske},
  {Mantovani}, {Marchesoni}, {Marion}, {M{\'a}rka}, {M{\'a}rka}, {Markosyan},
  {Maros}, {Martelli}, {Martellini}, {Martin}, {Martin}, {Martynov}, {Marx},
  {Mason}, {Masserot}, {Massinger}, {Masso-Reid}, {Matichard}, {Matone},
  {Mavalvala}, {Mazumder}, {Mazzolo}, {McCarthy}, {McClelland}, {McCormick},
  {McGuire}, {McIntyre}, {McIver}, {McManus}, {McWilliams}, {Meacher},
  {Meadors}, {Meidam}, {Melatos}, {Mendell}, {Mendoza-Gandara}, {Mercer},
  {Merilh}, {Merzougui}, {Meshkov}, {Messenger}, {Messick}, {Meyers},
  {Mezzani}, {Miao}, {Michel}, {Middleton}, {Mikhailov}, {Milano}, {Miller},
  {Millhouse}, {Minenkov}, {Ming}, {Mirshekari}, {Mishra}, {Mitra},
  {Mitrofanov}, {Mitselmakher}, {Mittleman}, {Moggi}, {Mohan}, {Mohapatra},
  {Montani}, {Moore}, {Moore}, {Moraru}, {Moreno}, {Morriss}, {Mossavi},
  {Mours}, {Mow-Lowry}, {Mueller}, {Mueller}, {Muir}, {Mukherjee}, {Mukherjee},
  {Mukherjee}, {Mukund}, {Mullavey}, {Munch}, {Murphy}, {Murray}, {Mytidis},
  {Nardecchia}, {Naticchioni}, {Nayak}, {Necula}, {Nedkova}, {Nelemans},
  {Neri}, {Neunzert}, {Newton}, {Nguyen}, {Nielsen}, {Nissanke}, {Nitz},
  {Nocera}, {Nolting}, {Normandin}, {Nuttall}, {Oberling}, {Ochsner}, {O'Dell},
  {Oelker}, {Ogin}, {Oh}, {Oh}, {Ohme}, {Oliver}, {Oppermann}, {Oram},
  {O'Reilly}, {O'Shaughnessy}, {Ottaway}, {Ottens}, {Overmier}, {Owen}, {Pai},
  {Pai}, {Palamos}, {Palashov}, {Palomba}, {Pal-Singh}, {Pan}, {Pankow},
  {Pannarale}, {Pant}, {Paoletti}, {Paoli}, {Papa}, {Paris}, {Parker},
  {Pascucci}, {Pasqualetti}, {Passaquieti}, {Passuello}, {Patricelli},
  {Patrick}, {Pearlstone}, {Pedraza}, {Pedurand}, {Pekowsky}, {Pele}, {Penn},
  {Perreca}, {Phelps}, {Piccinni}, {Pichot}, {Piergiovanni}, {Pierro},
  {Pillant}, {Pinard}, {Pinto}, {Pitkin}, {Poggiani}, {Popolizio}, {Post},
  {Powell}, {Prasad}, {Predoi}, {Premachandra}, {Prestegard}, {Price},
  {Prijatelj}, {Principe}, {Privitera}, {Prodi}, {Prokhorov}, {Puncken},
  {Punturo}, {Puppo}, {P{\"u}rrer}, {Qi}, {Qin}, {Quetschke}, {Quintero},
  {Quitzow-James}, {Raab}, {Rabeling}, {Radkins}, {Raffai}, {Raja},
  {Rakhmanov}, {Rapagnani}, {Raymond}, {Razzano}, {Re}, {Read}, {Reed},
  {Regimbau}, {Rei}, {Reid}, {Reitze}, {Rew}, {Reyes}, {Ricci}, {Riles},
  {Robertson}, {Robie}, {Robinet}, {Rocchi}, {Rolland}, {Rollins}, {Roma},
  {Romano}, {Romanov}, {Romie}, {Rosi{\'n}ska}, {Rowan}, {R{\"u}diger},
  {Ruggi}, {Ryan}, {Sachdev}, {Sadecki}, {Sadeghian}, {Salconi}, {Saleem},
  {Salemi}, {Samajdar}, {Sammut}, {Sanchez}, {Sand berg}, {Sandeen}, {Sanders},
  {Sassolas}, {Sathyaprakash}, {Saulson}, {Sauter}, {Savage}, {Sawadsky},
  {Schale}, {Schilling}, {Schmidt}, {Schmidt}, {Schnabel}, {Schofield},
  {Sch{\"o}nbeck}, {Schreiber}, {Schuette}, {Schutz}, {Scott}, {Scott},
  {Sellers}, {Sengupta}, {Sentenac}, {Sequino}, {Sergeev}, {Serna},
  {Setyawati}, {Sevigny}, {Shaddock}, {Shah}, {Shahriar}, {Shaltev}, {Shao},
  {Shapiro}, {Shawhan}, {Sheperd}, {Shoemaker}, {Shoemaker}, {Siellez},
  {Siemens}, {Sigg}, {Silva}, {Simakov}, {Singer}, {Singer}, {Singh}, {Singh},
  {Singhal}, {Sintes}, {Slagmolen}, {Smith}, {Smith}, {Smith}, {Son}, {Sorazu},
  {Sorrentino}, {Souradeep}, {Srivastava}, {Staley}, {Steinke}, {Steinlechner},
  {Steinlechner}, {Steinmeyer}, {Stephens}, {Stone}, {Strain}, {Straniero},
  {Stratta}, {Strauss}, {Strigin}, {Sturani}, {Stuver}, {Summerscales}, {Sun},
  {Sutton}, {Swinkels}, {Szczepa{\'n}czyk}, {Tacca}, {Talukder}, {Tanner},
  {T{\'a}pai}, {Tarabrin}, {Taracchini}, {Taylor}, {Theeg},
  {Thirugnanasambandam}, {Thomas}, {Thomas}, {Thomas}, {Thorne}, {Thorne},
  {Thrane}, {Tiwari}, {Tiwari}, {Tokmakov}, {Tomlinson}, {Tonelli}, {Torres},
  {Torrie}, {T{\"o}yr{\"a}}, {Travasso}, {Traylor}, {Trifir{\`o}}, {Tringali},
  {Trozzo}, {Tse}, {Turconi}, {Tuyenbayev}, {Ugolini}, {Unnikrishnan}, {Urban},
  {Usman}, {Vahlbruch}, {Vajente}, {Valdes}, {van Bakel}, {van Beuzekom}, {van
  den Brand}, {Van Den Broeck}, {Vander-Hyde}, {van der Schaaf}, {van
  Heijningen}, {van Veggel}, {Vardaro}, {Vass}, {Vas{\'u}th}, {Vaulin},
  {Vecchio}, {Vedovato}, {Veitch}, {Veitch}, {Venkateswara}, {Verkindt},
  {Vetrano}, {Vicer{\'e}}, {Vinciguerra}, {Vine}, {Vinet}, {Vitale}, {Vo},
  {Vocca}, {Vorvick}, {Voss}, {Vousden}, {Vyatchanin}, {Wade}, {Wade}, {Wade},
  {Walker}, {Wallace}, {Walsh}, {Wang}, {Wang}, {Wang}, {Wang}, {Wang}, {Ward},
  {Warner}, {Was}, {Weaver}, {Wei}, {Weinert}, {Weinstein}, {Weiss}, {Welborn},
  {Wen}, {We{\ss}els}, {Westphal}, {Wette}, {Whelan}, {Whitcomb}, {White},
  {Whiting}, {Williams}, {Williamson}, {Willis}, {Willke}, {Wimmer}, {Winkler},
  {Wipf}, {Wittel}, {Woan}, {Worden}, {Wright}, {Wu}, {Yablon}, {Yam},
  {Yamamoto}, {Yancey}, {Yap}, {Yu}, {Yvert}, {Zadro{\.Z}ny}, {Zangrando},
  {Zanolin}, {Zendri}, {Zevin}, {Zhang}, {Zhang}, {Zhang}, {Zhang}, {Zhao},
  {Zhou}, {Zhou}, {Zhu}, {Zucker}, {Zuraw}, {Zweizig}, {LIGO Scientific
  Collaboration}, \& {Virgo Collaboration}}]{2016PhRvL.116m1103A}
---. 2016{\natexlab{c}}, \prl, 116, 131103,
  \dodoi{10.1103/PhysRevLett.116.131103}

\bibitem[{{Abbott} {et~al.}(2016{\natexlab{d}}){Abbott}, {Abbott}, {Abbott},
  {Abernathy}, {Acernese}, {Ackley}, {Adams}, {Adams}, {Addesso}, {Adhikari},
  {Adya}, {Affeldt}, {Agathos}, {Agatsuma}, {Aggarwal}, {Aguiar}, {Aiello},
  {Ain}, {Ajith}, {Allen}, {Allocca}, {Altin}, {Anderson}, {Anderson}, {Arai},
  {Araya}, {Arceneaux}, {Areeda}, {Arnaud}, {Arun}, {Ascenzi}, {Ashton}, {Ast},
  {Aston}, {Astone}, {Aufmuth}, {Aulbert}, {Babak}, {Bacon}, {Bader}, {Baker},
  {Baldaccini}, {Ballardin}, {Ballmer}, {Barayoga}, {Barclay}, {Barish},
  {Barker}, {Barone}, {Barr}, {Barsotti}, {Barsuglia}, {Barta}, {Bartlett},
  {Bartos}, {Bassiri}, {Basti}, {Batch}, {Baune}, {Bavigadda}, {Bazzan},
  {Behnke}, {Bejger}, {Belczynski}, {Bell}, {Bell}, {Berger}, {Bergman},
  {Bergmann}, {Berry}, {Bersanetti}, {Bertolini}, {Betzwieser}, {Bhagwat},
  {Bhandare}, {Bilenko}, {Billingsley}, {Birch}, {Birney}, {Biscans}, {Bisht},
  {Bitossi}, {Biwer}, {Bizouard}, {Blackburn}, {Blair}, {Blair}, {Blair},
  {Bloemen}, {Bock}, {Bodiya}, {Boer}, {Bogaert}, {Bogan}, {Bohe}, {Bojtos},
  {Bond}, {Bondu}, {Bonnand}, {Boom}, {Bork}, {Boschi}, {Bose}, {Bouffanais},
  {Bozzi}, {Bradaschia}, {Brady}, {Braginsky}, {Branchesi}, {Brau}, {Briant},
  {Brillet}, {Brinkmann}, {Brisson}, {Brockill}, {Brooks}, {Brown}, {Brown},
  {Brown}, {Buchanan}, {Buikema}, {Bulik}, {Bulten}, {Buonanno}, {Buskulic},
  {Buy}, {Byer}, {Cadonati}, {Cagnoli}, {Cahillane}, {Calder{\'o}n Bustillo},
  {Callister}, {Calloni}, {Camp}, {Cannon}, {Cao}, {Capano}, {Capocasa},
  {Carbognani}, {Caride}, {Casanueva Diaz}, {Casentini}, {Caudill},
  {Cavagli{\`a}}, {Cavalier}, {Cavalieri}, {Cella}, {Cepeda}, {Cerboni
  Baiardi}, {Cerretani}, {Cesarini}, {Chakraborty}, {Chalermsongsak},
  {Chamberlin}, {Chan}, {Chao}, {Charlton}, {Chassande-Mottin}, {Chen}, {Chen},
  {Cheng}, {Chincarini}, {Chiummo}, {Cho}, {Cho}, {Chow}, {Christensen}, {Chu},
  {Chua}, {Chung}, {Ciani}, {Clara}, {Clark}, {Cleva}, {Coccia}, {Cohadon},
  {Colla}, {Collette}, {Cominsky}, {Constancio}, {Conte}, {Conti}, {Cook},
  {Corbitt}, {Cornish}, {Corsi}, {Cortese}, {Costa}, {Coughlin}, {Coughlin},
  {Coulon}, {Countryman}, {Couvares}, {Cowan}, {Coward}, {Cowart}, {Coyne},
  {Coyne}, {Craig}, {Creighton}, {Cripe}, {Crowder}, {Cumming}, {Cunningham},
  {Cuoco}, {Dal Canton}, {Danilishin}, {D'Antonio}, {Danzmann}, {Darman},
  {Dattilo}, {Dave}, {Daveloza}, {Davier}, {Davies}, {Daw}, {Day}, {DeBra},
  {Debreczeni}, {Degallaix}, {De Laurentis}, {Del{\'e}glise}, {Del Pozzo},
  {Denker}, {Dent}, {Dereli}, {Dergachev}, {DeRosa}, {DeRosa}, {DeSalvo},
  {Dhurandhar}, {D{\'\i}az}, {Di Fiore}, {Di Giovanni}, {Di Lieto}, {Di Pace},
  {Di Palma}, {Di Virgilio}, {Dojcinoski}, {Dolique}, {Donovan}, {Dooley},
  {Doravari}, {Douglas}, {Downes}, {Drago}, {Drever}, {Driggers}, {Du},
  {Ducrot}, {Dwyer}, {Edo}, {Edwards}, {Effler}, {Eggenstein}, {Ehrens},
  {Eichholz}, {Eikenberry}, {Engels}, {Essick}, {Etzel}, {Evans}, {Evans},
  {Everett}, {Factourovich}, {Fafone}, {Fair}, {Fairhurst}, {Fan}, {Fang},
  {Farinon}, {Farr}, {Farr}, {Favata}, {Fays}, {Fehrmann}, {Fejer}, {Ferrante},
  {Ferreira}, {Ferrini}, {Fidecaro}, {Fiori}, {Fiorucci}, {Fisher}, {Flaminio},
  {Fletcher}, {Fournier}, {Franco}, {Frasca}, {Frasconi}, {Frei}, {Freise},
  {Frey}, {Frey}, {Fricke}, {Fritschel}, {Frolov}, {Fulda}, {Fyffe}, {Gabbard},
  {Gair}, {Gammaitoni}, {Gaonkar}, {Garufi}, {Gatto}, {Gaur}, {Gehrels},
  {Gemme}, {Gendre}, {Genin}, {Gennai}, {George}, {Gergely}, {Germain},
  {Ghosh}, {Ghosh}, {Giaime}, {Giardina}, {Giazotto}, {Gill}, {Glaefke},
  {Goetz}, {Goetz}, {Gondan}, {Gonz{\'a}lez}, {Gonzalez Castro}, {Gopakumar},
  {Gordon}, {Gorodetsky}, {Gossan}, {Gosselin}, {Gouaty}, {Graef}, {Graff},
  {Granata}, {Grant}, {Gras}, {Gray}, {Greco}, {Green}, {Groot}, {Grote},
  {Grunewald}, {Guidi}, {Guo}, {Gupta}, {Gupta}, {Gushwa}, {Gustafson},
  {Gustafson}, {Hacker}, {Hall}, {Hall}, {Hammond}, {Haney}, {Hanke}, {Hanks},
  {Hanna}, {Hannam}, {Hanson}, {Hardwick}, {Harms}, {Harry}, {Harry}, {Hart},
  {Hartman}, {Haster}, {Haughian}, {Heidmann}, {Heintze}, {Heitmann}, {Hello},
  {Hemming}, {Hendry}, {Heng}, {Hennig}, {Heptonstall}, {Heurs}, {Hild},
  {Hoak}, {Hodge}, {Hofman}, {Hollitt}, {Holt}, {Holz}, {Hopkins}, {Hosken},
  {Hough}, {Houston}, {Howell}, {Hu}, {Huang}, {Huerta}, {Huet}, {Hughey},
  {Husa}, {Huttner}, {Huynh-Dinh}, {Idrisy}, {Indik}, {Ingram}, {Inta}, {Isa},
  {Isac}, {Isi}, {Islas}, {Isogai}, {Iyer}, {Izumi}, {Jacqmin}, {Jang}, {Jani},
  {Jaranowski}, {Jawahar}, {Jim{\'e}nez-Forteza}, {Johnson}, {Jones}, {Jones},
  {Jonker}, {Ju}, {K}, {Kalaghatgi}, {Kalogera}, {Kandhasamy}, {Kang},
  {Kanner}, {Karki}, {Kasprzack}, {Katsavounidis}, {Katzman}, {Kaufer}, {Kaur},
  {Kawabe}, {Kawazoe}, {K{\'e}f{\'e}lian}, {Kehl}, {Keitel}, {Kelley}, {Kells},
  {Kennedy}, {Key}, {Khalaidovski}, {Khalili}, {Khan}, {Khan}, {Khan},
  {Khazanov}, {Kijbunchoo}, {Kim}, {Kim}, {Kim}, {Kim}, {Kim}, {Kim}, {King},
  {King}, {Kinzel}, {Kissel}, {Kleybolte}, {Klimenko}, {Koehlenbeck},
  {Kokeyama}, {Koley}, {Kondrashov}, {Kontos}, {Korobko}, {Korth}, {Kowalska},
  {Kozak}, {Kringel}, {Krishnan}, {Kr{\'o}lak}, {Krueger}, {Kuehn}, {Kumar},
  {Kuo}, {Kutynia}, {Lackey}, {Landry}, {Lange}, {Lantz}, {Lasky}, {Lazzarini},
  {Lazzaro}, {Leaci}, {Leavey}, {Lebigot}, {Lee}, {Lee}, {Lee}, {Lee}, {Lenon},
  {Leonardi}, {Leong}, {Leroy}, {Letendre}, {Levin}, {Levine}, {Li}, {Libson},
  {Littenberg}, {Lockerbie}, {Logue}, {Lombardi}, {Lord}, {Lorenzini},
  {Loriette}, {Lormand}, {Losurdo}, {Lough}, {L{\"u}ck}, {Lundgren}, {Luo},
  {Lynch}, {Ma}, {MacDonald}, {Machenschalk}, {MacInnis}, {Macleod},
  {Maga{\~n}a-Sandoval}, {Magee}, {Mageswaran}, {Majorana}, {Maksimovic},
  {Malvezzi}, {Man}, {Mandel}, {Mandic}, {Mangano}, {Mansell}, {Manske},
  {Mantovani}, {Marchesoni}, {Marion}, {M{\'a}rka}, {M{\'a}rka}, {Markosyan},
  {Maros}, {Martelli}, {Martellini}, {Martin}, {Martin}, {Martynov}, {Marx},
  {Mason}, {Masserot}, {Massinger}, {Masso-Reid}, {Matichard}, {Matone},
  {Mavalvala}, {Mazumder}, {Mazzolo}, {McCarthy}, {McClelland}, {McCormick},
  {McGuire}, {McIntyre}, {McIver}, {McManus}, {McWilliams}, {Meacher},
  {Meadors}, {Meidam}, {Melatos}, {Mendell}, {Mendoza-Gandara}, {Mercer},
  {Merilh}, {Merzougui}, {Meshkov}, {Messenger}, {Messick}, {Meyers},
  {Mezzani}, {Miao}, {Michel}, {Middleton}, {Mikhailov}, {Milano}, {Miller},
  {Millhouse}, {Minenkov}, {Ming}, {Mirshekari}, {Mishra}, {Mitra},
  {Mitrofanov}, {Mitselmakher}, {Mittleman}, {Moggi}, {Mohan}, {Mohapatra},
  {Montani}, {Moore}, {Moore}, {Moraru}, {Moreno}, {Morriss}, {Mossavi},
  {Mours}, {Mow-Lowry}, {Mueller}, {Mueller}, {Muir}, {Mukherjee}, {Mukherjee},
  {Mukherjee}, {Mukund}, {Mullavey}, {Munch}, {Murphy}, {Murray}, {Mytidis},
  {Nardecchia}, {Naticchioni}, {Nayak}, {Necula}, {Nedkova}, {Nelemans},
  {Neri}, {Neunzert}, {Newton}, {Nguyen}, {Nielsen}, {Nissanke}, {Nitz},
  {Nocera}, {Nolting}, {Normandin}, {Nuttall}, {Oberling}, {Ochsner}, {O'Dell},
  {Oelker}, {Ogin}, {Oh}, {Oh}, {Ohme}, {Oliver}, {Oppermann}, {Oram},
  {O'Reilly}, {O'Shaughnessy}, {Ottaway}, {Ottens}, {Overmier}, {Owen}, {Pai},
  {Pai}, {Palamos}, {Palashov}, {Palomba}, {Pal-Singh}, {Pan}, {Pankow},
  {Pannarale}, {Pant}, {Paoletti}, {Paoli}, {Papa}, {Paris}, {Parker},
  {Pascucci}, {Pasqualetti}, {Passaquieti}, {Passuello}, {Patricelli},
  {Patrick}, {Pearlstone}, {Pedraza}, {Pedurand}, {Pekowsky}, {Pele}, {Penn},
  {Perreca}, {Phelps}, {Piccinni}, {Pichot}, {Piergiovanni}, {Pierro},
  {Pillant}, {Pinard}, {Pinto}, {Pitkin}, {Poggiani}, {Popolizio}, {Post},
  {Powell}, {Prasad}, {Predoi}, {Premachandra}, {Prestegard}, {Price},
  {Prijatelj}, {Principe}, {Privitera}, {Prix}, {Prodi}, {Prokhorov},
  {Puncken}, {Punturo}, {Puppo}, {P{\"u}rrer}, {Qi}, {Qin}, {Quetschke},
  {Quintero}, {Quitzow-James}, {Raab}, {Rabeling}, {Radkins}, {Raffai}, {Raja},
  {Rakhmanov}, {Rapagnani}, {Raymond}, {Razzano}, {Re}, {Read}, {Reed},
  {Regimbau}, {Rei}, {Reid}, {Reitze}, {Rew}, {Reyes}, {Ricci}, {Riles},
  {Robertson}, {Robie}, {Robinet}, {Rocchi}, {Rolland}, {Rollins}, {Roma},
  {Romano}, {Romano}, {Romanov}, {Romie}, {Rosi{\'n}ska}, {Rowan},
  {R{\"u}diger}, {Ruggi}, {Ryan}, {Sachdev}, {Sadecki}, {Sadeghian}, {Salconi},
  {Saleem}, {Salemi}, {Samajdar}, {Sammut}, {Sanchez}, {Sandberg}, {Sandeen},
  {Sanders}, {Sassolas}, {Sathyaprakash}, {Saulson}, {Sauter}, {Savage},
  {Sawadsky}, {Schale}, {Schilling}, {Schmidt}, {Schmidt}, {Schnabel},
  {Schofield}, {Sch{\"o}nbeck}, {Schreiber}, {Schuette}, {Schutz}, {Scott},
  {Scott}, {Sellers}, {Sentenac}, {Sequino}, {Sergeev}, {Serna}, {Setyawati},
  {Sevigny}, {Shaddock}, {Shah}, {Shahriar}, {Shaltev}, {Shao}, {Shapiro},
  {Shawhan}, {Sheperd}, {Shoemaker}, {Shoemaker}, {Siellez}, {Siemens}, {Sigg},
  {Silva}, {Simakov}, {Singer}, {Singer}, {Singh}, {Singh}, {Singhal},
  {Sintes}, {Slagmolen}, {Smith}, {Smith}, {Smith}, {Son}, {Sorazu},
  {Sorrentino}, {Souradeep}, {Srivastava}, {Staley}, {Steinke}, {Steinlechner},
  {Steinlechner}, {Steinmeyer}, {Stephens}, {Stevenson}, {Stone}, {Strain},
  {Straniero}, {Stratta}, {Strauss}, {Strigin}, {Sturani}, {Stuver},
  {Summerscales}, {Sun}, {Sutton}, {Swinkels}, {Szczepa{\'n}czyk}, {Tacca},
  {Talukder}, {Tanner}, {T{\'a}pai}, {Tarabrin}, {Taracchini}, {Taylor},
  {Theeg}, {Thirugnanasambandam}, {Thomas}, {Thomas}, {Thomas}, {Thorne},
  {Thorne}, {Thrane}, {Tiwari}, {Tiwari}, {Tokmakov}, {Tomlinson}, {Tonelli},
  {Torres}, {Torrie}, {T{\"o}yr{\"a}}, {Travasso}, {Traylor}, {Trifir{\`o}},
  {Tringali}, {Trozzo}, {Tse}, {Turconi}, {Tuyenbayev}, {Ugolini},
  {Unnikrishnan}, {Urban}, {Usman}, {Vahlbruch}, {Vajente}, {Valdes}, {van
  Bakel}, {van Beuzekom}, {van den Brand}, {van den Broeck}, {Vander-Hyde},
  {van der Schaaf}, {van Heijningen}, {van Veggel}, {Vardaro}, {Vass},
  {Vas{\'u}th}, {Vaulin}, {Vecchio}, {Vedovato}, {Veitch}, {Veitch},
  {Venkateswara}, {Verkindt}, {Vetrano}, {Vicer{\'e}}, {Vinciguerra}, {Vine},
  {Vinet}, {Vitale}, {Vo}, {Vocca}, {Vorvick}, {Voss}, {Vousden}, {Vyatchanin},
  {Wade}, {Wade}, {Wade}, {Walker}, {Wallace}, {Walsh}, {Wang}, {Wang}, {Wang},
  {Wang}, {Wang}, {Ward}, {Warner}, {Was}, {Weaver}, {Wei}, {Weinert},
  {Weinstein}, {Weiss}, {Welborn}, {Wen}, {We{\ss}els}, {Westphal}, {Wette},
  {Whelan}, {White}, {Whiting}, {Williams}, {Williamson}, {Willis}, {Willke},
  {Wimmer}, {Winkler}, {Wipf}, {Wittel}, {Woan}, {Worden}, {Wright}, {Wu},
  {Yablon}, {Yam}, {Yamamoto}, {Yancey}, {Yap}, {Yu}, {Yvert}, {Zadro{\.z}ny},
  {Zangrando}, {Zanolin}, {Zendri}, {Zevin}, {Zhang}, {Zhang}, {Zhang},
  {Zhang}, {Zhao}, {Zhou}, {Zhou}, {Zhu}, {Zucker}, {Zuraw}, {and}, {Zweizig},
  {LIGO Scientific Collaboration}, \& {Virgo
  Collaboration}}]{2016ApJ...818L..22A}
---. 2016{\natexlab{d}}, \apjl, 818, L22, \dodoi{10.3847/2041-8205/818/2/L22}

\bibitem[{{Abbott} {et~al.}(2017){Abbott}, {Abbott}, {Abbott}, {Acernese},
  {Ackley}, {Adams}, {Adams}, {Addesso}, {Adhikari}, {Adya}, {Affeldt},
  {Afrough}, {Agarwal}, {Agathos}, {Agatsuma}, {Aggarwal}, {Aguiar}, {Aiello},
  {Ain}, {Ajith}, {Allen}, {Allen}, {Allocca}, {Altin}, {Amato}, {Ananyeva},
  {Anderson}, {Anderson}, {Angelova}, {Antier}, {Appert}, {Arai}, {Araya},
  {Areeda}, {Arnaud}, {Arun}, {Ascenzi}, {Ashton}, {Ast}, {Aston}, {Astone},
  {Atallah}, {Aufmuth}, {Aulbert}, {AultONeal}, {Austin}, {Avila-Alvarez},
  {Babak}, {Bacon}, {Bader}, {Bae}, {Bailes}, {Baker}, {Baldaccini},
  {Ballardin}, {Ballmer}, {Banagiri}, {Barayoga}, {Barclay}, {Barish},
  {Barker}, {Barkett}, {Barone}, {Barr}, {Barsotti}, {Barsuglia}, {Barta},
  {Barthelmy}, {Bartlett}, {Bartos}, {Bassiri}, {Basti}, {Batch}, {Bawaj},
  {Bayley}, {Bazzan}, {B{\'e}csy}, {Beer}, {Bejger}, {Belahcene}, {Bell},
  {Berger}, {Bergmann}, {Bernuzzi}, {Bero}, {Berry}, {Bersanetti}, {Bertolini},
  {Betzwieser}, {Bhagwat}, {Bhandare}, {Bilenko}, {Billingsley}, {Billman},
  {Birch}, {Birney}, {Birnholtz}, {Biscans}, {Biscoveanu}, {Bisht}, {Bitossi},
  {Biwer}, {Bizouard}, {Blackburn}, {Blackman}, {Blair}, {Blair}, {Blair},
  {Bloemen}, {Bock}, {Bode}, {Boer}, {Bogaert}, {Bohe}, {Bondu}, {Bonilla},
  {Bonnand}, {Boom}, {Bork}, {Boschi}, {Bose}, {Bossie}, {Bouffanais}, {Bozzi},
  {Bradaschia}, {Brady}, {Branchesi}, {Brau}, {Briant}, {Brillet}, {Brinkmann},
  {Brisson}, {Brockill}, {Broida}, {Brooks}, {Brown}, {Brown}, {Brunett},
  {Buchanan}, {Buikema}, {Bulik}, {Bulten}, {Buonanno}, {Buskulic}, {Buy},
  {Byer}, {Cabero}, {Cadonati}, {Cagnoli}, {Cahillane}, {Calder{\'o}n
  Bustillo}, {Callister}, {Calloni}, {Camp}, {Canepa}, {Canizares}, {Cannon},
  {Cao}, {Cao}, {Capano}, {Capocasa}, {Carbognani}, {Caride}, {Carney},
  {Carullo}, {Casanueva Diaz}, {Casentini}, {Caudill}, {Cavagli{\`a}},
  {Cavalier}, {Cavalieri}, {Cella}, {Cepeda}, {Cerd{\'a}-Dur{\'a}n},
  {Cerretani}, {Cesarini}, {Chamberlin}, {Chan}, {Chao}, {Charlton}, {Chase},
  {Chassande-Mottin}, {Chatterjee}, {Chatziioannou}, {Cheeseboro}, {Chen},
  {Chen}, {Chen}, {Cheng}, {Chia}, {Chincarini}, {Chiummo}, {Chmiel}, {Cho},
  {Cho}, {Chow}, {Christensen}, {Chu}, {Chua}, {Chua}, {Chung}, {Chung},
  {Ciani}, {Ciolfi}, {Cirelli}, {Cirone}, {Clara}, {Clark}, {Clearwater},
  {Cleva}, {Cocchieri}, {Coccia}, {Cohadon}, {Cohen}, {Colla}, {Collette},
  {Cominsky}, {Constancio}, {Conti}, {Cooper}, {Corban}, {Corbitt},
  {Cordero-Carri{\'o}n}, {Corley}, {Cornish}, {Corsi}, {Cortese}, {Costa},
  {Coughlin}, {Coughlin}, {Coulon}, {Countryman}, {Couvares}, {Covas}, {Cowan},
  {Coward}, {Cowart}, {Coyne}, {Coyne}, {Creighton}, {Creighton}, {Cripe},
  {Crowder}, {Cullen}, {Cumming}, {Cunningham}, {Cuoco}, {Dal Canton},
  {D{\'a}lya}, {Danilishin}, {D'Antonio}, {Danzmann}, {Dasgupta}, {Da Silva
  Costa}, {Dattilo}, {Dave}, {Davier}, {Davis}, {Daw}, {Day}, {De}, {DeBra},
  {Degallaix}, {De Laurentis}, {Del{\'e}glise}, {Del Pozzo}, {Demos}, {Denker},
  {Dent}, {De Pietri}, {Dergachev}, {De Rosa}, {DeRosa}, {De Rossi}, {DeSalvo},
  {de Varona}, {Devenson}, {Dhurandhar}, {D{\'\i}az}, {Dietrich}, {Di Fiore},
  {Di Giovanni}, {Di Girolamo}, {Di Lieto}, {Di Pace}, {Di Palma}, {Di Renzo},
  {Doctor}, {Dolique}, {Donovan}, {Dooley}, {Doravari}, {Dorrington},
  {Douglas}, {Dovale {\'A}lvarez}, {Downes}, {Drago}, {Dreissigacker},
  {Driggers}, {Du}, {Ducrot}, {Dudi}, {Dupej}, {Dwyer}, {Edo}, {Edwards},
  {Effler}, {Eggenstein}, {Ehrens}, {Eichholz}, {Eikenberry}, {Eisenstein},
  {Essick}, {Estevez}, {Etienne}, {Etzel}, {Evans}, {Evans}, {Factourovich},
  {Fafone}, {Fair}, {Fairhurst}, {Fan}, {Farinon}, {Farr}, {Farr},
  {Fauchon-Jones}, {Favata}, {Fays}, {Fee}, {Fehrmann}, {Feicht}, {Fejer},
  {Fernandez-Galiana}, {Ferrante}, {Ferreira}, {Ferrini}, {Fidecaro},
  {Finstad}, {Fiori}, {Fiorucci}, {Fishbach}, {Fisher}, {Fitz-Axen},
  {Flaminio}, {Fletcher}, {Fong}, {Font}, {Forsyth}, {Forsyth}, {Fournier},
  {Frasca}, {Frasconi}, {Frei}, {Freise}, {Frey}, {Frey}, {Fries}, {Fritschel},
  {Frolov}, {Fulda}, {Fyffe}, {Gabbard}, {Gadre}, {Gaebel}, {Gair},
  {Gammaitoni}, {Ganija}, {Gaonkar}, {Garcia-Quiros}, {Garufi}, {Gateley},
  {Gaudio}, {Gaur}, {Gayathri}, {Gehrels}, {Gemme}, {Genin}, {Gennai},
  {George}, {George}, {Gergely}, {Germain}, {Ghonge}, {Ghosh}, {Ghosh},
  {Ghosh}, {Giaime}, {Giardina}, {Giazotto}, {Gill}, {Glover}, {Goetz},
  {Goetz}, {Gomes}, {Goncharov}, {Gonz{\'a}lez}, {Gonzalez Castro},
  {Gopakumar}, {Gorodetsky}, {Gossan}, {Gosselin}, {Gouaty}, {Grado}, {Graef},
  {Granata}, {Grant}, {Gras}, {Gray}, {Greco}, {Green}, {Gretarsson}, {Groot},
  {Grote}, {Grunewald}, {Gruning}, {Guidi}, {Guo}, {Gupta}, {Gupta}, {Gushwa},
  {Gustafson}, {Gustafson}, {Halim}, {Hall}, {Hall}, {Hamilton}, {Hammond},
  {Haney}, {Hanke}, {Hanks}, {Hanna}, {Hannam}, {Hannuksela}, {Hanson},
  {Hardwick}, {Harms}, {Harry}, {Harry}, {Hart}, {Haster}, {Haughian}, {Healy},
  {Heidmann}, {Heintze}, {Heitmann}, {Hello}, {Hemming}, {Hendry}, {Heng},
  {Hennig}, {Heptonstall}, {Heurs}, {Hild}, {Hinderer}, {Ho}, {Hoak}, {Hofman},
  {Holt}, {Holz}, {Hopkins}, {Horst}, {Hough}, {Houston}, {Howell}, {Hreibi},
  {Hu}, {Huerta}, {Huet}, {Hughey}, {Husa}, {Huttner}, {Huynh-Dinh}, {Indik},
  {Inta}, {Intini}, {Isa}, {Isac}, {Isi}, {Iyer}, {Izumi}, {Jacqmin}, {Jani},
  {Jaranowski}, {Jawahar}, {Jim{\'e}nez-Forteza}, {Johnson},
  {Johnson-McDaniel}, {Jones}, {Jones}, {Jonker}, {Ju}, {Junker}, {Kalaghatgi},
  {Kalogera}, {Kamai}, {Kand hasamy}, {Kang}, {Kanner}, {Kapadia}, {Karki},
  {Karvinen}, {Kasprzack}, {Kastaun}, {Katolik}, {Katsavounidis}, {Katzman},
  {Kaufer}, {Kawabe}, {K{\'e}f{\'e}lian}, {Keitel}, {Kemball}, {Kennedy},
  {Kent}, {Key}, {Khalili}, {Khan}, {Khan}, {Khan}, {Khazanov}, {Kijbunchoo},
  {Kim}, {Kim}, {Kim}, {Kim}, {Kim}, {Kim}, {Kimbrell}, {King}, {King},
  {Kinley-Hanlon}, {Kirchhoff}, {Kissel}, {Kleybolte}, {Klimenko}, {Knowles},
  {Koch}, {Koehlenbeck}, {Koley}, {Kondrashov}, {Kontos}, {Korobko}, {Korth},
  {Kowalska}, {Kozak}, {Kr{\"a}mer}, {Kringel}, {Krishnan}, {Kr{\'o}lak},
  {Kuehn}, {Kumar}, {Kumar}, {Kumar}, {Kuo}, {Kutynia}, {Kwang}, {Lackey},
  {Lai}, {Landry}, {Lang}, {Lange}, {Lantz}, {Lanza}, {Larson},
  {Lartaux-Vollard}, {Lasky}, {Laxen}, {Lazzarini}, {Lazzaro}, {Leaci},
  {Leavey}, {Lee}, {Lee}, {Lee}, {Lee}, {Lee}, {Lehmann}, {Lenon}, {Leon},
  {Leonardi}, {Leroy}, {Letendre}, {Levin}, {Li}, {Linker}, {Littenberg},
  {Liu}, {Liu}, {Lo}, {Lockerbie}, {London}, {Lord}, {Lorenzini}, {Loriette},
  {Lormand}, {Losurdo}, {Lough}, {Lousto}, {Lovelace}, {L{\"u}ck}, {Lumaca},
  {Lundgren}, {Lynch}, {Ma}, {Macas}, {Macfoy}, {Machenschalk}, {MacInnis},
  {Macleod}, {Maga{\~n}a Hernandez}, {Maga{\~n}a-Sandoval}, {Maga{\~n}a
  Zertuche}, {Magee}, {Majorana}, {Maksimovic}, {Man}, {Mandic}, {Mangano},
  {Mansell}, {Manske}, {Mantovani}, {Marchesoni}, {Marion}, {M{\'a}rka},
  {M{\'a}rka}, {Markakis}, {Markosyan}, {Markowitz}, {Maros}, {Marquina},
  {Marsh}, {Martelli}, {Martellini}, {Martin}, {Martin}, {Martynov}, {Marx},
  {Mason}, {Massera}, {Masserot}, {Massinger}, {Masso-Reid}, {Mastrogiovanni},
  {Matas}, {Matichard}, {Matone}, {Mavalvala}, {Mazumder}, {McCarthy},
  {McClelland}, {McCormick}, {McCuller}, {McGuire}, {McIntyre}, {McIver},
  {McManus}, {McNeill}, {McRae}, {McWilliams}, {Meacher}, {Meadors}, {Mehmet},
  {Meidam}, {Mejuto-Villa}, {Melatos}, {Mendell}, {Mercer}, {Merilh},
  {Merzougui}, {Meshkov}, {Messenger}, {Messick}, {Metzdorff}, {Meyers},
  {Miao}, {Michel}, {Middleton}, {Mikhailov}, {Milano}, {Miller}, {Miller},
  {Miller}, {Millhouse}, {Milovich-Goff}, {Minazzoli}, {Minenkov}, {Ming},
  {Mishra}, {Mitra}, {Mitrofanov}, {Mitselmakher}, {Mittleman}, {Moffa},
  {Moggi}, {Mogushi}, {Mohan}, {Mohapatra}, {Molina}, {Montani}, {Moore},
  {Moraru}, {Moreno}, {Morisaki}, {Morriss}, {Mours}, {Mow-Lowry}, {Mueller},
  {Muir}, {Mukherjee}, {Mukherjee}, {Mukherjee}, {Mukund}, {Mullavey}, {Munch},
  {Mu{\~n}iz}, {Muratore}, {Murray}, {Nagar}, {Napier}, {Nardecchia},
  {Naticchioni}, {Nayak}, {Neilson}, {Nelemans}, {Nelson}, {Nery}, {Neunzert},
  {Nevin}, {Newport}, {Newton}, {Ng}, {Nguyen}, {Nguyen}, {Nichols}, {Nielsen},
  {Nissanke}, {Nitz}, {Noack}, {Nocera}, {Nolting}, {North}, {Nuttall},
  {Oberling}, {O'Dea}, {Ogin}, {Oh}, {Oh}, {Ohme}, {Okada}, {Oliver},
  {Oppermann}, {Oram}, {O'Reilly}, {Ormiston}, {Ortega}, {O'Shaughnessy},
  {Ossokine}, {Ottaway}, {Overmier}, {Owen}, {Pace}, {Page}, {Page}, {Pai},
  {Pai}, {Palamos}, {Palashov}, {Palomba}, {Pal-Singh}, {Pan}, {Pan}, {Pang},
  {Pang}, {Pankow}, {Pannarale}, {Pant}, {Paoletti}, {Paoli}, {Papa}, {Parida},
  {Parker}, {Pascucci}, {Pasqualetti}, {Passaquieti}, {Passuello}, {Patil},
  {Patricelli}, {Pearlstone}, {Pedraza}, {Pedurand}, {Pekowsky}, {Pele},
  {Penn}, {Perez}, {Perreca}, {Perri}, {Pfeiffer}, {Phelps}, {Piccinni},
  {Pichot}, {Piergiovanni}, {Pierro}, {Pillant}, {Pinard}, {Pinto}, {Pirello},
  {Pitkin}, {Poe}, {Poggiani}, {Popolizio}, {Porter}, {Post}, {Powell},
  {Prasad}, {Pratt}, {Pratten}, {Predoi}, {Prestegard}, {Prijatelj},
  {Principe}, {Privitera}, {Prix}, {Prodi}, {Prokhorov}, {Puncken}, {Punturo},
  {Puppo}, {P{\"u}rrer}, {Qi}, {Quetschke}, {Quintero}, {Quitzow-James},
  {Raab}, {Rabeling}, {Radkins}, {Raffai}, {Raja}, {Rajan}, {Rajbhandari},
  {Rakhmanov}, {Ramirez}, {Ramos-Buades}, {Rapagnani}, {Raymond}, {Razzano},
  {Read}, {Regimbau}, {Rei}, {Reid}, {Reitze}, {Ren}, {Reyes}, {Ricci},
  {Ricker}, {Rieger}, {Riles}, {Rizzo}, {Robertson}, {Robie}, {Robinet},
  {Rocchi}, {Rolland}, {Rollins}, {Roma}, {Romano}, {Romano}, {Romel}, {Romie},
  {Rosi{\'n}ska}, {Ross}, {Rowan}, {R{\"u}diger}, {Ruggi}, {Rutins}, {Ryan},
  {Sachdev}, {Sadecki}, {Sadeghian}, {Sakellariadou}, {Salconi}, {Saleem},
  {Salemi}, {Samajdar}, {Sammut}, {Sampson}, {Sanchez}, {Sanchez},
  {Sanchis-Gual}, {Sand berg}, {Sanders}, {Sassolas}, {Sathyaprakash},
  {Saulson}, {Sauter}, {Savage}, {Sawadsky}, {Schale}, {Scheel}, {Scheuer},
  {Schmidt}, {Schmidt}, {Schnabel}, {Schofield}, {Sch{\"o}nbeck}, {Schreiber},
  {Schuette}, {Schulte}, {Schutz}, {Schwalbe}, {Scott}, {Scott}, {Seidel},
  {Sellers}, {Sengupta}, {Sentenac}, {Sequino}, {Sergeev}, {Shaddock},
  {Shaffer}, {Shah}, {Shahriar}, {Shaner}, {Shao}, {Shapiro}, {Shawhan},
  {Sheperd}, {Shoemaker}, {Shoemaker}, {Siellez}, {Siemens}, {Sieniawska},
  {Sigg}, {Silva}, {Singer}, {Singh}, {Singhal}, {Sintes}, {Slagmolen},
  {Smith}, {Smith}, {Smith}, {Somala}, {Son}, {Sonnenberg}, {Sorazu},
  {Sorrentino}, {Souradeep}, {Spencer}, {Srivastava}, {Staats}, {Staley},
  {Steinke}, {Steinlechner}, {Steinlechner}, {Steinmeyer}, {Stevenson},
  {Stone}, {Stops}, {Strain}, {Stratta}, {Strigin}, {Strunk}, {Sturani},
  {Stuver}, {Summerscales}, {Sun}, {Sunil}, {Suresh}, {Sutton}, {Swinkels},
  {Szczepa{\'n}czyk}, {Tacca}, {Tait}, {Talbot}, {Talukder}, {Tanner},
  {T{\'a}pai}, {Taracchini}, {Tasson}, {Taylor}, {Taylor}, {Tewari}, {Theeg},
  {Thies}, {Thomas}, {Thomas}, {Thomas}, {Thorne}, {Thorne}, {Thrane},
  {Tiwari}, {Tiwari}, {Tokmakov}, {Toland}, {Tonelli}, {Tornasi},
  {Torres-Forn{\'e}}, {Torrie}, {T{\"o}yr{\"a}}, {Travasso}, {Traylor},
  {Trinastic}, {Tringali}, {Trozzo}, {Tsang}, {Tse}, {Tso}, {Tsukada}, {Tsuna},
  {Tuyenbayev}, {Ueno}, {Ugolini}, {Unnikrishnan}, {Urban}, {Usman},
  {Vahlbruch}, {Vajente}, {Valdes}, {Vallisneri}, {van Bakel}, {van Beuzekom},
  {van den Brand}, {Van Den Broeck}, {Vand er-Hyde}, {van der Schaaf}, {van
  Heijningen}, {van Veggel}, {Vardaro}, {Varma}, {Vass}, {Vas{\'u}th},
  {Vecchio}, {Vedovato}, {Veitch}, {Veitch}, {Venkateswara}, {Venugopalan},
  {Verkindt}, {Vetrano}, {Vicer{\'e}}, {Viets}, {Vinciguerra}, {Vine}, {Vinet},
  {Vitale}, {Vo}, {Vocca}, {Vorvick}, {Vyatchanin}, {Wade}, {Wade}, {Wade},
  {Walet}, {Walker}, {Wallace}, {Walsh}, {Wang}, {Wang}, {Wang}, {Wang},
  {Wang}, {Ward}, {Warner}, {Was}, {Watchi}, {Weaver}, {Wei}, {Weinert},
  {Weinstein}, {Weiss}, {Wen}, {Wessel}, {We{\ss}els}, {Westerweck},
  {Westphal}, {Wette}, {Whelan}, {Whitcomb}, {Whiting}, {Whittle}, {Wilken},
  {Williams}, {Williams}, {Williamson}, {Willis}, {Willke}, {Wimmer},
  {Winkler}, {Wipf}, {Wittel}, {Woan}, {Woehler}, {Wofford}, {Wong}, {Worden},
  {Wright}, {Wu}, {Wysocki}, {Xiao}, {Yamamoto}, {Yancey}, {Yang}, {Yap},
  {Yazback}, {Yu}, {Yu}, {Yvert}, {Zadro{\.Z}ny}, {Zanolin}, {Zelenova},
  {Zendri}, {Zevin}, {Zhang}, {Zhang}, {Zhang}, {Zhang}, {Zhao}, {Zhou},
  {Zhou}, {Zhu}, {Zhu}, {Zimmerman}, {Zucker}, {Zweizig}, {LIGO Scientific
  Collaboration}, \& {Virgo Collaboration}}]{2017PhRvL.119p1101A}
---. 2017, \prl, 119, 161101, \dodoi{10.1103/PhysRevLett.119.161101}

\bibitem[{{Abbott} {et~al.}(2019){Abbott}, {Abbott}, {Abbott}, {Abraham},
  {Acernese}, {Ackley}, {Adams}, {Adhikari}, {Adya}, {Affeldt}, {Agathos},
  {Agatsuma}, {Aggarwal}, {Aguiar}, {Aiello}, {Ain}, {Ajith}, {Allen},
  {Allocca}, {Aloy}, {Altin}, {Amato}, {Anand}, {Ananyeva}, {Anderson},
  {Anderson}, {Angelova}, {Antier}, {Appert}, {Arai}, {Araya}, {Areeda},
  {Ar{\`e}ne}, {Arnaud}, {Aronson}, {Ascenzi}, {Ashton}, {Aston}, {Astone},
  {Aubin}, {Aufmuth}, {AultONeal}, {Austin}, {Avendano}, {Avila-Alvarez},
  {Babak}, {Bacon}, {Badaracco}, {Bader}, {Bae}, {Baird}, {Baker},
  {Baldaccini}, {Ballardin}, {Ballmer}, {Bals}, {Banagiri}, {Barayoga},
  {Barbieri}, {Barclay}, {Barish}, {Barker}, {Barkett}, {Barnum}, {Barone},
  {Barr}, {Barsotti}, {Barsuglia}, {Barta}, {Bartlett}, {Bartos}, {Bassiri},
  {Basti}, {Bawaj}, {Bayley}, {Bazzan}, {B{\'e}csy}, {Bejger}, {Belahcene},
  {Bell}, {Beniwal}, {Benjamin}, {Berger}, {Bergmann}, {Bernuzzi}, {Berry},
  {Bersanetti}, {Bertolini}, {Betzwieser}, {Bhandare}, {Bidler}, {Biggs},
  {Bilenko}, {Bilgili}, {Billingsley}, {Birney}, {Birnholtz}, {Biscans},
  {Bischi}, {Biscoveanu}, {Bisht}, {Bitossi}, {Bizouard}, {Blackburn},
  {Blackman}, {Blair}, {Blair}, {Blair}, {Bloemen}, {Bobba}, {Bode}, {Boer},
  {Boetzel}, {Bogaert}, {Bondu}, {Bonnand}, {Booker}, {Boom}, {Bork}, {Boschi},
  {Bose}, {Bossilkov}, {Bosveld}, {Bouffanais}, {Bozzi}, {Bradaschia}, {Brady},
  {Bramley}, {Branchesi}, {Brau}, {Breschi}, {Briant}, {Briggs}, {Brighenti},
  {Brillet}, {Brinkmann}, {Brockill}, {Brooks}, {Brooks}, {Brown}, {Brunett},
  {Buikema}, {Bulik}, {Bulten}, {Buonanno}, {Buskulic}, {Buy}, {Byer},
  {Cabero}, {Cadonati}, {Cagnoli}, {Cahillane}, {Calder{\'o}n Bustillo},
  {Callister}, {Calloni}, {Camp}, {Campbell}, {Canepa}, {Cannon}, {Cao}, {Cao},
  {Carapella}, {Carbognani}, {Caride}, {Carney}, {Carullo}, {Casanueva Diaz},
  {Casentini}, {Caudill}, {Cavagli{\`a}}, {Cavalier}, {Cavalieri}, {Cella},
  {Cerd{\'a}-Dur{\'a}n}, {Cesarini}, {Chaibi}, {Chakravarti}, {Chamberlin},
  {Chan}, {Chao}, {Charlton}, {Chase}, {Chassande-Mottin}, {Chatterjee},
  {Chaturvedi}, {Chatziioannou}, {Cheeseboro}, {Chen}, {Chen}, {Chen}, {Cheng},
  {Cheong}, {Chia}, {Chiadini}, {Chincarini}, {Chiummo}, {Cho}, {Cho}, {Cho},
  {Christensen}, {Chu}, {Chua}, {Chung}, {Chung}, {Ciani}, {Cie{\'s}lar},
  {Ciobanu}, {Ciolfi}, {Cipriano}, {Cirone}, {Clara}, {Clark}, {Clearwater},
  {Cleva}, {Coccia}, {Cohadon}, {Cohen}, {Colleoni}, {Collette}, {Collins},
  {Colpi}, {Cominsky}, {Constancio}, {Conti}, {Cooper}, {Corban}, {Corbitt},
  {Cordero-Carri{\'o}n}, {Corezzi}, {Corley}, {Cornish}, {Corre}, {Corsi},
  {Cortese}, {Costa}, {Cotesta}, {Coughlin}, {Coughlin}, {Coulon},
  {Countryman}, {Couvares}, {Covas}, {Cowan}, {Coward}, {Cowart}, {Coyne},
  {Coyne}, {Creighton}, {Creighton}, {Cripe}, {Croquette}, {Crowder}, {Cullen},
  {Cumming}, {Cunningham}, {Cuoco}, {Dal Canton}, {D{\'a}lya}, {D'Angelo},
  {Danilishin}, {D'Antonio}, {Danzmann}, {Dasgupta}, {Da Silva Costa},
  {Datrier}, {Dattilo}, {Dave}, {Davier}, {Davis}, {Daw}, {DeBra},
  {Deenadayalan}, {Degallaix}, {De Laurentis}, {Del{\'e}glise}, {Del Pozzo},
  {DeMarchi}, {Demos}, {Dent}, {De Pietri}, {De Rosa}, {De Rossi}, {DeSalvo},
  {de Varona}, {Dhurandhar}, {D{\'\i}az}, {Dietrich}, {Di Fiore}, {DiFronzo},
  {Di Giorgio}, {Di Giovanni}, {Di Giovanni}, {Di Girolamo}, {Di Lieto},
  {Ding}, {Di Pace}, {Di Palma}, {Di Renzo}, {Divakarla}, {Dmitriev}, {Doctor},
  {Donovan}, {Dooley}, {Doravari}, {Dorrington}, {Downes}, {Drago}, {Driggers},
  {Du}, {Ducoin}, {Dupej}, {Durante}, {Dwyer}, {Easter}, {Eddolls}, {Edo},
  {Effler}, {Ehrens}, {Eichholz}, {Eikenberry}, {Eisenmann}, {Eisenstein},
  {Errico}, {Essick}, {Estelles}, {Estevez}, {Etienne}, {Etzel}, {Evans},
  {Evans}, {Fafone}, {Fairhurst}, {Fan}, {Farinon}, {Farr}, {Farr},
  {Fauchon-Jones}, {Favata}, {Fays}, {Fazio}, {Fee}, {Feicht}, {Fejer}, {Feng},
  {Fernandez-Galiana}, {Ferrante}, {Ferreira}, {Ferreira}, {Fidecaro}, {Fiori},
  {Fiorucci}, {Fishbach}, {Fisher}, {Fishner}, {Fittipaldi}, {Fitz-Axen},
  {Fiumara}, {Flaminio}, {Fletcher}, {Floden}, {Flynn}, {Fong}, {Font},
  {Forsyth}, {Fournier}, {Hernandez Vivanco}, {Frasca}, {Frasconi}, {Frei},
  {Freise}, {Frey}, {Frey}, {Fritschel}, {Frolov}, {Fronz{\`e}}, {Fulda},
  {Fyffe}, {Gabbard}, {Gadre}, {Gaebel}, {Gair}, {Gammaitoni}, {Gaonkar},
  {Garc{\'\i}a-Quir{\'o}s}, {Garufi}, {Gateley}, {Gaudio}, {Gaur}, {Gayathri},
  {Gemme}, {Genin}, {Gennai}, {George}, {George}, {Gergely}, {Ghonge}, {Ghosh},
  {Ghosh}, {Ghosh}, {Giacomazzo}, {Giaime}, {Giardina}, {Gibson}, {Gill},
  {Glover}, {Gniesmer}, {Godwin}, {Goetz}, {Goetz}, {Goncharov},
  {Gonz{\'a}lez}, {Gonzalez Castro}, {Gopakumar}, {Gossan}, {Gosselin},
  {Gouaty}, {Grace}, {Grado}, {Granata}, {Grant}, {Gras}, {Grassia}, {Gray},
  {Gray}, {Greco}, {Green}, {Green}, {Gretarsson}, {Grimaldi}, {Grimm},
  {Groot}, {Grote}, {Grunewald}, {Gruning}, {Guidi}, {Gulati}, {Guo}, {Gupta},
  {Gupta}, {Gupta}, {Gustafson}, {Gustafson}, {Haegel}, {Halim}, {Hall},
  {Hall}, {Hamilton}, {Hammond}, {Haney}, {Hanke}, {Hanks}, {Hanna}, {Hannam},
  {Hannuksela}, {Hansen}, {Hanson}, {Harder}, {Hardwick}, {Haris}, {Harms},
  {Harry}, {Harry}, {Hasskew}, {Haster}, {Haughian}, {Hayes}, {Healy},
  {Heidmann}, {Heintze}, {Heitmann}, {Hellman}, {Hello}, {Hemming}, {Hendry},
  {Heng}, {Hennig}, {Heurs}, {Hild}, {Hinderer}, {Hochheim}, {Hofman},
  {Holgado}, {Holland}, {Holt}, {Holz}, {Hopkins}, {Horst}, {Hough}, {Howell},
  {Hoy}, {Huang}, {H{\"u}bner}, {Huerta}, {Huet}, {Hughey}, {Hui}, {Husa},
  {Huttner}, {Huynh-Dinh}, {Idzkowski}, {Iess}, {Inchauspe}, {Ingram}, {Inta},
  {Intini}, {Irwin}, {Isa}, {Isac}, {Isi}, {Iyer}, {Jacqmin}, {Jadhav}, {Jani},
  {Janthalur}, {Jaranowski}, {Jariwala}, {Jenkins}, {Jiang}, {Johnson},
  {Jones}, {Jones}, {Jones}, {Jones}, {Jonker}, {Ju}, {Junker}, {Kalaghatgi},
  {Kalogera}, {Kamai}, {Kandhasamy}, {Kang}, {Kanner}, {Kapadia}, {Karki},
  {Kashyap}, {Kasprzack}, {Katsanevas}, {Katsavounidis}, {Katzman}, {Kaufer},
  {Kawabe}, {Keerthana}, {K{\'e}f{\'e}lian}, {Keitel}, {Kennedy}, {Key},
  {Khalili}, {Khan}, {Khan}, {Khazanov}, {Khetan}, {Khursheed}, {Kijbunchoo},
  {Kim}, {Kim}, {Kim}, {Kim}, {Kim}, {Kim}, {Kimball}, {King}, {Kinley-Hanlon},
  {Kirchhoff}, {Kissel}, {Kleybolte}, {Klika}, {Klimenko}, {Knowles}, {Koch},
  {Koehlenbeck}, {Koekoek}, {Koley}, {Kondrashov}, {Kontos}, {Koper},
  {Korobko}, {Korth}, {Kovalam}, {Kozak}, {Kr{\"a}mer}, {Kringel},
  {Krishnendu}, {Kr{\'o}lak}, {Krupinski}, {Kuehn}, {Kumar}, {Kumar}, {Kumar},
  {Kumar}, {Kuo}, {Kutynia}, {Kwang}, {Lackey}, {Laghi}, {Lai}, {Lam},
  {Landry}, {Lane}, {Lang}, {Lange}, {Lantz}, {Lanza}, {Lartaux-Vollard},
  {Lasky}, {Laxen}, {Lazzarini}, {Lazzaro}, {Leaci}, {Leavey}, {Lecoeuche},
  {Lee}, {Lee}, {Lee}, {Lee}, {Lee}, {Lee}, {Lehmann}, {Lenon}, {Leroy},
  {Letendre}, {Levin}, {Li}, {Li}, {Li}, {Li}, {Li}, {Lin}, {Linde}, {Linker},
  {Littenberg}, {Liu}, {Liu}, {Llorens-Monteagudo}, {Lo}, {London}, {Longo},
  {Lorenzini}, {Loriette}, {Lormand}, {Losurdo}, {Lough}, {Lousto}, {Lovelace},
  {Lower}, {L{\"u}ck}, {Lumaca}, {Lundgren}, {Lynch}, {Ma}, {Macas}, {Macfoy},
  {MacInnis}, {Macleod}, {Macquet}, {Maga{\~n}a Hernandez},
  {Maga{\~n}a-Sandoval}, {Magee}, {Majorana}, {Maksimovic}, {Malik}, {Man},
  {Mandic}, {Mangano}, {Mansell}, {Manske}, {Mantovani}, {Mapelli},
  {Marchesoni}, {Marion}, {M{\'a}rka}, {M{\'a}rka}, {Markakis}, {Markosyan},
  {Markowitz}, {Maros}, {Marquina}, {Marsat}, {Martelli}, {Martin}, {Martin},
  {Martinez}, {Martynov}, {Masalehdan}, {Mason}, {Massera}, {Masserot},
  {Massinger}, {Masso-Reid}, {Mastrogiovanni}, {Matas}, {Matichard}, {Matone},
  {Mavalvala}, {McCann}, {McCarthy}, {McClelland}, {McCormick}, {McCuller},
  {McGuire}, {McIsaac}, {McIver}, {McManus}, {McRae}, {McWilliams}, {Meacher},
  {Meadors}, {Mehmet}, {Mehta}, {Meidam}, {Mejuto Villa}, {Melatos}, {Mendell},
  {Mercer}, {Mereni}, {Merfeld}, {Merilh}, {Merzougui}, {Meshkov}, {Messenger},
  {Messick}, {Messina}, {Metzdorff}, {Meyers}, {Meylahn}, {Miani}, {Miao},
  {Michel}, {Middleton}, {Milano}, {Miller}, {Millhouse}, {Mills},
  {Milovich-Goff}, {Minazzoli}, {Minenkov}, {Mishkin}, {Mishra}, {Mistry},
  {Mitra}, {Mitrofanov}, {Mitselmakher}, {Mittleman}, {Mo}, {Moffa}, {Mogushi},
  {Mohapatra}, {Molina-Ruiz}, {Mondin}, {Montani}, {Moore}, {Moraru},
  {Morawski}, {Moreno}, {Morisaki}, {Mours}, {Mow-Lowry}, {Muciaccia},
  {Mukherjee}, {Mukherjee}, {Mukherjee}, {Mukherjee}, {Mukund}, {Mullavey},
  {Munch}, {Mu{\~n}iz}, {Muratore}, {Murray}, {Nagar}, {Nardecchia},
  {Naticchioni}, {Nayak}, {Neil}, {Neilson}, {Nelemans}, {Nelson}, {Nery},
  {Neunzert}, {Nevin}, {Ng}, {Ng}, {Nguyen}, {Nguyen}, {Nichols}, {Nichols},
  {Nissanke}, {Nocera}, {North}, {Nuttall}, {Obergaulinger}, {Oberling},
  {O'Brien}, {Oganesyan}, {Ogin}, {Oh}, {Oh}, {Ohme}, {Ohta}, {Okada},
  {Oliver}, {Oppermann}, {Oram}, {O'Reilly}, {Ormiston}, {Ortega},
  {O'Shaughnessy}, {Ossokine}, {Ottaway}, {Overmier}, {Owen}, {Pace}, {Pagano},
  {Page}, {Pagliaroli}, {Pai}, {Pai}, {Palamos}, {Palashov}, {Palomba}, {Pan},
  {Panda}, {Pang}, {Pankow}, {Pannarale}, {Pant}, {Paoletti}, {Paoli},
  {Parida}, {Parker}, {Pascucci}, {Pasqualetti}, {Passaquieti}, {Passuello},
  {Patil}, {Patricelli}, {Payne}, {Pearlstone}, {Pechsiri}, {Pedersen},
  {Pedraza}, {Pedurand}, {Pele}, {Penn}, {Perego}, {Perez}, {P{\'e}rigois},
  {Perreca}, {Petermann}, {Pfeiffer}, {Phelps}, {Phukon}, {Piccinni}, {Pichot},
  {Piergiovanni}, {Pierro}, {Pillant}, {Pinard}, {Pinto}, {Pirello}, {Pitkin},
  {Plastino}, {Poggiani}, {Pong}, {Ponrathnam}, {Popolizio}, {Porter},
  {Powell}, {Prajapati}, {Prasad}, {Prasai}, {Prasanna}, {Pratten},
  {Prestegard}, {Principe}, {Prodi}, {Prokhorov}, {Punturo}, {Puppo},
  {P{\"u}rrer}, {Qi}, {Quetschke}, {Quinonez}, {Raab}, {Raaijmakers},
  {Radkins}, {Radulesco}, {Raffai}, {Raja}, {Rajan}, {Rajbhandari},
  {Rakhmanov}, {Ramirez}, {Ramos-Buades}, {Rana}, {Rao}, {Rapagnani},
  {Raymond}, {Razzano}, {Read}, {Regimbau}, {Rei}, {Reid}, {Reitze},
  {Rettegno}, {Ricci}, {Richardson}, {Richardson}, {Ricker}, {Riemenschneider},
  {Riles}, {Rizzo}, {Robertson}, {Robinet}, {Rocchi}, {Rolland}, {Rollins},
  {Roma}, {Romanelli}, {Romano}, {Romel}, {Romie}, {Rose}, {Rose}, {Rose},
  {Rosi{\'n}ska}, {Rosofsky}, {Ross}, {Rowan}, {R{\"u}diger}, {Ruggi},
  {Rutins}, {Ryan}, {Sachdev}, {Sadecki}, {Sakellariadou}, {Salafia},
  {Salconi}, {Saleem}, {Samajdar}, {Sammut}, {Sanchez}, {Sanchez},
  {Sanchis-Gual}, {Sanders}, {Santiago}, {Santos}, {Sarin}, {Sassolas},
  {Sauter}, {Savage}, {Schale}, {Scheel}, {Scheuer}, {Schmidt}, {Schnabel},
  {Schofield}, {Sch{\"o}nbeck}, {Schreiber}, {Schulte}, {Schutz}, {Scott},
  {Scott}, {Seidel}, {Sellers}, {Sengupta}, {Sennett}, {Sentenac}, {Sequino},
  {Sergeev}, {Setyawati}, {Shaddock}, {Shaffer}, {Shahriar}, {Shaner},
  {Sharma}, {Sharma}, {Shawhan}, {Shen}, {Shink}, {Shoemaker}, {Shoemaker},
  {Shukla}, {ShyamSundar}, {Siellez}, {Sieniawska}, {Sigg}, {Singer}, {Singh},
  {Singh}, {Singhal}, {Sintes}, {Sitmukhambetov}, {Skliris}, {Slagmolen},
  {Slaven-Blair}, {Smith}, {Smith}, {Somala}, {Son}, {Soni}, {Sorazu},
  {Sorrentino}, {Souradeep}, {Sowell}, {Spencer}, {Spera}, {Srivastava},
  {Srivastava}, {Staats}, {Stachie}, {Standke}, {Steer}, {Steinke},
  {Steinlechner}, {Steinlechner}, {Steinmeyer}, {Stevenson}, {Stocks}, {Stone},
  {Stops}, {Strain}, {Stratta}, {Strigin}, {Strunk}, {Sturani}, {Stuver},
  {Sudhir}, {Summerscales}, {Sun}, {Sunil}, {Sur}, {Suresh}, {Sutton},
  {Swinkels}, {Szczepa{\'n}czyk}, {Tacca}, {Tait}, {Talbot}, {Tanner}, {Tao},
  {T{\'a}pai}, {Tapia}, {Tasson}, {Taylor}, {Tenorio}, {Terkowski}, {Thomas},
  {Thomas}, {Thondapu}, {Thorne}, {Thrane}, {Tiwari}, {Tiwari}, {Tiwari},
  {Toland}, {Tonelli}, {Tornasi}, {Torres-Forn{\'e}}, {Torrie},
  {T{\"o}yr{\"a}}, {Travasso}, {Traylor}, {Tringali}, {Tripathee}, {Trovato},
  {Trozzo}, {Tsang}, {Tse}, {Tso}, {Tsukada}, {Tsuna}, {Tsutsui}, {Tuyenbayev},
  {Ueno}, {Ugolini}, {Unnikrishnan}, {Urban}, {Usman}, {Vahlbruch}, {Vajente},
  {Valdes}, {Valentini}, {van Bakel}, {van Beuzekom}, {van den Brand}, {Van Den
  Broeck}, {Vander-Hyde}, {van der Schaaf}, {VanHeijningen}, {van Veggel},
  {Vardaro}, {Varma}, {Vass}, {Vas{\'u}th}, {Vecchio}, {Vedovato}, {Veitch},
  {Veitch}, {Venkateswara}, {Venugopalan}, {Verkindt}, {Vetrano}, {Vicer{\'e}},
  {Viets}, {Vinciguerra}, {Vine}, {Vinet}, {Vitale}, {Vo}, {Vocca}, {Vorvick},
  {Vyatchanin}, {Wade}, {Wade}, {Wade}, {Walet}, {Walker}, {Wallace}, {Walsh},
  {Wang}, {Wang}, {Wang}, {Wang}, {Wang}, {Ward}, {Warden}, {Warner}, {Was},
  {Watchi}, {Weaver}, {Wei}, {Weinert}, {Weinstein}, {Weiss}, {Wellmann},
  {Wen}, {Wessel}, {We{\ss}els}, {Westhouse}, {Wette}, {Whelan}, {Whiting},
  {Whittle}, {Wilken}, {Williams}, {Williamson}, {Willis}, {Willke}, {Winkler},
  {Wipf}, {Wittel}, {Woan}, {Woehler}, {Wofford}, {Wright}, {Wu}, {Wysocki},
  {Xiao}, {Xu}, {Yamamoto}, {Yancey}, {Yang}, {Yang}, {Yang}, {Yap}, {Yazback},
  {Yeeles}, {Yu}, {Yu}, {Yuen}, {Zadro{\.z}ny}, {Zadro{\.z}ny}, {Zanolin},
  {Zelenova}, {Zendri}, {Zevin}, {Zhang}, {Zhang}, {Zhang}, {Zhao}, {Zhao},
  {Zhou}, {Zhou}, {Zhu}, {Zucker}, {Zweizig}, {Papa}, \&
  {Salemi}}]{2019arXiv190503457T}
---. 2019, PhRvD, 100, 024017, \dodoi{10.1103/PhysRevD.100.024017}

\bibitem[{{Abbott} {et~al.}(2020){Abbott}, {Abbott}, {Abbott}, {Abraham},
  {Acernese}, {Ackley}, {Adams}, {Adya}, {Affeldt}, {Agathos}, {Agatsuma},
  {Aggarwal}, {Aguiar}, {Aiello}, {Ain}, {Ajith}, {Allen}, {Allocca}, {Aloy},
  {Altin}, {Amato}, {Anand}, {Ananyeva}, {Anderson}, {Anderson}, {Angelova},
  {Antier}, {Appert}, {Arai}, {Araya}, {Areeda}, {Ar{\`e}ne}, {Arnaud},
  {Aronson}, {Ascenzi}, {Ashton}, {Aston}, {Astone}, {Aubin}, {Aufmuth},
  {AultONeal}, {Austin}, {Avendano}, {Avila-Alvarez}, {Babak}, {Bacon},
  {Badaracco}, {Bader}, {Bae}, {Baird}, {Baker}, {Baldaccini}, {Ballardin},
  {Ballmer}, {Bals}, {Banagiri}, {Barayoga}, {Barbieri}, {Barclay}, {Barish},
  {Barker}, {Barkett}, {Barnum}, {Barone}, {Barr}, {Barsotti}, {Barsuglia},
  {Barta}, {Bartlett}, {Bartos}, {Bassiri}, {Basti}, {Bawaj}, {Bayley},
  {Bazzan}, {B{\'e}csy}, {Bejger}, {Belahcene}, {Bell}, {Beniwal}, {Benjamin},
  {Bergmann}, {Bernuzzi}, {Berry}, {Bersanetti}, {Bertolini}, {Betzwieser},
  {Bhandare}, {Bidler}, {Biggs}, {Bilenko}, {Bilgili}, {Billingsley}, {Birney},
  {Birnholtz}, {Biscans}, {Bischi}, {Biscoveanu}, {Bisht}, {Bitossi},
  {Bizouard}, {Blackburn}, {Blackman}, {Blair}, {Blair}, {Blair}, {Bloemen},
  {Bobba}, {Bode}, {Boer}, {Boetzel}, {Bogaert}, {Bondu}, {Bonnand}, {Booker},
  {Boom}, {Bork}, {Boschi}, {Bose}, {Bossilkov}, {Bosveld}, {Bouffanais},
  {Bozzi}, {Bradaschia}, {Brady}, {Bramley}, {Branchesi}, {Brau}, {Breschi},
  {Briant}, {Briggs}, {Brighenti}, {Brillet}, {Brinkmann}, {Brockill},
  {Brooks}, {Brooks}, {Brown}, {Brunett}, {Buikema}, {Bulik}, {Bulten},
  {Buonanno}, {Buskulic}, {Buy}, {Byer}, {Cabero}, {Cadonati}, {Cagnoli},
  {Cahillane}, {Bustillo}, {Callister}, {Calloni}, {Camp}, {Campbell},
  {Canepa}, {Cannon}, {Cao}, {Cao}, {Carapella}, {Carbognani}, {Caride},
  {Carney}, {Carullo}, {Diaz}, {Casentini}, {Caudill}, {Cavagli{\`a}},
  {Cavalier}, {Cavalieri}, {Cella}, {Cerd{\'a}-Dur{\'a}n}, {Cesarini},
  {Chaibi}, {Chakravarti}, {Chamberlin}, {Chan}, {Chao}, {Charlton}, {Chase},
  {Chassande-Mottin}, {Chatterjee}, {Chaturvedi}, {Cheeseboro}, {Chen}, {Chen},
  {Chen}, {Cheng}, {Cheong}, {Chia}, {Chiadini}, {Chincarini}, {Chiummo},
  {Cho}, {Cho}, {Cho}, {Christensen}, {Chu}, {Chua}, {Chung}, {Chung}, {Ciani},
  {Cie{\'s}lar}, {Ciobanu}, {Ciolfi}, {Cipriano}, {Cirone}, {Clara}, {Clark},
  {Clearwater}, {Cleva}, {Coccia}, {Cohadon}, {Cohen}, {Colleoni}, {Collette},
  {Collins}, {Colpi}, {Cominsky}, {Constancio}, {Conti}, {Cooper}, {Corban},
  {Corbitt}, {Cordero-Carri{\'o}n}, {Corezzi}, {Corley}, {Cornish}, {Corre},
  {Corsi}, {Cortese}, {Costa}, {Cotesta}, {Coughlin}, {Coughlin}, {Coulon},
  {Countryman}, {Couvares}, {Covas}, {Cowan}, {Coward}, {Cowart}, {Coyne},
  {Coyne}, {Creighton}, {Creighton}, {Cripe}, {Croquette}, {Crowder}, {Cullen},
  {Cumming}, {Cunningham}, {Cuoco}, {Canton}, {D{\'a}lya}, {D'Angelo},
  {Danilishin}, {D'Antonio}, {Danzmann}, {Dasgupta}, {Costa}, {Datrier},
  {Dattilo}, {Dave}, {Davier}, {Davis}, {Daw}, {DeBra}, {Deenadayalan},
  {Degallaix}, {De Laurentis}, {Del{\'e}glise}, {Del Pozzo}, {DeMarchi},
  {Demos}, {Dent}, {De Pietri}, {De Rosa}, {De Rossi}, {DeSalvo}, {de Varona},
  {Dhurandhar}, {D{\'\i}az}, {Dietrich}, {Di Fiore}, {DiFronzo}, {Di Giorgio},
  {Di Giovanni}, {Di Giovanni}, {Di Girolamo}, {Di Lieto}, {Ding}, {Di Pace},
  {Di Palma}, {Di Renzo}, {Divakarla}, {Dmitriev}, {Doctor}, {Donovan},
  {Dooley}, {Doravari}, {Dorrington}, {Downes}, {Drago}, {Driggers}, {Du},
  {Ducoin}, {Dupej}, {Durante}, {Dwyer}, {Easter}, {Eddolls}, {Edo}, {Effler},
  {Ehrens}, {Eichholz}, {Eikenberry}, {Eisenmann}, {Eisenstein}, {Errico},
  {Essick}, {Estelles}, {Estevez}, {Etienne}, {Etzel}, {Evans}, {Evans},
  {Fafone}, {Fairhurst}, {Fan}, {Farinon}, {Farr}, {Farr}, {Fauchon-Jones},
  {Favata}, {Fays}, {Fazio}, {Fee}, {Feicht}, {Fejer}, {Feng}, {Fernand
  ez-Galiana}, {Ferrante}, {Ferreira}, {Ferreira}, {Fidecaro}, {Fiori},
  {Fiorucci}, {Fishbach}, {Fisher}, {Fishner}, {Fittipaldi}, {Fitz-Axen},
  {Fiumara}, {Flaminio}, {Fletcher}, {Floden}, {Flynn}, {Fong}, {Font},
  {Forsyth}, {Fournier}, {Vivanco}, {Frasca}, {Frasconi}, {Frei}, {Freise},
  {Frey}, {Frey}, {Fritschel}, {Frolov}, {Fronz{\`e}}, {Fulda}, {Fyffe},
  {Gabbard}, {Gadre}, {Gaebel}, {Gair}, {Gammaitoni}, {Gaonkar},
  {Garc{\'\i}a-Quir{\'o}s}, {Garufi}, {Gateley}, {Gaudio}, {Gaur}, {Gayathri},
  {Gemme}, {Genin}, {Gennai}, {George}, {George}, {Gergely}, {Ghonge}, {Ghosh},
  {Ghosh}, {Ghosh}, {Giacomazzo}, {Giaime}, {Giardina}, {Gibson}, {Gill},
  {Glover}, {Gniesmer}, {Godwin}, {Goetz}, {Goetz}, {Goncharov},
  {Gonz{\'a}lez}, {Castro}, {Gopakumar}, {Gossan}, {Gosselin}, {Gouaty},
  {Grace}, {Grado}, {Granata}, {Grant}, {Gras}, {Grassia}, {Gray}, {Gray},
  {Greco}, {Green}, {Green}, {Gretarsson}, {Grimaldi}, {Grimm}, {Groot},
  {Grote}, {Grunewald}, {Gruning}, {Guidi}, {Gulati}, {Guo}, {Gupta}, {Gupta},
  {Gupta}, {Gustafson}, {Gustafson}, {Haegel}, {Halim}, {Hall}, {Hall},
  {Hamilton}, {Hammond}, {Haney}, {Hanke}, {Hanks}, {Hanna}, {Hannam},
  {Hannuksela}, {Hansen}, {Hanson}, {Harder}, {Hardwick}, {Haris}, {Harms},
  {Harry}, {Harry}, {Hasskew}, {Haster}, {Haughian}, {Hayes}, {Healy},
  {Heidmann}, {Heintze}, {Heitmann}, {Hellman}, {Hello}, {Hemming}, {Hendry},
  {Heng}, {Hennig}, {Heurs}, {Hild}, {Hinderer}, {Hochheim}, {Hofman},
  {Holgado}, {Holland}, {Holt}, {Holz}, {Hopkins}, {Horst}, {Hough}, {Howell},
  {Hoy}, {Huang}, {H{\"u}bner}, {Huerta}, {Huet}, {Hughey}, {Hui}, {Husa},
  {Huttner}, {Huynh-Dinh}, {Idzkowski}, {Iess}, {Inchauspe}, {Ingram}, {Inta},
  {Intini}, {Irwin}, {Isa}, {Isac}, {Isi}, {Iyer}, {Jacqmin}, {Jadhav}, {Jani},
  {Janthalur}, {Jaranowski}, {Jariwala}, {Jenkins}, {Jiang}, {Johnson},
  {Jones}, {Jones}, {Jones}, {Jones}, {Jonker}, {Ju}, {Junker}, {Kalaghatgi},
  {Kalogera}, {Kamai}, {Kandhasamy}, {Kang}, {Kanner}, {Kapadia}, {Karki},
  {Kashyap}, {Kasprzack}, {Katsanevas}, {Katsavounidis}, {Katzman}, {Kaufer},
  {Kawabe}, {Keerthana}, {K{\'e}f{\'e}lian}, {Keitel}, {Kennedy}, {Key},
  {Khalili}, {Khan}, {Khan}, {Khazanov}, {Khetan}, {Khursheed}, {Kijbunchoo},
  {Kim}, {Kim}, {Kim}, {Kim}, {Kim}, {Kim}, {Kimball}, {King}, {Kinley-Hanlon},
  {Kirchhoff}, {Kissel}, {Kleybolte}, {Klika}, {Klimenko}, {Knowles}, {Koch},
  {Koehlenbeck}, {Koekoek}, {Koley}, {Kondrashov}, {Kontos}, {Koper},
  {Korobko}, {Korth}, {Kovalam}, {Kozak}, {Kr{\"a}mer}, {Kringel},
  {Krishnendu}, {Kr{\'o}lak}, {Krupinski}, {Kuehn}, {Kumar}, {Kumar}, {Kumar},
  {Kumar}, {Kuo}, {Kutynia}, {Kwang}, {Lackey}, {Laghi}, {Lai}, {Lam},
  {Landry}, {Lane}, {Lang}, {Lange}, {Lantz}, {Lanza}, {Lartaux-Vollard},
  {Lasky}, {Laxen}, {Lazzarini}, {Lazzaro}, {Leaci}, {Leavey}, {Lecoeuche},
  {Lee}, {Lee}, {Lee}, {Lee}, {Lee}, {Lee}, {Lehmann}, {Lenon}, {Leroy},
  {Letendre}, {Levin}, {Li}, {Li}, {Li}, {Li}, {Li}, {Lin}, {Linde}, {Linker},
  {Littenberg}, {Liu}, {Liu}, {Llorens-Monteagudo}, {Lo}, {London}, {Longo},
  {Lorenzini}, {Loriette}, {Lormand}, {Losurdo}, {Lough}, {Lousto}, {Lovelace},
  {Lower}, {L{\"u}ck}, {Lumaca}, {Lundgren}, {Lynch}, {Ma}, {Macas}, {Macfoy},
  {MacInnis}, {Macleod}, {Macquet}, {Hernandez}, {Maga{\~n}a-Sandoval},
  {Magee}, {Majorana}, {Maksimovic}, {Malik}, {Man}, {Mandic}, {Mangano},
  {Mansell}, {Manske}, {Mantovani}, {Mapelli}, {Marchesoni}, {Marion},
  {M{\'a}rka}, {M{\'a}rka}, {Markakis}, {Markosyan}, {Markowitz}, {Maros},
  {Marquina}, {Marsat}, {Martelli}, {Martin}, {Martin}, {Martinez}, {Martynov},
  {Masalehdan}, {Mason}, {Massera}, {Masserot}, {Massinger}, {Masso-Reid},
  {Mastrogiovanni}, {Matas}, {Matichard}, {Matone}, {Mavalvala}, {McCann},
  {McCarthy}, {McClelland}, {McCormick}, {McCuller}, {McGuire}, {McIsaac},
  {McIver}, {McManus}, {McRae}, {McWilliams}, {Meacher}, {Meadors}, {Mehmet},
  {Mehta}, {Meidam}, {Villa}, {Melatos}, {Mendell}, {Mercer}, {Mereni},
  {Merfeld}, {Merilh}, {Merzougui}, {Meshkov}, {Messenger}, {Messick},
  {Messina}, {Metzdorff}, {Meyers}, {Meylahn}, {Miani}, {Miao}, {Michel},
  {Middleton}, {Milano}, {Miller}, {Millhouse}, {Mills}, {Milovich-Goff},
  {Minazzoli}, {Minenkov}, {Mishkin}, {Mishra}, {Mistry}, {Mitra},
  {Mitrofanov}, {Mitselmakher}, {Mittleman}, {Mo}, {Moffa}, {Mogushi},
  {Mohapatra}, {Molina-Ruiz}, {Mondin}, {Montani}, {Moore}, {Moraru},
  {Morawski}, {Moreno}, {Morisaki}, {Mours}, {Mow-Lowry}, {Muciaccia},
  {Mukherjee}, {Mukherjee}, {Mukherjee}, {Mukherjee}, {Mukund}, {Mullavey},
  {Munch}, {Mu{\~n}iz}, {Muratore}, {Murray}, {Nardecchia}, {Naticchioni},
  {Nayak}, {Neil}, {Neilson}, {Nelemans}, {Nelson}, {Nery}, {Neunzert},
  {Nevin}, {Ng}, {Ng}, {Nguyen}, {Nguyen}, {Nichols}, {Nichols}, {Nissanke},
  {Nocera}, {North}, {Nuttall}, {Obergaulinger}, {Oberling}, {O'Brien},
  {Oganesyan}, {Ogin}, {Oh}, {Oh}, {Ohme}, {Ohta}, {Okada}, {Oliver},
  {Oppermann}, {Oram}, {O'Reilly}, {Ormiston}, {Ortega}, {O'Shaughnessy},
  {Ossokine}, {Ottaway}, {Overmier}, {Owen}, {Pace}, {Pagano}, {Page},
  {Pagliaroli}, {Pai}, {Pai}, {Palamos}, {Palashov}, {Palomba}, {Pan}, {Panda},
  {Pang}, {Pankow}, {Pannarale}, {Pant}, {Paoletti}, {Paoli}, {Parida},
  {Parker}, {Pascucci}, {Pasqualetti}, {Passaquieti}, {Passuello}, {Patil},
  {Patricelli}, {Payne}, {Pearlstone}, {Pechsiri}, {Pedersen}, {Pedraza},
  {Pedurand }, {Pele}, {Penn}, {Perego}, {Perez}, {P{\'e}rigois}, {Perreca},
  {Petermann}, {Pfeiffer}, {Phelps}, {Phukon}, {Piccinni}, {Pichot},
  {Piergiovanni}, {Pierro}, {Pillant}, {Pinard}, {Pinto}, {Pirello}, {Pitkin},
  {Plastino}, {Poggiani}, {Pong}, {Ponrathnam}, {Popolizio}, {Porter},
  {Powell}, {Prajapati}, {Prasad}, {Prasai}, {Prasanna}, {Pratten},
  {Prestegard}, {Principe}, {Prodi}, {Prokhorov}, {Punturo}, {Puppo},
  {P{\"u}rrer}, {Qi}, {Quetschke}, {Quinonez}, {Raab}, {Raaijmakers},
  {Radkins}, {Radulesco}, {Raffai}, {Raja}, {Rajan}, {Rajbhandari},
  {Rakhmanov}, {Ramirez}, {Ramos-Buades}, {Rana}, {Rao}, {Rapagnani},
  {Raymond}, {Razzano}, {Read}, {Regimbau}, {Rei}, {Reid}, {Reitze},
  {Rettegno}, {Ricci}, {Richardson}, {Richardson}, {Ricker}, {Riemenschneider},
  {Riles}, {Rizzo}, {Robertson}, {Robinet}, {Rocchi}, {Rolland }, {Rollins},
  {Roma}, {Romanelli}, {Romano}, {Romel}, {Romie}, {Rose}, {Rose}, {Rose},
  {Rosi{\'n}ska}, {Rosofsky}, {Ross}, {Rowan}, {R{\"u}diger}, {Ruggi},
  {Rutins}, {Ryan}, {Sachdev}, {Sadecki}, {Sakellariadou}, {Salafia},
  {Salconi}, {Saleem}, {Samajdar}, {Sammut}, {Sanchez}, {Sanchez},
  {Sanchis-Gual}, {Sanders}, {Santiago}, {Santos}, {Sarin}, {Sassolas},
  {Sauter}, {Savage}, {Schale}, {Scheel}, {Scheuer}, {Schmidt}, {Schnabel},
  {Schofield}, {Sch{\"o}nbeck}, {Schreiber}, {Schulte}, {Schutz}, {Scott},
  {Scott}, {Seidel}, {Sellers}, {Sengupta}, {Sennett}, {Sentenac}, {Sequino},
  {Sergeev}, {Setyawati}, {Shaddock}, {Shaffer}, {Shahriar}, {Shaner},
  {Sharma}, {Sharma}, {Shawhan}, {Shen}, {Shink}, {Shoemaker}, {Shoemaker},
  {Shukla}, {ShyamSundar}, {Siellez}, {Sieniawska}, {Sigg}, {Singer}, {Singh},
  {Singh}, {Singhal}, {Sintes}, {Sitmukhambetov}, {Skliris}, {Slagmolen},
  {Slaven-Blair}, {Smith}, {Smith}, {Somala}, {Son}, {Soni}, {Sorazu},
  {Sorrentino}, {Souradeep}, {Sowell}, {Spencer}, {Spera}, {Srivastava},
  {Srivastava}, {Staats}, {Stachie}, {Standke}, {Steer}, {Steinke},
  {Steinlechner}, {Steinlechner}, {Steinmeyer}, {Stevenson}, {Stocks}, {Stone},
  {Stops}, {Strain}, {Stratta}, {Strigin}, {Strunk}, {Sturani}, {Stuver},
  {Sudhir}, {Summerscales}, {Sun}, {Sunil}, {Sur}, {Suresh}, {Sutton},
  {Swinkels}, {Szczepa{\'n}czyk}, {Tacca}, {Tait}, {Talbot}, {Tanner}, {Tao},
  {T{\'a}pai}, {Tapia}, {Tasson}, {Taylor}, {Tenorio}, {Terkowski}, {Thomas},
  {Thomas}, {Thondapu}, {Thorne}, {Thrane}, {Tiwari}, {Tiwari}, {Tiwari},
  {Toland}, {Tonelli}, {Tornasi}, {Torres-Forn{\'e}}, {Torrie},
  {T{\"o}yr{\"a}}, {Travasso}, {Traylor}, {Tringali}, {Tripathee}, {Trovato},
  {Trozzo}, {Tsang}, {Tse}, {Tso}, {Tsukada}, {Tsuna}, {Tsutsui}, {Tuyenbayev},
  {Ueno}, {Ugolini}, {Unnikrishnan}, {Urban}, {Usman}, {Vahlbruch}, {Vajente},
  {Valdes}, {Valentini}, {van Bakel}, {van Beuzekom}, {van den Brand}, {Van Den
  Broeck}, {Vander-Hyde}, {van der Schaaf}, {VanHeijningen}, {van Veggel},
  {Vardaro}, {Varma}, {Vass}, {Vas{\'u}th}, {Vecchio}, {Vedovato}, {Veitch},
  {Veitch}, {Venkateswara}, {Venugopalan}, {Verkindt}, {Vetrano}, {Vicer{\'e}},
  {Viets}, {Vinciguerra}, {Vine}, {Vinet}, {Vitale}, {Vo}, {Vocca}, {Vorvick},
  {Vyatchanin}, {Wade}, {Wade}, {Wade}, {Walet}, {Walker}, {Wallace}, {Walsh},
  {Wang}, {Wang}, {Wang}, {Wang}, {Wang}, {Ward}, {Warden}, {Warner}, {Was},
  {Watchi}, {Weaver}, {Wei}, {Weinert}, {Weinstein}, {Weiss}, {Wellmann},
  {Wen}, {Wessel}, {We{\ss}els}, {Westhouse}, {Wette}, {Whelan}, {Whiting},
  {Whittle}, {Wilken}, {Williams}, {Williamson}, {Willis}, {Willke}, {Winkler},
  {Wipf}, {Wittel}, {Woan}, {Woehler}, {Wofford}, {Wright}, {Wu}, {Wysocki},
  {Xiao}, {Xu}, {Yamamoto}, {Yancey}, {Yang}, {Yang}, {Yang}, {Yap}, {Yazback},
  {Yeeles}, {Yu}, {Yu}, {Yuen}, {Zadro{\.Z}ny}, {Zadro{\.Z}ny}, {Zanolin},
  {Zelenova}, {Zendri}, {Zevin}, {Zhang}, {Zhang}, {Zhang}, {Zhao}, {Zhao},
  {Zhou}, {Zhou}, {Zhu}, {Zucker}, {Zweizig}, {Holoien}, {Kochanek}, {Prieto},
  {Shappee}, {Stanek}, {Haislip}, {Kouprianov}, {Reichart}, {Sand },
  {Tartaglia}, {Valenti}, {Wyatt}, {Yang}, {Salemi}, {LIGO Scientific
  Collaboration}, \& {Virgo Collaboration}}]{oosnsearch}
---. 2020, \prd, 101, 084002, \dodoi{10.1103/PhysRevD.101.084002}

\bibitem[{{Abdikamalov} {et~al.}(2015){Abdikamalov}, {Ott}, {Radice},
  {Roberts}, {Haas}, {Reisswig}, {M{\"o}sta}, {Klion}, \&
  {Schnetter}}]{2015ApJ...808...70A}
{Abdikamalov}, E., {Ott}, C.~D., {Radice}, D., {et~al.} 2015, \apj, 808, 70,
  \dodoi{10.1088/0004-637X/808/1/70}

\bibitem[{{Acernese} {et~al.}(2015){Acernese}, {Agathos}, {Agatsuma}, {Aisa},
  {Allemandou}, {Allocca}, {Amarni}, {Astone}, {Balestri}, {Ballardin},
  {Barone}, {Baronick}, {Barsuglia}, {Basti}, {Basti}, {Bauer}, {Bavigadda},
  {Bejger}, {Beker}, {Belczynski}, {Bersanetti}, {Bertolini}, {Bitossi},
  {Bizouard}, {Bloemen}, {Blom}, {Boer}, {Bogaert}, {Bondi}, {Bondu},
  {Bonelli}, {Bonnand}, {Boschi}, {Bosi}, {Bouedo}, {Bradaschia}, {Branchesi},
  {Briant}, {Brillet}, {Brisson}, {Bulik}, {Bulten}, {Buskulic}, {Buy},
  {Cagnoli}, {Calloni}, {Campeggi}, {Canuel}, {Carbognani}, {Cavalier},
  {Cavalieri}, {Cella}, {Cesarini}, {Chassande-Mottin}, {Chincarini},
  {Chiummo}, {Chua}, {Cleva}, {Coccia}, {Cohadon}, {Colla}, {Colombini},
  {Conte}, {Coulon}, {Cuoco}, {Dalmaz}, {D'Antonio}, {Dattilo}, {Davier},
  {Day}, {Debreczeni}, {Degallaix}, {Del{\'e}glise}, {Del Pozzo}, {Dereli}, {De
  Rosa}, {Di Fiore}, {Di Lieto}, {Di Virgilio}, {Doets}, {Dolique}, {Drago},
  {Ducrot}, {Endr{\H{o}}czi}, {Fafone}, {Farinon}, {Ferrante}, {Ferrini},
  {Fidecaro}, {Fiori}, {Flaminio}, {Fournier}, {Franco}, {Frasca}, {Frasconi},
  {Gammaitoni}, {Garufi}, {Gaspard}, {Gatto}, {Gemme}, {Gendre}, {Genin},
  {Gennai}, {Ghosh}, {Giacobone}, {Giazotto}, {Gouaty}, {Granata}, {Greco},
  {Groot}, {Guidi}, {Harms}, {Heidmann}, {Heitmann}, {Hello}, {Hemming},
  {Hennes}, {Hofman}, {Jaranowski}, {Jonker}, {Kasprzack}, {K{\'e}f{\'e}lian},
  {Kowalska}, {Kraan}, {Kr{\'o}lak}, {Kutynia}, {Lazzaro}, {Leonardi}, {Leroy},
  {Letendre}, {Li}, {Lieunard}, {Lorenzini}, {Loriette}, {Losurdo},
  {Magazz{\`u}}, {Majorana}, {Maksimovic}, {Malvezzi}, {Man}, {Mangano},
  {Mantovani}, {Marchesoni}, {Marion}, {Marque}, {Martelli}, {Martellini},
  {Masserot}, {Meacher}, {Meidam}, {Mezzani}, {Michel}, {Milano}, {Minenkov},
  {Moggi}, {Mohan}, {Montani}, {Morgado}, {Mours}, {Mul}, {Nagy}, {Nardecchia},
  {Naticchioni}, {Nelemans}, {Neri}, {Neri}, {Nocera}, {Pacaud}, {Palomba},
  {Paoletti}, {Paoli}, {Pasqualetti}, {Passaquieti}, {Passuello}, {Perciballi},
  {Petit}, {Pichot}, {Piergiovanni}, {Pillant}, {Piluso}, {Pinard}, {Poggiani},
  {Prijatelj}, {Prodi}, {Punturo}, {Puppo}, {Rabeling}, {R{\'a}cz},
  {Rapagnani}, {Razzano}, {Re}, {Regimbau}, {Ricci}, {Robinet}, {Rocchi},
  {Rolland}, {Romano}, {Rosi{\'n}ska}, {Ruggi}, {Saracco}, {Sassolas},
  {Schimmel}, {Sentenac}, {Sequino}, {Shah}, {Siellez}, {Straniero},
  {Swinkels}, {Tacca}, {Tonelli}, {Travasso}, {Turconi}, {Vajente}, {van
  Bakel}, {van Beuzekom}, {van den Brand}, {Van Den Broeck}, {van der Sluys},
  {van Heijningen}, {Vas{\'u}th}, {Vedovato}, {Veitch}, {Verkindt}, {Vetrano},
  {Vicer{\'e}}, {Vinet}, {Visser}, {Vocca}, {Ward}, {Was}, {Wei}, {Yvert},
  {Zadro {\.z}ny}, \& {Zendri}}]{2015CQGra..32b4001A}
{Acernese}, F., {Agathos}, M., {Agatsuma}, K., {et~al.} 2015, Classical and
  Quantum Gravity, 32, 024001, \dodoi{10.1088/0264-9381/32/2/024001}

\bibitem[{{Ando} {et~al.}(2005){Ando}, Beacom, \& Y\"uksel}]{ando:05}
{Ando}, S., Beacom, F., \& Y\"uksel, H. 2005, Phys. Rev. Lett., 95, 171101

\bibitem[{{Andrae} {et~al.}(2010){Andrae}, {Schulze-Hartung}, \&
  {Melchior}}]{2010arXiv1012.3754A}
{Andrae}, R., {Schulze-Hartung}, T., \& {Melchior}, P. 2010, arXiv e-prints,
  arXiv:1012.3754.
\newblock \doarXiv{1012.3754}

\bibitem[{{Andresen} {et~al.}(2017){Andresen}, {M{\"u}ller}, {M{\"u}ller}, \&
  {Janka}}]{andresen_17}
{Andresen}, H., {M{\"u}ller}, B., {M{\"u}ller}, E., \& {Janka}, H.-T. 2017,
  \mnras, 468, 2032, \dodoi{10.1093/mnras/stx618}

\bibitem[{{Andresen} {et~al.}(2019){Andresen}, {M{\"u}ller}, {Janka}, {Summa},
  {Gill}, \& {Zanolin}}]{andresen_19}
{Andresen}, H., {M{\"u}ller}, E., {Janka}, H.~T., {et~al.} 2019, Monthly
  Notices of the Royal Astronomical Society, 486, 2238,
  \dodoi{10.1093/mnras/stz990}

\bibitem[{Andresen {et~al.}(2017)Andresen, Müller, Müller, \&
  Janka}]{10.1093/mnras/stx618}
Andresen, H., Müller, B., Müller, E., \& Janka, H.-T. 2017, Monthly Notices
  of the Royal Astronomical Society, 468, 2032, \dodoi{10.1093/mnras/stx618}

\bibitem[{Andresen {et~al.}(2019)Andresen, Müller, Janka, Summa, Gill, \&
  Zanolin}]{10.1093/mnras/stz990}
Andresen, H., Müller, E., Janka, H.-T., {et~al.} 2019, Monthly Notices of the
  Royal Astronomical Society, 486, 2238, \dodoi{10.1093/mnras/stz990}

\bibitem[{{Babak} {et~al.}(2006){Babak}, {Balasubramanian}, {Churches},
  {Cokelaer}, \& {Sathyaprakash}}]{2006CQGra..23.5477B}
{Babak}, S., {Balasubramanian}, R., {Churches}, D., {Cokelaer}, T., \&
  {Sathyaprakash}, B.~S. 2006, Classical and Quantum Gravity, 23, 5477,
  \dodoi{10.1088/0264-9381/23/18/002}

\bibitem[{{Barker} {et~al.}(2021){Barker}, {Harris}, {Warren}, {O'Connor}, \&
  {Couch}}]{2021arXiv210201118B}
{Barker}, B.~L., {Harris}, C.~E., {Warren}, M.~L., {O'Connor}, E.~P., \&
  {Couch}, S.~M. 2021, arXiv e-prints, arXiv:2102.01118.
\newblock \doarXiv{2102.01118}

\bibitem[{{Blinnikov}(2017)}]{2017hsn..book..843B}
{Blinnikov}, S. 2017, in Handbook of Supernovae, ed. A.~W. {Alsabti} \&
  P.~{Murdin} (Cham: Springer), 843, \dodoi{10.1007/978-3-319-21846-5\_31}

\bibitem[{{Bose} \& {Kumar}(2014)}]{2014ApJ...782...98B}
{Bose}, S., \& {Kumar}, B. 2014, \apj, 782, 98,
  \dodoi{10.1088/0004-637X/782/2/98}

\bibitem[{{Botticella} {et~al.}(2012){Botticella}, {Smartt}, {Kennicutt},
  {Cappellaro}, {Sereno}, \& {Lee}}]{botticella:11}
{Botticella}, M.~T., {Smartt}, S.~J., {Kennicutt}, R.~C., {et~al.} 2012, \aap,
  537, A132.
\newblock \doarXiv{1111.1692}

\bibitem[{{Brown} {et~al.}(2013){Brown}, {Baliber}, {Bianco}, {Bowman},
  {Burleson}, {Conway}, {Crellin}, {Depagne}, {De Vera}, {Dilday}, {Dragomir},
  {Dubberley}, {Eastman}, {Elphick}, {Falarski}, {Foale}, {Ford}, {Fulton},
  {Garza}, {Gomez}, {Graham}, {Greene}, {Haldeman}, {Hawkins}, {Haworth},
  {Haynes}, {Hidas}, {Hjelstrom}, {Howell}, {Hygelund}, {Lister}, {Lobdill},
  {Martinez}, {Mullins}, {Norbury}, {Parrent}, {Paulson}, {Petry}, {Pickles},
  {Posner}, {Rosing}, {Ross}, {Sand}, {Saunders}, {Shobbrook}, {Shporer},
  {Street}, {Thomas}, {Tsapras}, {Tufts}, {Valenti}, {Vander Horst}, {Walker},
  {White}, \& {Willis}}]{2013PASP..125.1031B}
{Brown}, T.~M., {Baliber}, N., {Bianco}, F.~B., {et~al.} 2013, \pasp, 125,
  1031, \dodoi{10.1086/673168}

\bibitem[{{Calzavara} \& {Matzner}(2004)}]{2004MNRAS.351..694C}
{Calzavara}, A.~J., \& {Matzner}, C.~D. 2004, \mnras, 351, 694,
  \dodoi{10.1111/j.1365-2966.2004.07818.x}

\bibitem[{{Cappellaro} {et~al.}(1999){Cappellaro}, {Evans}, \&
  {Turatto}}]{cappellaro:99}
{Cappellaro}, E., {Evans}, R., \& {Turatto}, M. 1999, \aap, 351, 459

\bibitem[{{Cappellaro} {et~al.}(1993{\natexlab{a}}){Cappellaro}, {Turatto},
  {Benetti}, {Tsvetkov}, {Bartunov}, \& {Makarova}}]{cappellaro:93}
{Cappellaro}, E., {Turatto}, M., {Benetti}, S., {et~al.} 1993{\natexlab{a}},
  \aap, 273, 383

\bibitem[{{Cappellaro} {et~al.}(1993{\natexlab{b}}){Cappellaro}, {Turatto},
  {Benetti}, {Tsvetkov}, {Bartunov}, \& {Makarova}}]{1993A&A...273..383C}
---. 1993{\natexlab{b}}, \aap, 273, 383

\bibitem[{{Cappellaro} {et~al.}(1997){Cappellaro}, {Turatto}, {Tsvetkov},
  {Bartunov}, {Pollas}, {Evans}, \& {Hamuy}}]{1997A&A...322..431C}
{Cappellaro}, E., {Turatto}, M., {Tsvetkov}, D.~Y., {et~al.} 1997, \aap, 322,
  431

\bibitem[{{Cappellaro} {et~al.}(2015){Cappellaro}, {Botticella}, {Pignata},
  {Grado}, {Greggio}, {Limatola}, {Vaccari}, {Baruffolo}, {Benetti}, {Bufano},
  {Capaccioli}, {Cascone}, {Covone}, {De Cicco}, {Falocco}, {Della Valle},
  {Jarvis}, {Marchetti}, {Napolitano}, {Paolillo}, {Pastorello}, {Radovich},
  {Schipani}, {Spiro}, {Tomasella}, \& {Turatto}}]{2015A&A...584A..62C}
{Cappellaro}, E., {Botticella}, M.~T., {Pignata}, G., {et~al.} 2015, \aap, 584,
  A62, \dodoi{10.1051/0004-6361/201526712}

\bibitem[{{Cardelli} {et~al.}(1989){Cardelli}, {Clayton}, \&
  {Mathis}}]{1989ApJ...345..245C}
{Cardelli}, J.~A., {Clayton}, G.~C., \& {Mathis}, J.~S. 1989, \apj, 345, 245,
  \dodoi{10.1086/167900}

\bibitem[{{Chambers} {et~al.}(2016){Chambers}, {Magnier}, {Metcalfe},
  {Flewelling}, {Huber}, {Waters}, {Denneau}, {Draper}, {Farrow}, {Finkbeiner},
  {Holmberg}, {Koppenhoefer}, {Price}, {Rest}, {Saglia}, {Schlafly}, {Smartt},
  {Sweeney}, {Wainscoat}, {Burgett}, {Chastel}, {Grav}, {Heasley}, {Hodapp},
  {Jedicke}, {Kaiser}, {Kudritzki}, {Luppino}, {Lupton}, {Monet}, {Morgan},
  {Onaka}, {Shiao}, {Stubbs}, {Tonry}, {White}, {Ba{\~n}ados}, {Bell},
  {Bender}, {Bernard}, {Boegner}, {Boffi}, {Botticella}, {Calamida},
  {Casertano}, {Chen}, {Chen}, {Cole}, {Deacon}, {Frenk}, {Fitzsimmons},
  {Gezari}, {Gibbs}, {Goessl}, {Goggia}, {Gourgue}, {Goldman}, {Grant},
  {Grebel}, {Hambly}, {Hasinger}, {Heavens}, {Heckman}, {Henderson}, {Henning},
  {Holman}, {Hopp}, {Ip}, {Isani}, {Jackson}, {Keyes}, {Koekemoer}, {Kotak},
  {Le}, {Liska}, {Long}, {Lucey}, {Liu}, {Martin}, {Masci}, {McLean}, {Mindel},
  {Misra}, {Morganson}, {Murphy}, {Obaika}, {Narayan}, {Nieto-Santisteban},
  {Norberg}, {Peacock}, {Pier}, {Postman}, {Primak}, {Rae}, {Rai}, {Riess},
  {Riffeser}, {Rix}, {R{\"o}ser}, {Russel}, {Rutz}, {Schilbach}, {Schultz},
  {Scolnic}, {Strolger}, {Szalay}, {Seitz}, {Small}, {Smith}, {Soderblom},
  {Taylor}, {Thomson}, {Taylor}, {Thakar}, {Thiel}, {Thilker}, {Unger},
  {Urata}, {Valenti}, {Wagner}, {Walder}, {Walter}, {Watters}, {Werner},
  {Wood-Vasey}, \& {Wyse}}]{2016arXiv161205560C}
{Chambers}, K.~C., {Magnier}, E.~A., {Metcalfe}, N., {et~al.} 2016, arXiv
  e-prints, arXiv:1612.05560.
\newblock \doarXiv{1612.05560}

\bibitem[{{Chevalier}(1992)}]{1992ApJ...394..599C}
{Chevalier}, R.~A. 1992, \apj, 394, 599, \dodoi{10.1086/171612}

\bibitem[{{Chevalier} \& {Irwin}(2011)}]{2011ApJ...729L...6C}
{Chevalier}, R.~A., \& {Irwin}, C.~M. 2011, \apjl, 729, L6,
  \dodoi{10.1088/2041-8205/729/1/L6}

\bibitem[{{Couch}(2013)}]{2013ApJ...775...35C}
{Couch}, S.~M. 2013, \apj, 775, 35, \dodoi{10.1088/0004-637X/775/1/35}

\bibitem[{{Dahlen} {et~al.}(2012){Dahlen}, {Strolger}, {Riess}, {Mattila},
  {Kankare}, \& {Mobasher}}]{2012ApJ...757...70D}
{Dahlen}, T., {Strolger}, L.-G., {Riess}, A.~G., {et~al.} 2012, \apj, 757, 70,
  \dodoi{10.1088/0004-637X/757/1/70}

\bibitem[{{Das} \& {Ray}(2017)}]{2017ApJ...851..138D}
{Das}, S., \& {Ray}, A. 2017, \apj, 851, 138, \dodoi{10.3847/1538-4357/aa97e1}

\bibitem[{Davies(2017)}]{Davies2017}
Davies, B. 2017, Philosophical Transactions of the Royal Society A:
  Mathematical, Physical and Engineering Sciences, 375, 20160270,
  \dodoi{10.1098/rsta.2016.0270}

\bibitem[{{Davies}(2017)}]{2017RSPTA.37560270D}
{Davies}, B. 2017, Philosophical Transactions of the Royal Society of London
  Series A, 375, 20160270, \dodoi{10.1098/rsta.2016.0270}

\bibitem[{{Davies} \& {Beasor}(2018)}]{2018MNRAS.474.2116D}
{Davies}, B., \& {Beasor}, E.~R. 2018, \mnras, 474, 2116,
  \dodoi{10.1093/mnras/stx2734}

\bibitem[{Dessart {et~al.}(2017)Dessart, Hillier, \& Audit}]{Dessart:2017pfi}
Dessart, L., Hillier, D.~J., \& Audit, E. 2017, Astron. Astrophys., 605, A83,
  \dodoi{10.1051/0004-6361/201730942}

\bibitem[{{Emilio Enriquez} {et~al.}(2011){Emilio Enriquez}, {Leonard},
  {Poznanski}, {Filippenko}, {Chornock}, {Foley}, {Ganeshalingam}, {Li}, \&
  {Silverman}}]{2011AAS...21733721E}
{Emilio Enriquez}, J., {Leonard}, D.~C., {Poznanski}, D., {et~al.} 2011, in
  Bulletin of the American Astronomical Society, Vol.~43, American Astronomical
  Society Meeting Abstracts \#217, 337.21

\bibitem[{{Evans} \& {Zanolin}(2017)}]{Evans2017}
{Evans}, M., \& {Zanolin}, M. 2017, in Handbook of Supernovae, ed. A.~W.
  {Alsabti} \& P.~{Murdin} (Cham: Springer), 1699,
  \dodoi{10.1007/978-3-319-21846-5\_10}

\bibitem[{{Foglizzo} {et~al.}(2009){Foglizzo}, {Guilet}, \&
  {Sato}}]{2009sf2a.conf..143F}
{Foglizzo}, T., {Guilet}, J., \& {Sato}, J. 2009, in SF2A-2009: Proceedings of
  the Annual meeting of the French Society of Astronomy and Astrophysics, ed.
  M.~{Heydari-Malayeri}, C.~{Reyl'E}, \& R.~{Samadi}, 143

\bibitem[{{Forster} {et~al.}(2020{\natexlab{a}}){Forster}, {Bauer}, {Pignata},
  {Arredondo}, {Cabrera-Vives}, {Carrasco-Davis}, {Estevez}, {Huijse}, {Reyes},
  {Reyes}, {Sanchez-Saez}, {Valenzuela}, {Castillo}, {Ruz-Mieres},
  {Rodriguez-Mancini}, {Bauer}, {Catelan}, {Eyheramendy}, \&
  {Graham}}]{2020TNSTR.914....1F}
{Forster}, F., {Bauer}, F.~E., {Pignata}, G., {et~al.} 2020{\natexlab{a}},
  TNSTR, 2020-914, 1

\bibitem[{{Forster} {et~al.}(2020{\natexlab{b}}){Forster}, {Pignata}, {Bauer},
  {Arredondo}, {Cabrera-Vives}, {Carrasco-Davis}, {Estevez}, {Huijse}, {Reyes},
  {Reyes}, {Sanchez-Saez}, {Valenzuela}, {Castillo}, {Ruz-Mieres},
  {Rodriguez-Mancini}, {Bauer}, {Catelan}, {Eyheramendy}, \&
  {Graham}}]{2020TNSTR..67....1F}
{Forster}, F., {Pignata}, G., {Bauer}, F.~E., {et~al.} 2020{\natexlab{b}},
  TNSTR, 2020-67, 1

\bibitem[{{Garnavich} {et~al.}(2016){Garnavich}, {Tucker}, {Rest}, {Shaya},
  {Olling}, {Kasen}, \& {Villar}}]{2016ApJ...820...23G}
{Garnavich}, P.~M., {Tucker}, B.~E., {Rest}, A., {et~al.} 2016, \apj, 820, 23,
  \dodoi{10.3847/0004-637X/820/1/23}

\bibitem[{Godoy-Rivera {et~al.}(2017)Godoy-Rivera, Brown, Shields, Kochanek,
  Stanek, Thompson, Holoien, Beacom, Bersier, Strader, Chomiuk, Falco, Morrell,
  Pojmanski, Stritzinger, WoÅºniak, Bock, Cacella, Conseil, Cruz, Fernandez,
  Kiyota, Koff, Krannich, Marples, Masi, Monard, Nicholls, Nicolas, Shappee,
  Post, Stone, Wiethoff, Prieto, Chen, Bose, Dong, Brimacombe, \&
  Bishop}]{10.1093/mnras/stx1544}
Godoy-Rivera, D., Brown, J.~S., Shields, J.~V., {et~al.} 2017, Monthly Notices
  of the Royal Astronomical Society, 471, 4966, \dodoi{10.1093/mnras/stx1544}

\bibitem[{{Groh}(2017)}]{2017RSPTA.37570219G}
{Groh}, J.~H. 2017, Philosophical Transactions of the Royal Society of London
  Series A, 375, 20170219, \dodoi{10.1098/rsta.2017.0219}

\bibitem[{{Grzegorzek}(2019)}]{2019TNSTR.666....1G}
{Grzegorzek}, J. 2019, TNSTR, 2019-666, 1

\bibitem[{{Guevel} \& {Hosseinzadeh}(2017)}]{david_guevel_2017_1043973}
{Guevel}, D., \& {Hosseinzadeh}, G. 2017, {Dguevel/Pyzogy: Initial Release},
  v0.0.1,  Zenodo, \dodoi{10.5281/zenodo.1043973}

\bibitem[{{Hanke} {et~al.}(2013){Hanke}, {M{\"u}ller}, {Wongwathanarat},
  {Marek}, \& {Janka}}]{2013ApJ...770...66H}
{Hanke}, F., {M{\"u}ller}, B., {Wongwathanarat}, A., {Marek}, A., \& {Janka},
  H.-T. 2013, \apj, 770, 66, \dodoi{10.1088/0004-637X/770/1/66}

\bibitem[{Hayama {et~al.}(2015)Hayama, Kuroda, Kotake, \&
  Takiwaki}]{PhysRevD.92.122001}
Hayama, K., Kuroda, T., Kotake, K., \& Takiwaki, T. 2015, Phys. Rev. D, 92,
  122001, \dodoi{10.1103/PhysRevD.92.122001}

\bibitem[{{Hayama} {et~al.}(2018{\natexlab{a}}){Hayama}, {Kuroda}, {Kotake}, \&
  {Takiwaki}}]{hayama_18}
{Hayama}, K., {Kuroda}, T., {Kotake}, K., \& {Takiwaki}, T. 2018{\natexlab{a}},
  \mnras, 477, L96, \dodoi{10.1093/mnrasl/sly055}

\bibitem[{{Hayama} {et~al.}(2018{\natexlab{b}}){Hayama}, {Kuroda}, {Kotake}, \&
  {Takiwaki}}]{2018MNRAS.477L..96H}
---. 2018{\natexlab{b}}, \mnras, 477, L96, \dodoi{10.1093/mnrasl/sly055}

\bibitem[{{Hicken} {et~al.}(2017){Hicken}, {Friedman}, {Blondin}, {Challis},
  {Berlind}, {Calkins}, {Esquerdo}, {Matheson}, {Modjaz}, {Rest}, \&
  {Kirshner}}]{2017ApJS..233....6H}
{Hicken}, M., {Friedman}, A.~S., {Blondin}, S., {et~al.} 2017, \apjs, 233, 6,
  \dodoi{10.3847/1538-4365/aa8ef4}

\bibitem[{Holoien {et~al.}(2019)Holoien, Bishop, Bersier, Strader, Chomiuk,
  Falco, Stritzinger, Holmbo, Morrell, Pojmanski, WoÅºniak, Bock, Shields,
  Vallely, Jayasinghe, Brown, Thompson, Kochanek, Stanek, Beacom, Cacella,
  Carballo, Cruz, Conseil, Farfan, Fernandez, Kiyota, Koff, Krannich, Marples,
  Masi, Monard, MuÃ±oz, Nicholls, Post, Stone, Trappett, Wiethoff, Shappee,
  Prieto, Chen, Bose, Dong, \& Brimacombe}]{10.1093/mnras/stz073}
Holoien, T. W.-S., Bishop, D.~W., Bersier, D., {et~al.} 2019, Monthly Notices
  of the Royal Astronomical Society, 484, 1899, \dodoi{10.1093/mnras/stz073}

\bibitem[{Hosseinzadeh {et~al.}(2020)Hosseinzadeh, Bravo, \&
  Pattnaik}]{griffin_hosseinzadeh_2020_specfit}
Hosseinzadeh, G., Bravo, T. E.~M., \& Pattnaik, R. 2020, Spectrum Fitting,
  v0.2.0,  Github.
\newblock \url{https://github.com/griffin-h/spectrum-fitting}

\bibitem[{Hosseinzadeh \& Gomez(2020)}]{griffin_hosseinzadeh_2020_4312178}
Hosseinzadeh, G., \& Gomez, S. 2020, Light Curve Fitting, v0.2.0,  Zenodo,
  \dodoi{10.5281/zenodo.4312178}

\bibitem[{{Ikeda} {et~al.}(2017){Ikeda}, {Kotake}, \&
  {Nakamura}}]{2017nuco.confb0109I}
{Ikeda}, E., {Kotake}, K., \& {Nakamura}, K. 2017, in 14th International
  Symposium on Nuclei in the Cosmos (NIC2016), ed. S.~{Kubono}, T.~{Kajino},
  S.~{Nishimura}, T.~{Isobe}, S.~{Nagataki}, T.~{Shima}, \& Y.~{Takeda},
  020109, \dodoi{10.7566/JPSCP.14.020109}

\bibitem[{{Janka}(2012)}]{janka_12}
{Janka}, H.-T. 2012, Annual Review of Nuclear and Particle Science, 62, 407,
  \dodoi{10.1146/annurev-nucl-102711-094901}

\bibitem[{Janka {et~al.}(2012)Janka, Hanke, HÃŒdepohl, Marek, MÃŒller, \&
  Obergaulinger}]{10.1093/ptep/pts067}
Janka, H.-T., Hanke, F., HÃŒdepohl, L., {et~al.} 2012, Progress of
  Theoretical and Experimental Physics, 2012, \dodoi{10.1093/ptep/pts067}

\bibitem[{{Janka} {et~al.}(2016){Janka}, {Melson}, \& {Summa}}]{janka_16}
{Janka}, H.-T., {Melson}, T., \& {Summa}, A. 2016, ArXiv e-prints.
\newblock \doarXiv{1602.05576}

\bibitem[{{Karachentsev} {et~al.}(2004){Karachentsev}, {Karachentseva},
  {Huchtmeier}, \& {Makarov}}]{karachentsev:04}
{Karachentsev}, I.~D., {Karachentseva}, V.~E., {Huchtmeier}, W.~K., \&
  {Makarov}, D.~I. 2004, \aj, 127, 2031

\bibitem[{{Kawahara} {et~al.}(2018){Kawahara}, {Kuroda}, {Takiwaki}, {Hayama},
  \& {Kotake}}]{kawahara_18}
{Kawahara}, H., {Kuroda}, T., {Takiwaki}, T., {Hayama}, K., \& {Kotake}, K.
  2018, \apj, 867, 126, \dodoi{10.3847/1538-4357/aae57b}

\bibitem[{Kawahara {et~al.}(2018)Kawahara, Kuroda, Takiwaki, Hayama, \&
  Kotake}]{Kawahara_2018}
Kawahara, H., Kuroda, T., Takiwaki, T., Hayama, K., \& Kotake, K. 2018, The
  Astrophysical Journal, 867, 126, \dodoi{10.3847/1538-4357/aae57b}

\bibitem[{Kistler {et~al.}(2013)Kistler, Haxton, \& YÃŒksel}]{Kistler_2013}
Kistler, M.~D., Haxton, W.~C., \& YÃŒksel, H. 2013, The Astrophysical
  Journal, 778, 81, \dodoi{10.1088/0004-637x/778/1/81}

\bibitem[{{Kotake}(2013)}]{2013CRPhy..14..318K}
{Kotake}, K. 2013, Comptes Rendus Physique, 14, 318,
  \dodoi{10.1016/j.crhy.2013.01.008}

\bibitem[{{Kotake} {et~al.}(2009){Kotake}, {Iwakami}, {Ohnishi}, \&
  {Yamada}}]{2009ApJ...704..951K}
{Kotake}, K., {Iwakami}, W., {Ohnishi}, N., \& {Yamada}, S. 2009, \apj, 704,
  951, \dodoi{10.1088/0004-637X/704/2/951}

\bibitem[{Kuroda {et~al.}(2017)Kuroda, Kotake, Hayama, \&
  Takiwaki}]{Kuroda_2017}
Kuroda, T., Kotake, K., Hayama, K., \& Takiwaki, T. 2017, The Astrophysical
  Journal, 851, 62, \dodoi{10.3847/1538-4357/aa988d}

\bibitem[{{Kuroda} {et~al.}(2012){Kuroda}, {Kotake}, \&
  {Takiwaki}}]{2012ApJ...755...11K}
{Kuroda}, T., {Kotake}, K., \& {Takiwaki}, T. 2012, \apj, 755, 11,
  \dodoi{10.1088/0004-637X/755/1/11}

\bibitem[{{Kuroda} {et~al.}(2016){Kuroda}, {Kotake}, \& {Takiwaki}}]{kuroda_16}
---. 2016, \apjl, 829, L14, \dodoi{10.3847/2041-8205/829/1/L14}

\bibitem[{Kuroda {et~al.}(2016)Kuroda, Kotake, \& Takiwaki}]{Kuroda_2016}
Kuroda, T., Kotake, K., \& Takiwaki, T. 2016, The Astrophysical Journal, 829,
  L14, \dodoi{10.3847/2041-8205/829/1/l14}

\bibitem[{{Li} {et~al.}(2011){Li}, {Leaman}, {Chornock}, {Filippenko},
  {Poznanski}, {Ganeshalingam}, {Wang}, {Modjaz}, {Jha}, {Foley}, \&
  {Smith}}]{li:11}
{Li}, W., {Leaman}, J., {Chornock}, R., {et~al.} 2011, \mnras, 412, 1441,
  \dodoi{10.1111/j.1365-2966.2011.18160.x}

\bibitem[{Lupton(2005)}]{lupton_transformations_2005}
Lupton, R. 2005, Transformations between {SDSS} magnitudes and other systems.
\newblock \url{https://www.sdss.org/dr16/algorithms/sdssubvritransform/}

\bibitem[{{Lynch} \& {LIGO/Virgo Collaboration}(2018)}]{2018APS..APRH14002L}
{Lynch}, R., \& {LIGO/Virgo Collaboration}. 2018, in APS Meeting Abstracts,
  Vol. 2018, APS April Meeting Abstracts, H14.002

\bibitem[{{Marek} {et~al.}(2009){Marek}, {Janka}, \&
  {M{\"u}ller}}]{2009A&A...496..475M}
{Marek}, A., {Janka}, H.~T., \& {M{\"u}ller}, E. 2009, \aap, 496, 475,
  \dodoi{10.1051/0004-6361/200810883}

\bibitem[{{Mattila} {et~al.}(2012){Mattila}, {Dahlen}, {Efstathiou}, {Kankare},
  {Melinder}, {Alonso-Herrero}, {P{\'e}rez-Torres}, {Ryder},
  {V{\"a}is{\"a}nen}, \& {{\"O}stlin}}]{mattila:12}
{Mattila}, S., {Dahlen}, T., {Efstathiou}, A., {et~al.} 2012, \apj, 756, 111,
  \dodoi{10.1088/0004-637X/756/2/111}

\bibitem[{{Matzner} \& {McKee}(1998)}]{Matzner:1998mg}
{Matzner}, C.~D., \& {McKee}, C.~F. 1998, in American Astronomical Society
  Meeting Abstracts, Vol. 193, American Astronomical Society Meeting Abstracts,
  118.03

\bibitem[{{Morozova} {et~al.}(2015){Morozova}, {Piro}, {Renzo}, {Ott},
  {Clausen}, {Couch}, {Ellis}, \& {Roberts}}]{morozova:2015}
{Morozova}, V., {Piro}, A.~L., {Renzo}, M., {et~al.} 2015, \apj, 814, 63

\bibitem[{{Morozova} {et~al.}(2017){Morozova}, {Piro}, \&
  {Valenti}}]{2017ApJ...838...28M}
{Morozova}, V., {Piro}, A.~L., \& {Valenti}, S. 2017, \apj, 838, 28,
  \dodoi{10.3847/1538-4357/aa6251}

\bibitem[{{Morozova} {et~al.}(2018){Morozova}, {Piro}, \&
  {Valenti}}]{2018ApJ...858...15M}
---. 2018, \apj, 858, 15, \dodoi{10.3847/1538-4357/aab9a6}

\bibitem[{Morozova {et~al.}(2018)Morozova, Radice, Burrows, \&
  Vartanyan}]{Morozova_2018}
Morozova, V., Radice, D., Burrows, A., \& Vartanyan, D. 2018, The Astrophysical
  Journal, 861, 10, \dodoi{10.3847/1538-4357/aac5f1}

\bibitem[{{M{\"u}ller} {et~al.}(2013){M{\"u}ller}, {Janka}, \&
  {Marek}}]{2013ApJ...766...43M}
{M{\"u}ller}, B., {Janka}, H.-T., \& {Marek}, A. 2013, \apj, 766, 43,
  \dodoi{10.1088/0004-637X/766/1/43}

\bibitem[{{M{\"u}ller} {et~al.}(2017){M{\"u}ller}, {Melson}, {Heger}, \&
  {Janka}}]{muller:2017}
{M{\"u}ller}, B., {Melson}, T., {Heger}, A., \& {Janka}, H.-T. 2017, \mnras,
  472, 491

\bibitem[{{Murphy} {et~al.}(2009){Murphy}, {Ott}, \&
  {Burrows}}]{2009ApJ...707.1173M}
{Murphy}, J.~W., {Ott}, C.~D., \& {Burrows}, A. 2009, \apj, 707, 1173,
  \dodoi{10.1088/0004-637X/707/2/1173}

\bibitem[{{Nagao} {et~al.}(2020){Nagao}, {Maeda}, \&
  {Ouchi}}]{2020MNRAS.497.5395N}
{Nagao}, T., {Maeda}, K., \& {Ouchi}, R. 2020, \mnras, 497, 5395,
  \dodoi{10.1093/mnras/staa2360}

\bibitem[{{Nakamura} {et~al.}(2014){Nakamura}, {Kuroda}, {Takiwaki}, \&
  {Kotake}}]{2014ApJ...793...45N}
{Nakamura}, K., {Kuroda}, T., {Takiwaki}, T., \& {Kotake}, K. 2014, \apj, 793,
  45, \dodoi{10.1088/0004-637X/793/1/45}

\bibitem[{{Negueruela} {et~al.}(2016){Negueruela}, {Dorda},
  {Gonz{\'a}lez-Fern{\'a}ndez}, \& {Marco}}]{2016IAUFM..29B.220N}
{Negueruela}, I., {Dorda}, R., {Gonz{\'a}lez-Fern{\'a}ndez}, C., \& {Marco}, A.
  2016, IAU Focus Meeting, 29B, 220, \dodoi{10.1017/S1743921316004993}

\bibitem[{{Nicholl} {et~al.}(2019{\natexlab{a}}){Nicholl}, {Short}, {Angus},
  {Muller}, \& {Yaron}}]{2019TNSCR.782....1N}
{Nicholl}, M., {Short}, P., {Angus}, C., {Muller}, T., \& {Yaron}, O.
  2019{\natexlab{a}}, TNSCR, 2019-782, 1

\bibitem[{{Nicholl} {et~al.}(2019{\natexlab{b}}){Nicholl}, {Short}, {Swann},
  {Angus}, {Smith}, \& {Yaron}}]{2019TNSCR.700....1N}
{Nicholl}, M., {Short}, P., {Swann}, E., {et~al.} 2019{\natexlab{b}}, TNSCR,
  2019-700, 1

\bibitem[{{O'Connor} \& {Couch}(2018)}]{oconnor_18}
{O'Connor}, E.~P., \& {Couch}, S.~M. 2018, The Astrophysical Journal, 865, 81,
  \dodoi{10.3847/1538-4357/aadcf7}

\bibitem[{{Ott}(2009)}]{2009CQGra..26f3001O}
{Ott}, C.~D. 2009, Classical and Quantum Gravity, 26, 063001,
  \dodoi{10.1088/0264-9381/26/6/063001}

\bibitem[{{Ott} {et~al.}(2012){Ott}, {Abdikamalov}, {O'Connor}, {Reisswig},
  {Haas}, {Kalmus}, {Drasco}, {Burrows}, \& {Schnetter}}]{2012PhRvD..86b4026O}
{Ott}, C.~D., {Abdikamalov}, E., {O'Connor}, E., {et~al.} 2012, \prd, 86,
  024026, \dodoi{10.1103/PhysRevD.86.024026}

\bibitem[{{Ott} {et~al.}(2013){Ott}, {Abdikamalov}, {M{\"o}sta}, {Haas},
  {Drasco}, {O'Connor}, {Reisswig}, {Meakin}, \&
  {Schnetter}}]{2013ApJ...768..115O}
{Ott}, C.~D., {Abdikamalov}, E., {M{\"o}sta}, P., {et~al.} 2013, \apj, 768,
  115, \dodoi{10.1088/0004-637X/768/2/115}

\bibitem[{{Pejcha} \& {Prieto}(2015{\natexlab{a}})}]{2015ApJ...799..215P}
{Pejcha}, O., \& {Prieto}, J.~L. 2015{\natexlab{a}}, \apj, 799, 215,
  \dodoi{10.1088/0004-637X/799/2/215}

\bibitem[{{Pejcha} \& {Prieto}(2015{\natexlab{b}})}]{2015ApJ...806..225P}
---. 2015{\natexlab{b}}, \apj, 806, 225, \dodoi{10.1088/0004-637X/806/2/225}

\bibitem[{{Powell} \& {M{\"u}ller}(2019)}]{powell_19}
{Powell}, J., \& {M{\"u}ller}, B. 2019, \mnras, 487, 1178,
  \dodoi{10.1093/mnras/stz1304}

\bibitem[{{Radice} {et~al.}(2019){Radice}, {Morozova}, {Burrows}, {Vartanyan},
  \& {Nagakura}}]{radice_19}
{Radice}, D., {Morozova}, V., {Burrows}, A., {Vartanyan}, D., \& {Nagakura}, H.
  2019, \apjl, 876, L9, \dodoi{10.3847/2041-8213/ab191a}

\bibitem[{Radice {et~al.}(2019)Radice, Morozova, Burrows, Vartanyan, \&
  Nagakura}]{Radice_2019}
Radice, D., Morozova, V., Burrows, A., Vartanyan, D., \& Nagakura, H. 2019, The
  Astrophysical Journal, 876, L9, \dodoi{10.3847/2041-8213/ab191a}

\bibitem[{Roulet {et~al.}(2019)Roulet, Dai, Venumadhav, Zackay, \&
  Zaldarriaga}]{PhysRevD.99.123022}
Roulet, J., Dai, L., Venumadhav, T., Zackay, B., \& Zaldarriaga, M. 2019, Phys.
  Rev. D, 99, 123022, \dodoi{10.1103/PhysRevD.99.123022}

\bibitem[{Roy {et~al.}(2019)Roy, Sengupta, \& Ajith}]{PhysRevD.99.024048}
Roy, S., Sengupta, A.~S., \& Ajith, P. 2019, Phys. Rev. D, 99, 024048,
  \dodoi{10.1103/PhysRevD.99.024048}

\bibitem[{Rozwadowska {et~al.}(2021)Rozwadowska, Vissani, \&
  Cappellaro}]{Rozwadowska:2020nab}
Rozwadowska, K., Vissani, F., \& Cappellaro, E. 2021, New Astron., 83, 101498,
  \dodoi{10.1016/j.newast.2020.101498}

\bibitem[{{Salvatier} {et~al.}(2015){Salvatier}, {Wiecki}, \&
  {Fonnesbeck}}]{2015arXiv150708050S}
{Salvatier}, J., {Wiecki}, T., \& {Fonnesbeck}, C. 2015, arXiv e-prints,
  arXiv:1507.08050.
\newblock \doarXiv{1507.08050}

\bibitem[{{Sand} {et~al.}(2018){Sand}, {Valenti}, {Tartaglia}, {Yang}, \&
  {Wyatt}}]{2018AAS...23124511S}
{Sand}, D., {Valenti}, S., {Tartaglia}, L., {Yang}, S., \& {Wyatt}, S. 2018, in
  American Astronomical Society Meeting Abstracts, Vol. 231, American
  Astronomical Society Meeting Abstracts \#231, 245.11

\bibitem[{{Sapir} \& {Waxman}(2017)}]{2017ApJ...838..130S}
{Sapir}, N., \& {Waxman}, E. 2017, \apj, 838, 130,
  \dodoi{10.3847/1538-4357/aa64df}

\bibitem[{Schawinski {et~al.}(2008)}]{SNLS:2008bsx}
Schawinski, K., {et~al.} 2008, Science, 321, 223,
  \dodoi{10.1126/science.1160456}

\bibitem[{{Scheidegger} {et~al.}(2010){Scheidegger}, {Whitehouse},
  {K{\"a}ppeli}, \& {Liebend{\"o}rfer}}]{2010CQGra..27k4101S}
{Scheidegger}, S., {Whitehouse}, S.~C., {K{\"a}ppeli}, R., \&
  {Liebend{\"o}rfer}, M. 2010, Classical and Quantum Gravity, 27, 114101,
  \dodoi{10.1088/0264-9381/27/11/114101}

\bibitem[{{Schlafly} \& {Finkbeiner}(2011)}]{2011ApJ...737..103S}
{Schlafly}, E.~F., \& {Finkbeiner}, D.~P. 2011, \apj, 737, 103,
  \dodoi{10.1088/0004-637X/737/2/103}

\bibitem[{{Schmidt} {et~al.}(1994){Schmidt}, {Kirshner}, {Eastman}, {Phillips},
  {Suntzeff}, {Hamuy}, {Maza}, \& {Aviles}}]{1994ApJ...432...42S}
{Schmidt}, B.~P., {Kirshner}, R.~P., {Eastman}, R.~G., {et~al.} 1994, \apj,
  432, 42, \dodoi{10.1086/174546}

\bibitem[{{Singh} {et~al.}(2019){Singh}, {Kumar}, {Moriya}, {Anupama}, {Sahu},
  {Brown}, {Andrews}, \& {Smith}}]{2019ApJ...882...68S}
{Singh}, A., {Kumar}, B., {Moriya}, T.~J., {et~al.} 2019, \apj, 882, 68,
  \dodoi{10.3847/1538-4357/ab3050}

\bibitem[{Sotani {et~al.}(2017)Sotani, Kuroda, Takiwaki, \&
  Kotake}]{PhysRevD.96.063005}
Sotani, H., Kuroda, T., Takiwaki, T., \& Kotake, K. 2017, Phys. Rev. D, 96,
  063005, \dodoi{10.1103/PhysRevD.96.063005}

\bibitem[{{Stanek}(2019)}]{2019TNSTR1030....1S}
{Stanek}, K.~Z. 2019, TNSTR, 2019-1030, 1

\bibitem[{{Stanek} {et~al.}(2019){Stanek}, {Bersier}, \&
  {Kochanek}}]{2019TNSTR.766....1S}
{Stanek}, K.~Z., {Bersier}, D., \& {Kochanek}, C.~S. 2019, TNSTR, 2019-766, 1

\bibitem[{{Szczepa{\'n}czyk} {et~al.}(2021){Szczepa{\'n}czyk}, {Antelis},
  {Benjamin}, {Cavagli{\`a}}, {Gondek-Rosi{\'n}ska}, {Hansen}, {Klimenko},
  {Morales}, {Moreno}, {Mukherjee}, {Nurbek}, {Powell}, {Singh},
  {Sitmukhambetov}, {Szewczyk}, {Valdez}, {Vedovato}, {Westhouse}, {Zanolin},
  \& {Zheng}}]{Szczepanczyk:2021bka}
{Szczepa{\'n}czyk}, M.~J., {Antelis}, J.~M., {Benjamin}, M., {et~al.} 2021,
  \prd, 104, 102002, \dodoi{10.1103/PhysRevD.104.102002}

\bibitem[{{Takiwaki} \& {Kotake}(2018)}]{takiwaki_18}
{Takiwaki}, T., \& {Kotake}, K. 2018, \mnras, 475, L91,
  \dodoi{10.1093/mnrasl/sly008}

\bibitem[{Takiwaki \& Kotake(2018)}]{10.1093/mnrasl/sly008}
Takiwaki, T., \& Kotake, K. 2018, Monthly Notices of the Royal Astronomical
  Society: Letters, 475, L91, \dodoi{10.1093/mnrasl/sly008}

\bibitem[{{Tammann} {et~al.}(1994){Tammann}, {Loeffler}, \&
  {Schroeder}}]{1994ApJS...92..487T}
{Tammann}, G.~A., {Loeffler}, W., \& {Schroeder}, A. 1994, \apjs, 92, 487,
  \dodoi{10.1086/192002}

\bibitem[{{Theureau} {et~al.}(2007){Theureau}, {Hanski}, {Coudreau}, {Hallet},
  \& {Martin}}]{2007A&A...465...71T}
{Theureau}, G., {Hanski}, M.~O., {Coudreau}, N., {Hallet}, N., \& {Martin},
  J.~M. 2007, \aap, 465, 71, \dodoi{10.1051/0004-6361:20066187}

\bibitem[{{Tonry} {et~al.}(2019{\natexlab{a}}){Tonry}, {Denneau}, {Heinze},
  {Weiland}, {Flewelling}, {Stalder}, {Rest}, {Stubbs}, {Smith}, {Smartt},
  {Young}, {Srivastav}, {McBrien}, {O'Neill}, {Clark}, {Fulton}, {Gillanders},
  {McCormack}, \& {Wright}}]{2019TNSTR.862....1T}
{Tonry}, J., {Denneau}, L., {Heinze}, A., {et~al.} 2019{\natexlab{a}}, TNSTR,
  2019-862, 1

\bibitem[{{Tonry} {et~al.}(2019{\natexlab{b}}){Tonry}, {Denneau}, {Heinze},
  {Weiland}, {Flewelling}, {Stalder}, {Rest}, {Stubbs}, {Smith}, {Smartt},
  {Young}, {Srivastav}, {McBrien}, {O'Neill}, {Clark}, {Fulton}, {Gillanders},
  {McCormack}, \& {Wright}}]{2019TNSTR.687....1T}
---. 2019{\natexlab{b}}, TNSTR, 2019-687, 1

\bibitem[{{Usman} {et~al.}(2016){Usman}, {Nitz}, {Harry}, {Biwer}, {Brown},
  {Cabero}, {Capano}, {Dal Canton}, {Dent}, {Fairhurst}, {Kehl}, {Keppel},
  {Krishnan}, {Lenon}, {Lundgren}, {Nielsen}, {Pekowsky}, {Pfeiffer},
  {Saulson}, {West}, \& {Willis}}]{2016CQGra..33u5004U}
{Usman}, S.~A., {Nitz}, A.~H., {Harry}, I.~W., {et~al.} 2016, Classical and
  Quantum Gravity, 33, 215004, \dodoi{10.1088/0264-9381/33/21/215004}

\bibitem[{{Valenti} {et~al.}(2014){Valenti}, {Sand}, {Pastorello}, {Graham},
  {Howell}, {Parrent}, {Tomasella}, {Ochner}, {Fraser}, {Benetti}, {Yuan},
  {Smartt}, {Maund}, {Arcavi}, {Gal-Yam}, {Inserra}, \&
  {Young}}]{2014MNRAS.438L.101V}
{Valenti}, S., {Sand}, D., {Pastorello}, A., {et~al.} 2014, \mnras, 438, L101,
  \dodoi{10.1093/mnrasl/slt171}

\bibitem[{{Valenti} {et~al.}(2016){Valenti}, {Howell}, {Stritzinger}, {Graham},
  {Hosseinzadeh}, {Arcavi}, {Bildsten}, {Jerkstrand}, {McCully}, {Pastorello},
  {Piro}, {Sand}, {Smartt}, {Terreran}, {Baltay}, {Benetti}, {Brown},
  {Filippenko}, {Fraser}, {Rabinowitz}, {Sullivan}, \&
  {Yuan}}]{2016MNRAS.459.3939V}
{Valenti}, S., {Howell}, D.~A., {Stritzinger}, M.~D., {et~al.} 2016, \mnras,
  459, 3939, \dodoi{10.1093/mnras/stw870}

\bibitem[{{Vallely} {et~al.}(2021){Vallely}, {Kochanek}, {Stanek}, {Fausnaugh},
  \& {Shappee}}]{2021MNRAS.500.5639V}
{Vallely}, P.~J., {Kochanek}, C.~S., {Stanek}, K.~Z., {Fausnaugh}, M., \&
  {Shappee}, B.~J. 2021, \mnras, 500, 5639, \dodoi{10.1093/mnras/staa3675}

\bibitem[{{van den Bergh} {et~al.}(1987){van den Bergh}, {McClure}, \&
  {Evans}}]{1987ApJ...323...44V}
{van den Bergh}, S., {McClure}, R.~D., \& {Evans}, R. 1987, \apj, 323, 44,
  \dodoi{10.1086/165806}

\bibitem[{{van den Bergh} \& {Tammann}(1991)}]{vandenbergh:91}
{van den Bergh}, S., \& {Tammann}, G.~A. 1991, \araa, 29, 363,
  \dodoi{10.1146/annurev.aa.29.090191.002051}

\bibitem[{{van den Broeck} {et~al.}(2009){van den Broeck}, {Brown}, {Cokelaer},
  {Harry}, {Jones}, {Sathyaprakash}, {Tagoshi}, \&
  {Takahashi}}]{2009PhRvD..80b4009V}
{van den Broeck}, C., {Brown}, D.~A., {Cokelaer}, T., {et~al.} 2009, \prd, 80,
  024009, \dodoi{10.1103/PhysRevD.80.024009}

\bibitem[{{Vartanyan} {et~al.}(2019){Vartanyan}, {Burrows}, \&
  {Radice}}]{vartanyan_19}
{Vartanyan}, D., {Burrows}, A., \& {Radice}, D. 2019, \mnras, 489, 2227,
  \dodoi{10.1093/mnras/stz2307}

\bibitem[{{Vink{\'o}} \& {Tak{\'a}ts}(2007)}]{2007AIPC..937..394V}
{Vink{\'o}}, J., \& {Tak{\'a}ts}, K. 2007, in American Institute of Physics
  Conference Series, Vol. 937, Supernova 1987A: 20 Years After: Supernovae and
  Gamma-Ray Bursters, ed. S.~{Immler}, K.~{Weiler}, \& R.~{McCray}, 394--398,
  \dodoi{10.1063/1.3682935}

\bibitem[{Waxman \& Katz(2017)}]{2017hsn..book..967W}
Waxman, E., \& Katz, B. 2017, Handbook of Supernovae, 967–1015,
  \dodoi{10.1007/978-3-319-21846-5_33}

\bibitem[{{Yakunin} {et~al.}(2010){Yakunin}, {Marronetti}, {Mezzacappa},
  {Bruenn}, {Lee}, {Chertkow}, {Hix}, {Blondin}, {Lentz}, {Messer}, \&
  {Yoshida}}]{2010CQGra..27s4005Y}
{Yakunin}, K.~N., {Marronetti}, P., {Mezzacappa}, A., {et~al.} 2010, Classical
  and Quantum Gravity, 27, 194005, \dodoi{10.1088/0264-9381/27/19/194005}

\bibitem[{Yakunin {et~al.}(2015)Yakunin, Mezzacappa, Marronetti, Yoshida,
  Bruenn, Hix, Lentz, Bronson~Messer, Harris, Endeve, Blondin, \&
  Lingerfelt}]{PhysRevD.92.084040}
Yakunin, K.~N., Mezzacappa, A., Marronetti, P., {et~al.} 2015, Phys. Rev. D,
  92, 084040, \dodoi{10.1103/PhysRevD.92.084040}

\bibitem[{{Yaron} \& {Gal-Yam}(2012)}]{2012PASP..124..668Y}
{Yaron}, O., \& {Gal-Yam}, A. 2012, \pasp, 124, 668, \dodoi{10.1086/666656}

\bibitem[{{Zackay} {et~al.}(2016){Zackay}, {Ofek}, \&
  {Gal-Yam}}]{2016ApJ...830...27Z}
{Zackay}, B., {Ofek}, E.~O., \& {Gal-Yam}, A. 2016, \apj, 830, 27,
  \dodoi{10.3847/0004-637X/830/1/27}

\end{thebibliography}

\end{document}